\documentclass[onecolumn,numberedappendix]{emulateapj}
\usepackage{graphicx}
\usepackage{appendix}

\def\er{$\pm$}

\begin{document}

\title{Systematic Uncertainties in the Spectroscopic Measurements of
  Neutron-Star Masses and Radii from Thermonuclear X-ray
  Bursts. I. Apparent Radii}

\author{Tolga G\"uver\altaffilmark{\dag}, Dimitrios     Psaltis,  \& Feryal
  \"Ozel}

\affil{Department  of  Astronomy, University  of  Arizona, 933
  N. Cherry Ave., Tucson, AZ 85721}

\altaffiltext{\dag}{Current Address : Sabanc\i ~University, Faculty
  of  Engineering and  Natural Sciences,  Orhanl\i ~- Tuzla,  Istanbul
  34956, Turkey}

\begin{abstract}  
  The masses and radii of low-magnetic field neutron stars can be
  measured by combining different observable quantities obtained from
  their X-ray spectra during thermonuclear X-ray bursts. One of these
  quantities is the apparent radius of each neutron star as inferred
  from the X-ray flux and spectral temperature measured during the
  cooling tails of bursts, when the thermonuclear flash is believed to
  have engulfed the entire star.  In this paper, we analyze 13,095
  X-ray spectra of 446 X-ray bursts observed from 12 sources in order
  to assess possible systematic effects in the measurements of the
  apparent radii of neutron stars. We first show that the vast
  majority of the observed X-ray spectra are consistent with blackbody
  functions to within a few percent.  We find that most X-ray bursts
  follow a very well determined correlation between X-ray flux and
  temperature, which is consistent with the whole neutron-star surface
  emitting uniformly during the cooling tails.  We develop a Bayesian
  Gaussian mixture algorithm to measure the apparent radii of the
  neutron stars in these sources, while detecting and excluding a
  small number of X-ray bursts that show irregular cooling behavior.
  This algorithm also provides us with a quantitative measure of the
  systematic uncertainties in the measurements.  We find that those
  errors in the spectroscopic determination of neutron-star radii that
  are introduced by systematic effects in the cooling tails of X-ray
  bursts are in the range $\simeq 3-8$\%. Such small errors are
  adequate to distinguish between different equations of state
  provided that uncertainties in the distance to each source and the
  absolute calibration of X-ray detectors do not dominate the error
  budget.
\end{abstract}


\section{Introduction}

The thermal spectra of neutron stars during thermonuclear X-ray bursts
have been used  during the last three decades  in numerous attempts to
measure the  neutron-star masses and  radii (e.g., van  Paradijs 1978,
1979; Foster, Fabian  \& Ross 1986; Sztajno et  al.\ 1987; van Paradijs
\&  Lewin 1987; Damen  et al.\  1989, 1990).  Such efforts  were often
hampered by large systematic uncertainties in the estimated distances
to the bursters and in the theoretical models for their X-ray spectra.
Moreover, the relatively small number of X-ray bursts observed by
early satellites from each individual source made it impossible to
assess systematic uncertainties related to the degree of anisotropy of
the thermonuclear burning on the neutron-star surface, or of the
obscuration by and the reflection off the accretion flow.

The situation has changed dramatically in the last few years. The
distances to several globular clusters that contain bursting neutron
stars has been narrowed down with the use of observations by the
Hubble space telescope (see, e.g., Kuulkers et al.\ 2003; \"Ozel,
G\"uver, \& Psaltis 2009; G\"uver et al.\ 2010b). The distances to
X-ray bursters in the Galactic disk and halo have also been determined
using alternate methods (e.g., G\"uver et al.\ 2010a). Theoretical
models of the X-ray spectra of bursting neutron stars have been
greatly improved and can account for the subtle effects of the
presence of heavy metals in their atmospheres (e.g., London, Taam, \&
Howard 1986; Foster, Fabian, \& Ross 1986; Ebisuzaki 1987; Madej,
Joss, \& R{\'o}{\.z}a{\'n}ska 2004; Majczyna et al.\ 2005; Suleimanov,
Poutanen, \& Werner 2011).  Finally, the high signal-to-noise
observations of the X-ray spectra of several hundreds of bursts with
the Rossi X-ray Timing Explorer (Galloway et al.\ 2008a) allow for the
formal uncertainties of individual measurements to be substantially
reduced.

The combination of these developments led recently to the measurement
of the masses and radii of several neutron stars (\"Ozel et al.\ 2009;
G\"uver et al.\ 2010a, 2010b; Steiner, Lattimer, \& Brown 2010) which
are already sufficient to provide broad constraints on the equation of
state of neutron-star matter (\"Ozel, Baym, \& G\"uver 2010). In some
cases, the formal uncertainties in the spectroscopic measurements are
as low as a few percent (G\"uver et al.\ 2010b) suggesting that their
accuracy might be limited by systematic effects rather than by
counting statistics.

In this series of articles, we use the large database of X-ray bursts
observed with the RXTE in order to assess the systematic effects in
various spectroscopic measurements of their properties.  In the first
article, we focus on the measurements of the apparent surface areas of
12 neutron stars as inferred from their X-ray spectra during the
cooling tails of the bursts.  Our goal is to quantify the degree to
which: {\em (i)\/} the X-ray burst spectra observed in the RXTE energy
range can be described by blackbody functions (the so-called color
correction arising from atmospheric effects are then applied a
posteriori); {\em (ii)} the entire surface area of each neutron star
burns practically uniformly during the cooling of the bursts; {\em
(iii)\/} the accretion flows make minor contributions to the emission
during the bursts.

\section{Sources, Observations, and Data Analysis}

\subsection{Source and Burst Selection}

\begin{deluxetable}{lccccc}
\tablecolumns{6}
\tablewidth{370pt}
\tablecaption{X-RAY BURSTERS}

\tablehead{\colhead{Name} & \colhead{RA} & \colhead{DEC} & \colhead{Number} &
\colhead{N$_{\rm H}$} & \colhead{ N$_{\rm H}$} \\
\colhead{} & \colhead{} & \colhead{} & \colhead{of Bursts} & \colhead{(10$^{22}$ cm$^{-2}$)} & \colhead{Method\tablenotemark{a}}}

\startdata
  4U~0513$-$40          & 05 14 06.60 & $-$40 02 37.0 & 6 & 0.014$^1$ &  GC\tablenotemark{b} \\
  4U~1608$-$52          & 16 12 43.00 & $-$52 25 23.0 & 26 & 1.08$\pm$0.16$^2$ & X-ray edges\tablenotemark{c}\\
  4U~1636$-$53          & 16 40 55.50 & $-$53 45 05.0 & 162 & 0.44$^3$ & X-ray edges\tablenotemark{c}\\
  4U~1702$-$429         & 17 06 15.31 & $-$43 02 08.7 & 46  & 1.95 & X-ray continuum\tablenotemark{d}\\
  4U~1705$-$44          & 17 08 54.47 & $-$44 06 07.4 & 44  & 2.44$\pm$0.09$^{4}$ & X-ray edges\tablenotemark{c}\\
  4U~1724$-$307         & 17 27 33.20 & $-$30 48 07.0 & 3  & 1.08$^1$ & GC\tablenotemark{b}\\
  4U~1728$-$34          & 17 31 57.40 & $-$33 50 05.0 & 90 & 2.49 $\pm$0.14$^4$ & X-ray edges\tablenotemark{c}\\
  KS~1731$-$260         & 17 34 12.70 & $-$26 05 48.5 & 24  & 2.98 & X-ray continuum\tablenotemark{d}\\
  4U~1735$-$44          & 17 38 58.30 & $-$44 27 00.0 & 6  & 0.28$^{3}$ & X-ray edges\tablenotemark{c}\\
  EXO~1745$-$248        & 17 48 56.00 & $-$24 53 42.0 & 22  & 1.4$\pm$0.45$^5$ & X-ray continuum\tablenotemark{d}\\
  4U~1746$-$37          & 17 50 12.7 &  $-$37 03 08.0 & 7 & 0.36$^6$ & GC\tablenotemark{b}\\
  SAX~J1748.9$-$2021    & 17 48 52.16 & $-$20 21 32.4 & 4 & 0.79$^6$ & GC\tablenotemark{b}\\
  SAX~J1750.8$-$2900    & 17 50 24.00 & $-$29 02 18.0 & 4  & 4.97 & X-ray continuum\tablenotemark{d}\\
  4U~1820$-$30          & 18 23 40.45 & $-$30 21 40.1 & 5  & 0.25 $\pm$0.03$^7$ & X-ray edges\tablenotemark{c}\\
  AQL~X$-$1             & 19 11 16.05 & $+$00 35 05.8 & 51  & 0.34$\pm$0.07$^8$ & Counterpart\tablenotemark{e}
\enddata
\tablenotetext{a}{References : (1)~Harris 1996; (2)~G\"uver et al.\ 2010a
(3)~Juett et al. (2004, 2006); (4)~Wroblewski et al. 2008;
(5)~Wijnands et al.\ 2005; (6)~Valenti et al.\ 2007;
(7)~G\"uver et al. 2010b; (8)~Chevalier et al.\ 1999}
\tablenotetext{b}{Optical/IR observations of the globular cluster}
\tablenotetext{c}{High resolution spectroscopy of X-ray absorption edges}
\tablenotetext{d}{Average of continuum X-ray spectroscopy}
\tablenotetext{e}{Optical spectroscopy of the counterpart}
\label{sourcestable}
\end{deluxetable}

We  base  our  study  on  the  X-ray  burst  catalog  of  Galloway  et
al.\ (2008a). We  chose 12 out of the 48 sources  in the catalog based
on the following criteria:

\noindent {\em  (i)\/} We  considered sources that  show at  least two
bursts with evidence for photospheric radius expansion, based on the
definition of the latter used by Galloway et al.\ (2008a).  This
requirement arises from our ultimate aim, which is to measure both the
mass and the radius of the neutron star in each system using a
combination of spectroscopic phenomena (as in, e.g., \"Ozel et
al. 2009).

\noindent  {\em (ii)\/}  We excluded  dippers, ADC  sources,  or known
high-inclination sources.  This list includes EXO~0748$-$676,
MXB~1659$-$298, 4U~1916$-$05, GRS~1747$-$312, 4U~1254$-$69, and
4U~1710$-$281, for which it was shown that geometric effects related
to obscuration or reflection significantly affect the flux from the
stellar surface that is measured by a distant observer (Galloway,
\"Ozel, \& Psaltis 2008b).

\noindent  {\em (iii)\/}  We did  not consider  the  known millisecond
pulsars SAX~J1808.4$-$3658 and HETE~J1900.1$-$2455 because the
presence of pulsations in their persistent emission implies that their
magnetic fields are dynamically important and, therefore, may affect
the properties of X-ray bursts.

\noindent  {\em (iv)}  We excluded  the sources  GRS~1741.9$-$2853 and
2E~1742.9$-$2929 as well as a small number of bursts from Aql~X-1,
4U~1728$-$34, and 4U~1746$-$37, for which there is substantial
evidence that their emission is affected by source confusion (Galloway
et al.\ 2008a; Keek et al. 2010).

\noindent {\em (v)\/}  For each source, we considered  only bursts for
which the flux in the persistent emission prior to the burst is at
most 10\% of the inferred Eddington flux for each source, i.e.,
$\gamma\equiv F_{\rm per}/F_{\rm Edd}<0.1$, as calculated by Galloway
et al.\ (2008a). Because we subtract the pre-burst persistent emission
from the decay spectrum of each X-ray burst, this requirement reduces
substantially the systematic uncertainties introduced by potential
changes in the emission from the accretion flow during the burst.

Table \ref{sourcestable} lists all the X-ray bursters that fulfill the
above requirements, together with the number of bursts observed by
RXTE for each source. This is the complete list of sources for which
the masses and radii can be measured, in principle, using the
spectroscopic method of \"Ozel et al.\ (2009), with currently
available data.  Our analyses of the bursts from EXO~1745$-$248,
4U~1608$-$52, and 4U~1820$-$30 have been reported elsewhere (\"Ozel et
al.\ 2009; G\"uver et al.\ 2010a, 2010b) and will not be repeated
here.

\subsection{Data Analysis}

We analyzed the burst data for the 12 sources shown in
Table~\ref{sourcestable} following the method outlined in Galloway et
al.\ (2008a; see also \"Ozel et al.  2009; G\"uver et al.\ 2010a,
2010b).

For each source, we extracted time resolved 2.5$-$25.0~keV X-ray
spectra using the {\it seextrct} ftool for the science event mode data
and the {\it saextrct} ftool for the science array mode data from all
of the RXTE/PCA layers.  Science mode observations provide high
count-rate data with a nominal time resolution of 125$\mu$s in 64
spectral channels over the whole energy range (2$-$60~keV) of the PCA
detector.  Following Galloway et al.\ 2008a, we extracted spectra
integrated over 0.25, 0.5, 1, and 2~s time intervals, depending on the
source count rate during the burst, so that the total number of counts
in each spectrum is roughly constant. (In a few cases, data gaps
during the observations result in X-ray spectra integrated over
shorter exposure times). We took the spectrum over a 16~s time
interval prior to the onset of each burst as the spectrum of the
persistent emission, which we subtracted from the burst spectra as
background.

We generated separate response matrix files for each burst using the
PCARSP version 11.7, HEASOFT release 6.7, and HEASARC's remote
calibration database and took into account the offset pointing of the
PCA during the creation of the response matrix files.  This latest
version of the PCA response matrix makes the instrument calibration
self-consistent over the PCA lifetime and yields a normalization of
the Crab pulsar that is within 1$-\sigma$ of the calibration
measurement of Toor \& Seward (1974) for that source.  In \S 4, we
discuss in some detail the effect of the uncertainties in the absolute
flux calibration on the measurement of the apparent surface area of
neutron stars. Finally, we corrected all of the X-ray spectra for PCA
deadtime following the method suggested by the RXTE
team\footnote[1]{ftp://legacy.gsfc.nasa.gov/xte/doc/cook\_book/pca\_deadtime.ps}.

To analyze the spectra, we used the Interactive Spectral
Interpretation System (ISIS), version 1.4.9-55 (Houck \& Denicola
2000).  For each fit, we included a systematic error of 0.5\% as
suggested by the RXTE calibration
team\footnote[2]{http://www.universe.nasa.gov/xrays/programs/rxte/pca/doc/rmf/pcarmf-11.7/}.

We fit each spectrum with a blackbody function using the {\it
bbodyrad} model (as defined in XSPEC, Arnaud 1996) and multiplied it
by the {\it tbabs} model (Wilms, Allen, \& McCray 2000) that takes
into account the interstellar extinction, assuming ISM abundances. The
model of the X-ray spectrum for each burst, therefore, depends on only
three parameters: the color temperature of the blackbody, $T_{\rm c}$,
the normalization of the blackbody spectrum, $A$, and the equivalent
hydrogen column density, $N_{\rm H}$ of the interstellar extinction.

Allowing the hydrogen column density $N_{\rm H}$ to be a free
parameter in the fitting procedure leads to correlated uncertainties
between the amount of interstellar extinction and the temperature of
the blackbody. Moreover, for every 10$^{22}$~cm$^{-2}$ overestimation
(underestimation) of the column density, the inferred flux of the
thermal emission is systematically larger (smaller) by $\approx$
~10$^{-9}$~erg~cm$^{-2}$~s$^{-1}$~(Galloway et al.\ 2008a). Reducing
the uncertainties related to the interstellar extinction for each
burster is, therefore, very important in controlling systematic
effects.

A recent analysis of high resolution grating spectra from a number of
X-ray binaries (Miller, Cackett, \& Reis 2009) showed that the
individual photoelectric absorption edges observed in the X-ray
spectra do not show significant variations with source luminosity or
spectral state.  This result strongly suggests that the neutral
hydrogen column density is dominated by absorption in the interstellar
medium and does not change on short timescales. In order to reduce
this systematic uncertainty and given the fact that there is no
evidence of variable neutral absorption for each system, we fixed the
hydrogen column density for each source to a constant value that we
obtained in one of the following ways.  {\it (i)} For a number of
sources, the equivalent hydrogen column density was inferred
independently using high-resolution spectrographs. {\it (ii)} In cases
where only an optical extinction or reddening measurement exists, we
used the relation given by G\"uver \& \"Ozel (2009) to convert it to
the equivalent hydrogen column density. {\it (iii)} Finally, for three
sources (4U~1702$-$429, KS~1731$-$260, SAX~J1750.8$-$2900) there are
no independent hydrogen column density measurements published in the
literature. In these cases, we fit the X-ray spectra obtained during
all the X-ray bursts of each source allowing the N$_{\rm H}$ value to
vary in a wide range.  We then found the resulting mean value of
N$_{\rm H}$ and used this as a constant in our second set of fits to
all the X-ray bursts. In Table~\ref{sourcestable}, we show the adopted
values for the hydrogen column density together with the measurement
uncertainties, the method with which the values were estimated, and
the appropriate references.  Future observations of these sources with
X-ray grating spectrometers onboard Chandra and XMM$-${\it Newton} can
help decrease the uncertainty arising from the lack of knowledge of
the properties of the interstellar matter towards these sources.

Our goal in this article is to study potential systematic
uncertainties in the measurement of the apparent area of each neutron
star during the cooling tails of X-ray bursts.  Hereafter, we adopt
the following working definition of the cooling tail.  It is the time
interval during which the inferred flux is lower than the peak flux of
the burst or the touchdown flux for photospheric radius expansion
bursts.  For the purpose of this definition, we use as a touchdown
point the first moment at which the blackbody temperature reaches its
highest value and the inferred apparent radius is lowest.  In order to
control the countrate statistics, we also set a lower limit on the
thermal flux during each cooling tail of
5$\times$10$^{-9}$~erg~s$^{-1}$~cm$^{-2}$ (or 5$\times$10$^{-10}$ for
the exceptionally faint sources 4U~1746$-$37 and 4U~0513$-$401).

In \S 3 and 4, we discuss in detail our approach of quantifying the
systematic uncertainties in the measurements of the apparent radii
during the cooling tails of thermonuclear bursts, using the sources
KS~1731$-$260, 4U~1728$-$34, and 4U~1636$-$536 as case studies.  For
another analysis of the cooling tails of X-ray bursts from
4U~1636$-$536, see Zhang, Mendez, \& Altamirano (2011). In \S5, we
repeat this procedure systematically for seven additional sources from
Table~\ref{sourcestable}.  Finally, the cooling tails of the bursts
observed from 4U~0513$-$40 and Aql~X-1 show irregular behavior, and we
report our analysis of them in the Appendix.

\section{Systematic Uncertainties in the Spectral Shapes}

Our first working hypothesis is that the spectra of neutron stars
during the cooling tails of thermonuclear bursts can be modeled by
blackbody functions in the observed energy range.  The fact that X-ray
spectra can be described well with blackbody functions has been
established since the first time resolved X-ray spectral studies of
thermonuclear bursts (see, e.g., Swank et al.\ 1977; Lewin, van
Paradijs, \& Taam 1993; Galloway et al.\ 2008a and references
therein).  Under that assumption, the temperature measured by fitting
blackbodies to the spectra are then corrected for the effects of the
atmosphere by applying a numerical prefactor called the color
correction factor $f_{\rm c}\equiv T_{\rm c}/T_{\rm eff}$, which is
the ratio of the color to the effective temperature.  This approach is
expected to be only an approximation for a number of reasons.

First, theoretical models of the atmospheres of neutron stars in
radiative equilibrium invariably show that the emerging spectra are
broader than blackbodies, especially at low temperatures and towards
low photon energies (e.g., Madej et al. 2004; Majczyna et al.\ 2005).
Moreover, for neutron stars that are rapidly spinning, the spectra
measured by an observer at infinity will be broader and more
asymmetric with respect to those at the stellar surface (\"Ozel \&
Psaltis 2003). The expected degree of broadening and asymmetry will be
at most comparable to
\begin{equation}
\frac{u}{c}\simeq \frac{2\pi \nu_{\rm s}R}{c}= 0.12
\left(\frac{\nu_{\rm s}}{600~{\rm Hz}}\right) \left(\frac{R}{10~{\rm
    km}}\right)\;,
\end{equation}
where $\nu_{\rm s}$ is the spin frequency of the neutron star and $R$
is its radius.  This is an uncertainty that can, in principle, be
corrected for by fitting the X-ray data with theoretical models of the
X-ray spectra emerging from the atmospheres of rapidly spinning
neutron stars.

Second, the accretion flow in the vicinity of the neutron star may
alter in a number of ways the observed spectrum. It may act as a
mirror, reflecting a fraction of the emerging radiation towards the
observer and, therefore, increasing the apparent surface area of
emission. It might occult a fraction of the stellar surface, reducing
the apparent surface area of emission (see Galloway et al.\ 2008b).
If the accretion flow is surrounded by a hot corona, Comptonization of
the surface emission will alter its spectrum as well as the relation
between energy flux and apparent surface area (e.g., Boutloukos,
Miller, \& Lamb 2010). Finally, the residual emission of the accretion
flow, which we attempt to remove by subtracting the pre-burst spectrum
of the source, might change during the duration of the burst,
introducing systematic changes in the net spectrum.

All of the above effects have the potential of changing the shape of
the X-ray spectrum of a burster and, perhaps more importantly, lead to
stochastic changes in the inferred apparent radii. Their overall
effect, however, is expected to be significantly reduced by the fact
that the intense radiation field during each X-ray burst should either
disrupt the inner accretion flow or cool any corona to the Compton
temperature of the radiation. Such a phenomenon has been observed
during a superburst from 4U~1820$-$30 (Ballantyne \& Strohmayer
2004). Moreover, these effects are expected to be the strongest for
high inclination systems (as was inferred by, e.g., Galloway et al.\
2008b). This is why we not only limited our sample to bursts for which
the ratio of the persistent flux prior to the burst to the Eddington
flux $\gamma \le 10\%$ in order to minimize accretion-related
uncertainties but also selected the known high inclination sources out
of our sample (see \S2).

\begin{figure}[t]
\centering
   \includegraphics[scale=0.29]{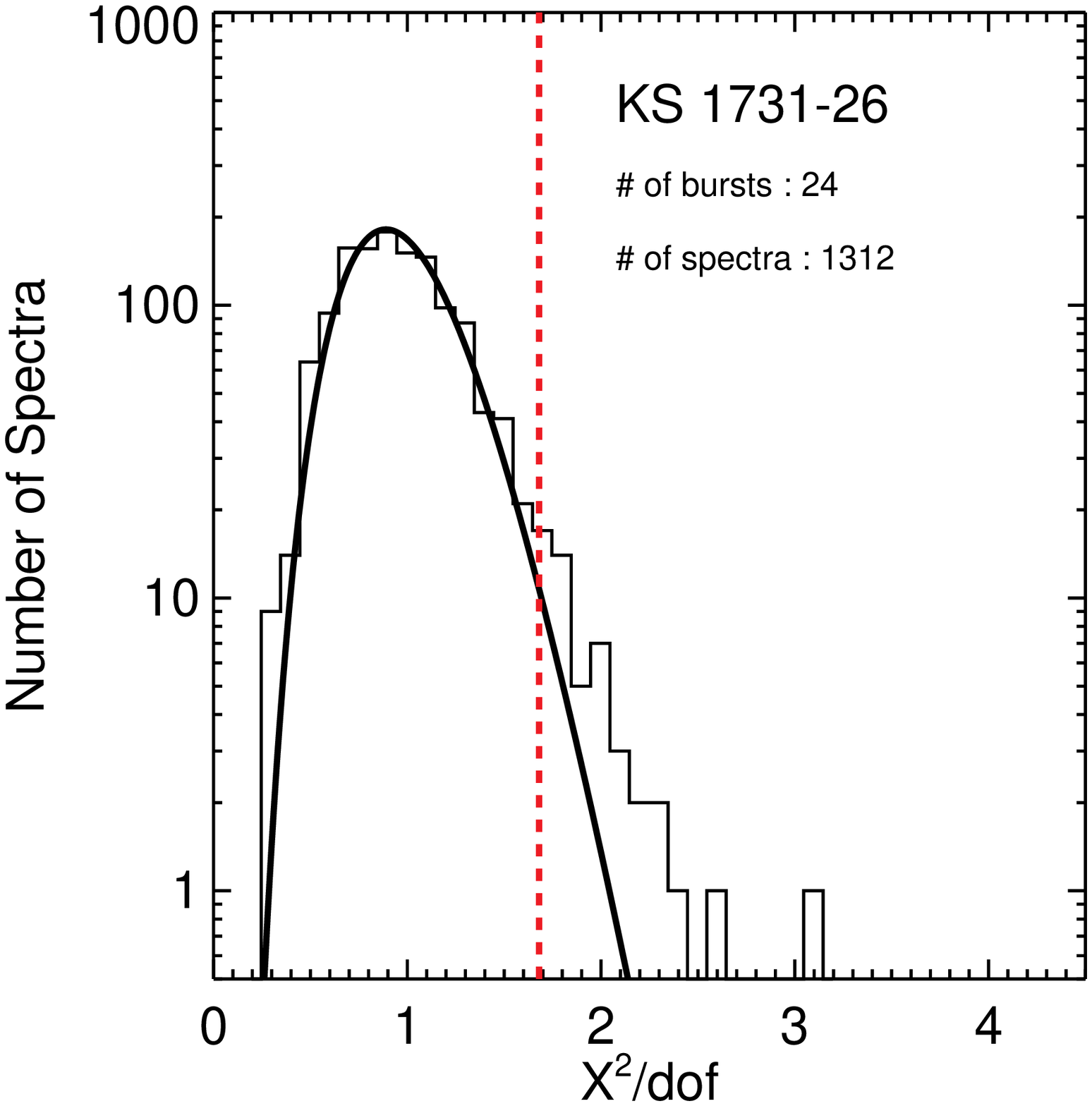}
   \includegraphics[scale=0.29]{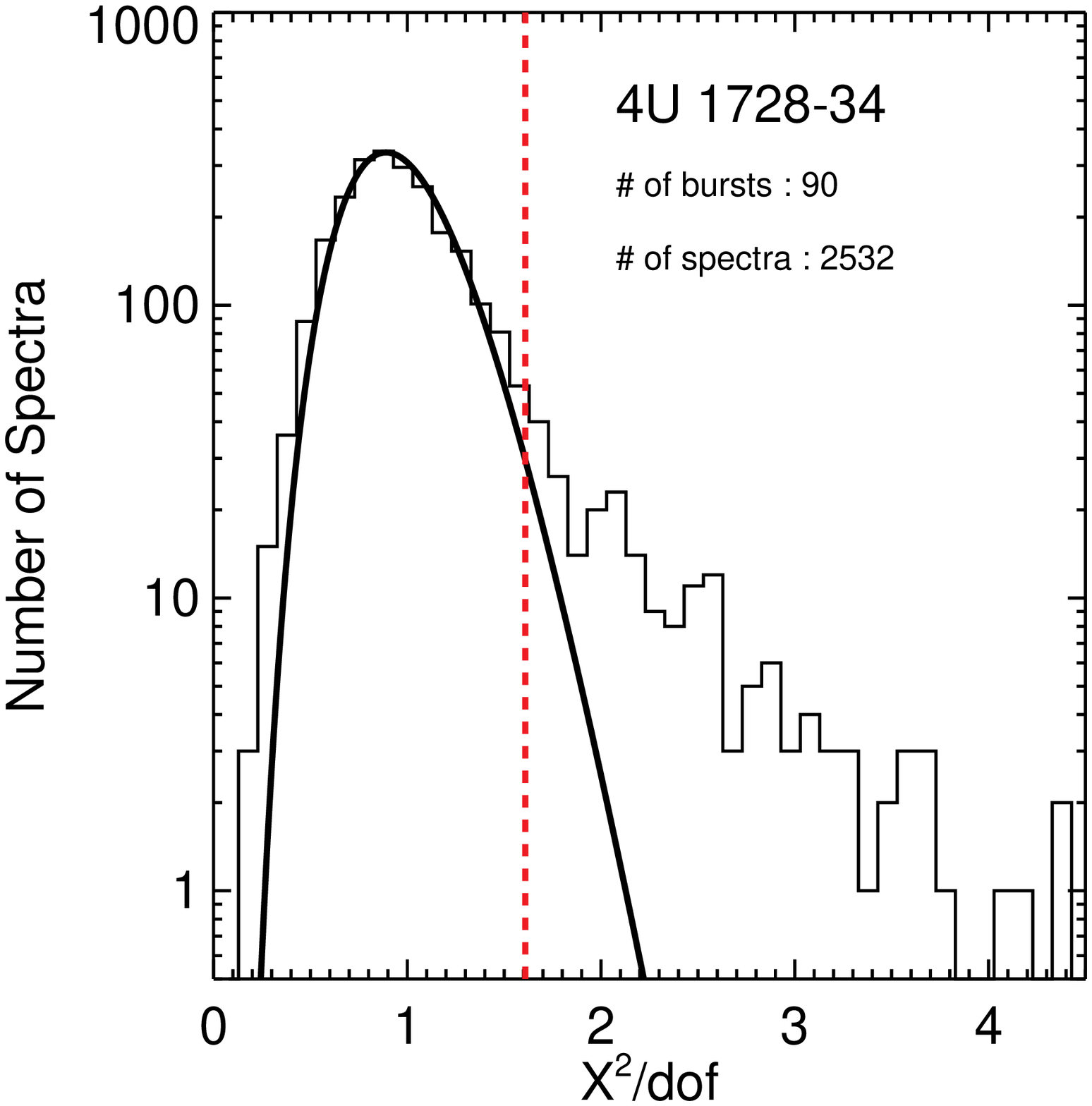}
   \includegraphics[scale=0.29]{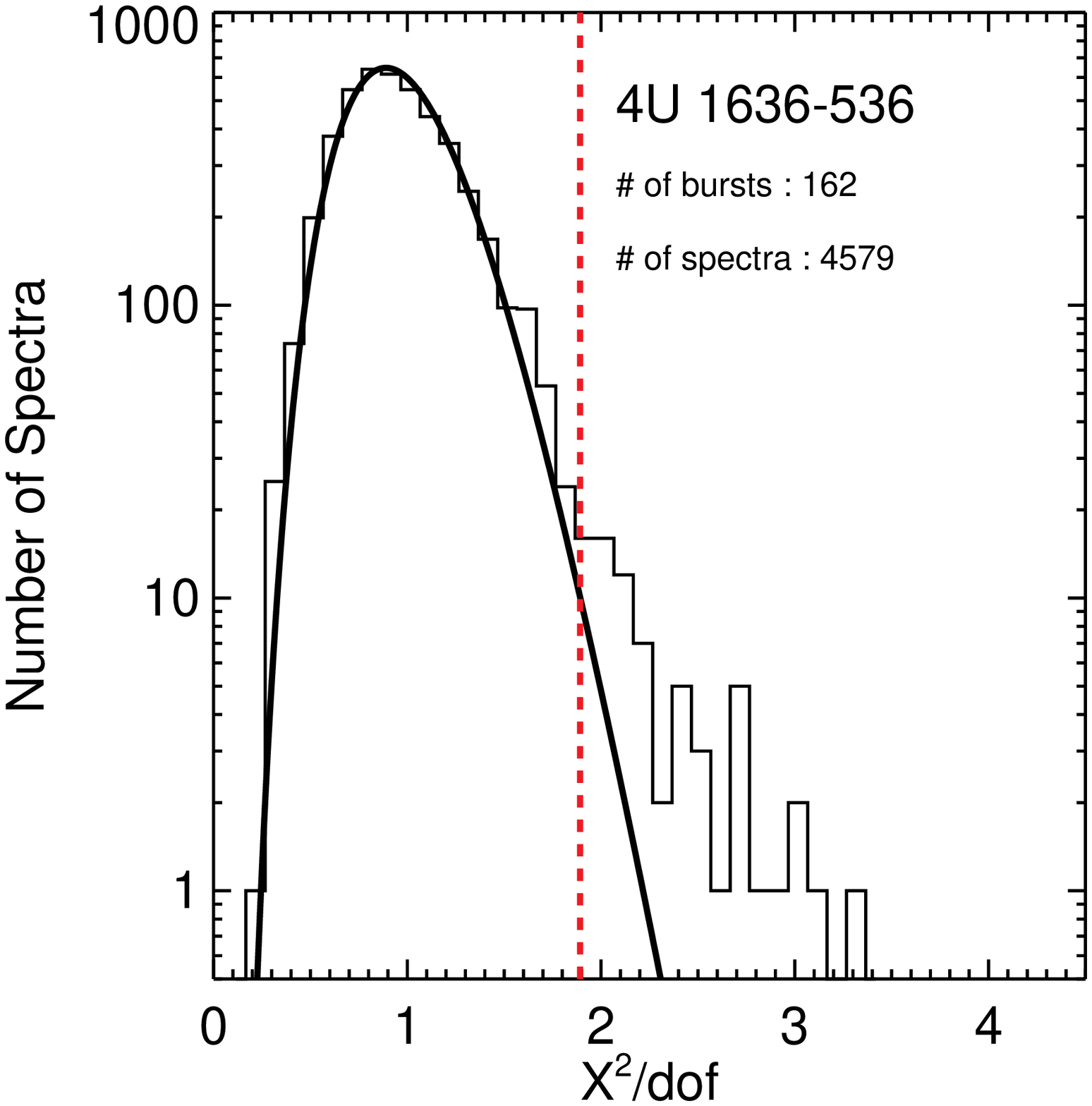}
 \caption{The histograms show the distributions of $X^2$/d.o.f. values
   obtained  from fitting  1309, 2519,  and 4596  X-ray  burst spectra
   observed   from  the   sources  KS~1731$-$260,   4U~1728$-$34,  and
   4U~1636$-$536,  respectively.  The solid  lines  show the  expected
   $\chi^2$/d.o.f. distributions for the  number of degrees of freedom
   used during the  fits. The vertical dashed lines  correspond to the
   highest values of $X^2$/d.o.f.  that we considered as statistically
   acceptable for each  source. The vast majority of  spectra are well
   described by blackbody functions.}
   \label{fig:chi2}
\end{figure}

A final source of systematic uncertainties in the determination of the
X-ray spectral shape of a burster is related to errors in the response
matrix of the detector.  These are expected to be $\lesssim 0.5$\%,
according to the RXTE calibration team (see \S2.2).

\subsection{Goodness-of-fit Measures}

In order to quantify the degree to which any of these effects
influence the spectra of X-ray bursters, we fit all cooling-tail
spectra from each source with a blackbody model, allowing for a level
of systematic uncertainty $\sigma_{\rm sys}$, determined in the
following way.  We reduced the degree of freedom in this procedure by
setting $\sigma_{\rm sys}$ in each spectral bin to be a constant
fraction $\xi$ of the Poisson error $\sigma_{\rm formal}$ and fixed
the value of $\xi$ to a constant for each source.  We then added this
systematic uncertainty to the formal Poisson error of each measurement
in quadrature, i.e., $\sigma^2=\sigma_{\rm formal}^2+ \sigma_{\rm
sys}^2$, and defined a statistic $X^2$ such that
\begin{equation}
X^2(\sigma_{\rm sys}^2)=
\frac{1}{1+\xi^2}
\chi^2\;.
\end{equation}
Here, $\chi^2$ is the standard statistic calculated for each fit.  The
posterior distribution of the new statistic $X^2$ may not be the same
as that of the $\chi^2$ statistic and will depend on the nature of the
systematic uncertainties. In order to explore the degree and nature of
systematic uncertainties, we fit for each source the $X^2$
distribution with the $\chi^2$ distribution expected for the number of
degrees of freedom used in the fits, with $\xi$ as a free parameter.
Note that the expected distribution that we use is formally correct if
the countrate data in the individual spectral bins have uncorrelated
errors.

The expected $\chi^2$ and the observed $X^2$ distributions for the
optimal $\xi$ value for KS~1731$-$260, 4U~1728$-$34, and 4U~1636$-$536
are shown in Figure~\ref{fig:chi2}. The $\xi$ value required to make
the observed $X^2$ distributions for KS~1731$-$260 consistent with the
expected ones is 0.55 (see Table~\ref{chi2table}).  This value
corresponds to systematic uncertainties that are very small.  As an
example, we consider a typical 0.25~s integration for KS~1731$-$260,
when its spectrum is characterized by a color temperature of
2.5~keV. In this case, the average number of counts in each spectral
bin in the 3-10~keV energy range is 110~ct and the corresponding
Poisson uncertainty is $\simeq 9.5$\%.  Multiplying this by the
inferred value $\xi=0.52$ for this source leads to the conclusion that
the systematic uncertainties required to render the observed spectrum
consistent with a blackbody are $\simeq 5$\%.  For the case of
4U~1728$-$34, which is brighter than KS~1731$-$260, the formal
uncertainties are $\simeq 6$\% and the systematic uncertainties amount
to $\simeq 3$\%.

Figure~\ref{fig:chi2} shows that the resulting $X^2$ distributions can
be well approximated by the $\chi^2$ distribution but have weak tails
extending to higher values.  These tails most likely arise from the
inconsistency of only a small number of spectra with blackbody
functions even though we cannot rule out the possibility that the
$X^2$ statistic follows a different posterior distribution than
$\chi^2$.

Figure~\ref{fig:chi2} allows us also to identify the X-ray spectra
that are statistically inconsistent with blackbodies. For each source,
there is a maximum value of $X^2$ per degree of freedom beyond which
the distribution of $X^2$ values deviates from the theoretical
expectation. For the case of KS~1731$-$260, this limiting value of the
reduced $X^2$ is 1.7, for 4U~1728$-$34 it is 1.7, and for
4U~1636$-$536 is 1.9 (see Table~\ref{chi2table}).  The spectra with
high reduced $X^2$ values often occur at the late stages of the
cooling tails, where the subtraction of the persistent emission is the
most problematic. However, unacceptable $X^2$ values may also be found
in other, seemingly random places of the cooling tails. Nevertheless,
the fraction of spectra that is inconsistent with blackbodies are
$\lesssim 3$\%, $\lesssim 6$\%, and $\lesssim 2$\% for KS~1731$-$260,
for 4U~1728$-$34, and for 4U~1636$-$536, respectively. Hereafter, we
consider only the spectra that we regard to be statistically
acceptable, given the values of the reduced $X^2$.

\renewcommand{\thefootnote}{\alph{footnote}}
\begin{table*}
\centering
\caption{Properties of $X^2$ distributions}
\begin{tabular}{lcccc}
  \hline\hline
Source & Number of & 
$\xi$& $X^2$/dof limit\tablenotemark{a} &
Fraction of \\ 
        &  Spectra  &  & & Acceptable Spectra \\
  \hline
  4U~0513$-$40      &   94 & 0.44 & 1.7 & 93.6\% \\
  4U~1636$-$53      & 4579 & 0.59 & 1.9 & 99.3\%\\
  4U~1702$-$429     &  285 & 0.21 & 2.0 & 99.3\%\\
  4U~1705$-$44      &  694 & \nodata\tablenotemark{b} & 1.5 & 87.5\%\\
  4U~1724$-$307     &   58 & 0.44 & 1.5 & 91.4\%\\
  4U~1728$-$34      & 2532 &0.57 & 1.7 & 93.4\%\\
  KS~1731$-$260     & 1312 & 0.52 & 1.7 & 98.2\%\\
  4U~1735$-$44      &   40 &\nodata\tablenotemark{b} & 2.0 &  97.5\%\\
  4U~1746$-$37      &  187 & \nodata\tablenotemark{b} & 1.6 & 85.0\%\\
  SAX~J1748.9$-$2021&  104 & 0.26 & 1.9 & 99.0\%\\
  SAX~J1750.8$-$2900&   82 & 0.54 & 1.7 & 100\%\\
  AQL~X$-$1         & 2191 & 0.45 & 1.6 & 92.1\%\\
  \hline
\footnotetext[1]{~Maximum value of $X^2$/dof beyond which we 
consider the blackbody fits of the spectra for each source 
to be statistically unacceptable.}
\footnotetext[2]{~The $\chi^2$ distributions for these sources
required no addition of systematic uncertainties.}
\end{tabular}
\label{chi2table}
\end{table*}
 \renewcommand{\thefootnote}{\arabic{footnote}}

\section{Systematic Uncertainties in the Inferred Emitting Area}

Our second working hypothesis is that, during the cooling tail of each
burst, the entire neutron star is emitting uniformly with negligible
lateral temperature variations. This assumption is again expected to
be violated at some level for a number of reasons. The non-uniformity
of accretion onto the neutron star (e.g., Inogamov \& Sunyaev 1999),
the finite time of propagation of the burning front around the star
(see, e.g., Nozakura, Ikeuchi, \& Fujimoto 1984; Bildsten 1995;
Spitkovsky, Levin, \& Ushomirsky 2002), as well as the excitation of
non-radial modes on the stellar surface (Heyl 2004; Piro \& Bildsten
2005; Narayan \& Cooper 2007) are all expected to lead to some
variations in the effective temperature of emission at different
latitudes and longitudes on the stellar surface.

The characteristics of burst oscillations observed during the cooling
tails of X-ray bursts, however, imply that any variations in the
surface temperatures of neutron stars can only be marginal.  Indeed,
any component of the variation in the surface temperature that is not
symmetric with respect to the rotation axis leads to oscillations of
the observed flux at the spin frequency of the neutron star. Such
oscillations have been observed in the tails of bursts from many
sources (Galloway et al.\ 2008a).  The r.m.s.\ amplitudes of burst
oscillations in the tails of bursts can be as large as 15\%, although
the typical amplitude is $\simeq 5$\% (Muno, \"Ozel, \& Chakrabarty
2002). The stringent upper limits on the observed amplitudes at the
harmonics of the spin frequencies can be accounted for only if the
temperature anisotropies are dominated by the $m=1$ mode in which
exactly half the neutron star is hotter than the other half (Muno et
al.\ 2002).  Moreover, the rather weak dependence of the r.m.s.\
amplitudes on photon energy (Muno, \"Ozel, \& Chakrabarty 2003)
requires that any temperature variation between the hotter and cooler
regions of the neutron star is $\lesssim 0.2$~keV, even for the bursts
that show the largest burst oscillation amplitudes. All of the above
strongly suggest that the expected flux anisotropy during the cooling
tail of an X-ray burst is $\lesssim 5-10$\% and, therefore, the
expected systematic uncertainties in the inferred apparent stellar
radius will be half of that value, since the latter scales as the
square root of the flux.

\begin{figure}[t]
\centering
   \includegraphics[scale=0.55]{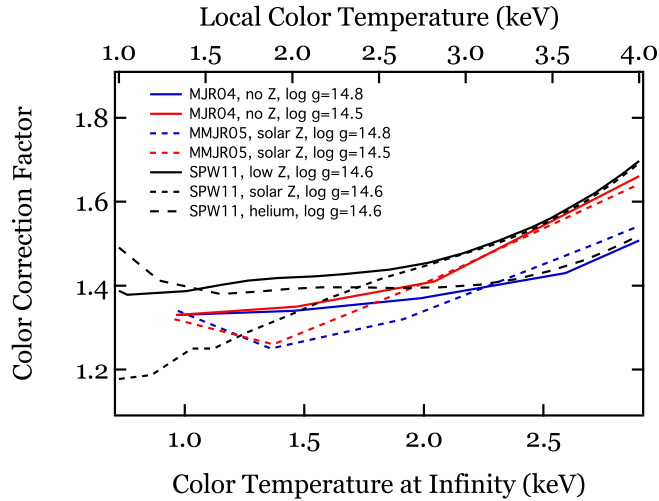}
\caption{The  dependence  of   the  color  correction  factor  $f_{\rm
    c}\equiv T_{\rm  eff}/T_{\rm c}$ on  the color temperature  of the
  X-ray spectrum for two sequences  of models by Madej et al.\ (2004),
  Majczyna et al.\ (2005), and Suleimanov et al. (2011) with different
  metal abundances  and surface gravities. Between  1~keV and 2.5~keV,
  the   color  correction   factor   depends  weakly   on  the   color
  temperature.\\
\\}
   \label{fig:fcolor}
\end{figure}

A final inherent systematic uncertainty in the spectroscopic
measurement of the apparent surface area of a neutron star arises from
the dependence of the color correction factor on the effective
temperature of the atmosphere. In Figure~\ref{fig:fcolor} we show the
predicted evolution of the color correction factor as a function of
the color temperature of the spectrum as measured by an observer at
infinity, for calculations by different groups for different neutron
star surface gravities and atmospheric compositions (Madej et al.\
2004; Majczyna et al.\ 2005; and Suleimanov et al.\ 2011).  To make
the models comparable to the observed evolution of the blackbody
normalizations presented in the next section, we plot the color
correction factor against the most directly observed quantity, i.e.,
the color temperature at infinity. When the color temperature at
infinity is less than 2.5~keV, purely helium or low metallicity models
show $0-8\%$ evolution of the color correction factor per keV of color
temperature at infinity, while above 2.5~keV, they show an evolution
of 12$-$20\% per keV. In contrast, solar metallicity models show a
steady increase with color temperature in the $\sim 1.5 - 3$~keV
range.\footnote{The color correction factor shows a turnover at
different Eddington ratios and correspondingly at different color
temperatures depending on composition, surface gravity, and
gravitational redshift. The rapid evolution presented in Suleimanov et
al.\ (2011) preferentially occurs at low Eddington ratios and color
temperatures smaller than 1 keV (which we cannot explore
observationally in the XTE burst data) and for very small surface
gravities ($\log g = 14.0$) which correspond only to neutron stars
with radii $\ge 15$~km.}

We infer the  apparent surface area $4\pi R_{\rm  app}^2$ of a neutron
star  by  measuring  the  X-ray  flux $F_{\rm  cool}$  and  the  color
temperature $T_{\rm  c}$ at different  intervals during the  burst, so
that
\begin{equation}
4\pi R_{\rm app}^2=\frac{4\pi D^2 F_{\rm cool}}{\sigma_{\rm SB} (T_{\rm
    c}/f_{\rm c})^4}\;.
\label{eq:rapp}
\end{equation}
Here, $\sigma_{\rm SB}$ is the Stefan-Boltzmann constant, $D$ is the
distance to the source, and $f_{\rm c} = T_{\rm c} / T_{\rm eff}$ is
the color correction factor. The value of the color correction factor
is dictated by the predominant source of absorption and emission in
the neutron-star atmosphere. It, therefore, depends on the effective
temperature, which determines primarily the ionization levels, and on
the effective gravitational acceleration, which determines the density
profiles of the atmospheric layers. Both these quantities decrease as
the cooling flux decreases. If we were to assume a constant value for
the color correction factor, as it is customary, we would obtain a
systematic change in the inferred apparent surface area as the cooling
flux of the burst decreases with time. Such variations have been
discussed in Damen et al.\ (1989) and in Bhattacharyya, Miller, \&
Galloway (2010). Note that this systematic effect can be corrected, in
principle, if the data are fit directly with detailed models of
neutron-star atmospheres (see, e.g., Majczyna \& Madej 2005).

A potentially important source of uncertainty in the measurements is
introduced by the errors in the absolute flux calibration of the
RXTE/PCA detector.  The current calibration of the PCA and the
cross-calibration between X-ray satellites have been carried out using
the Crab nebula as a standard candle (Jahoda et al.\ 2006; see also
Toor \& Seward 1974; Kirsch et al.\ 2005; Weisskopf et al.\ 2010). The
uncertainties in the flux calibration can be due to a potential
overall offset between the inferred and the true flux of the Crab
nebula, which may be as large as 10\% (Kirsch et al.\ 2005; Weisskopf
et al.\ 2010). This can only change the mean value of the inferred
apparent area in each source and does not alter the observed
dispersion. We will explore this issue as well as uncertainties
related to the variability of the Crab nebula itself (Wilson-Hodge et
al.\ 2011) in more detail in Paper~III of this series.

\begin{figure*}[t]
\centering \includegraphics[scale=0.3]{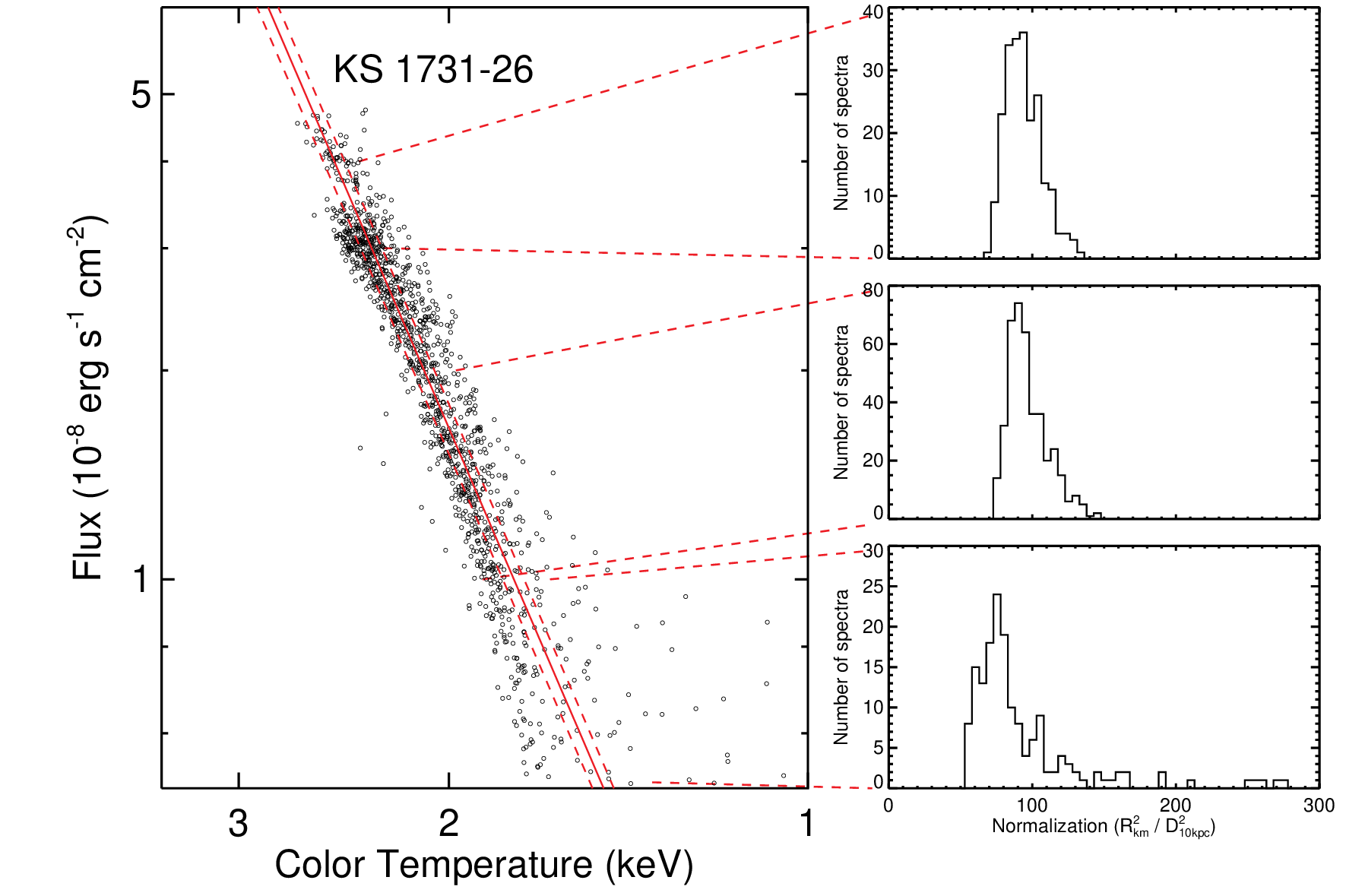}
   \includegraphics[scale=0.3]{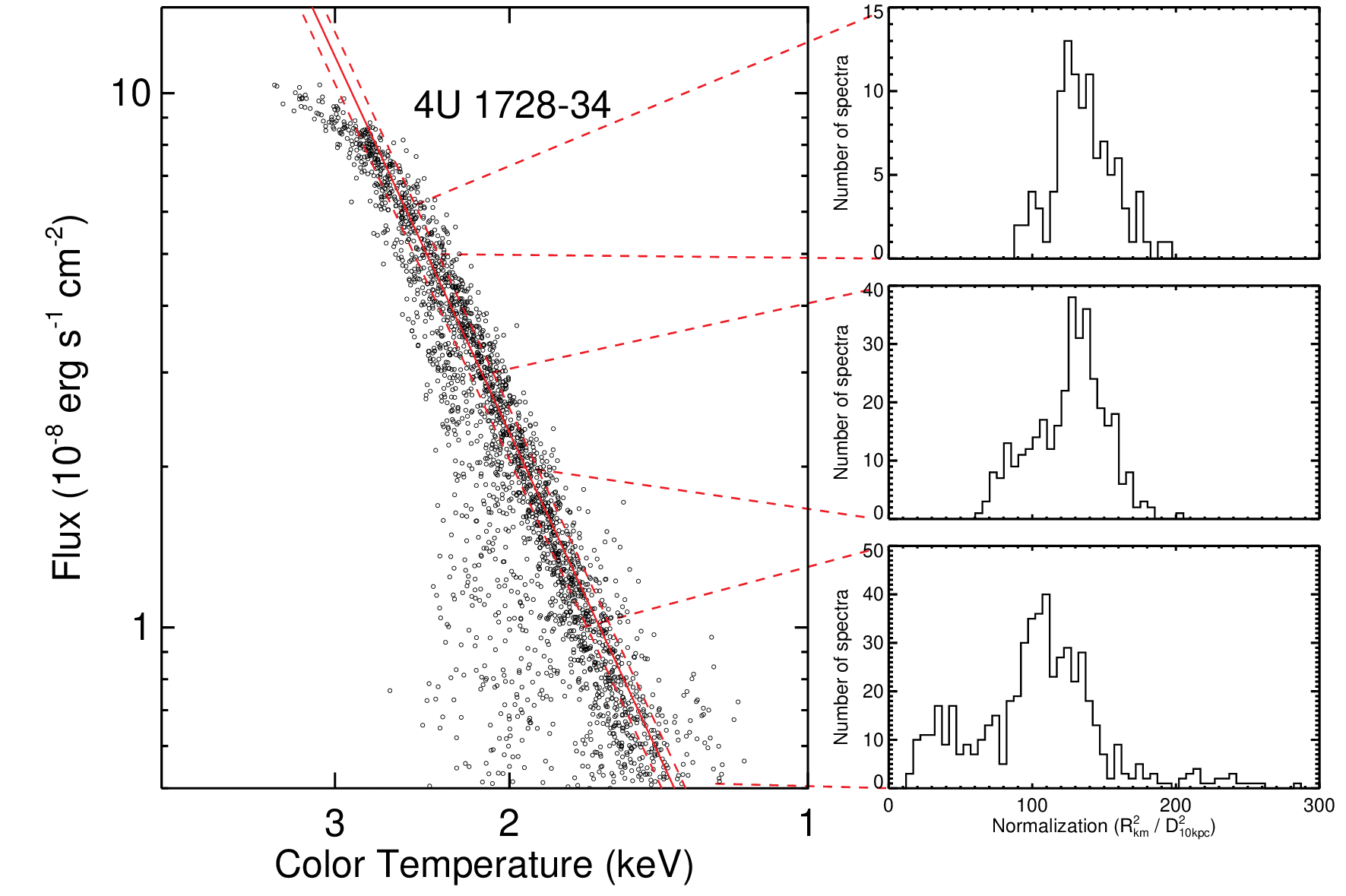}
   \includegraphics[scale=0.3]{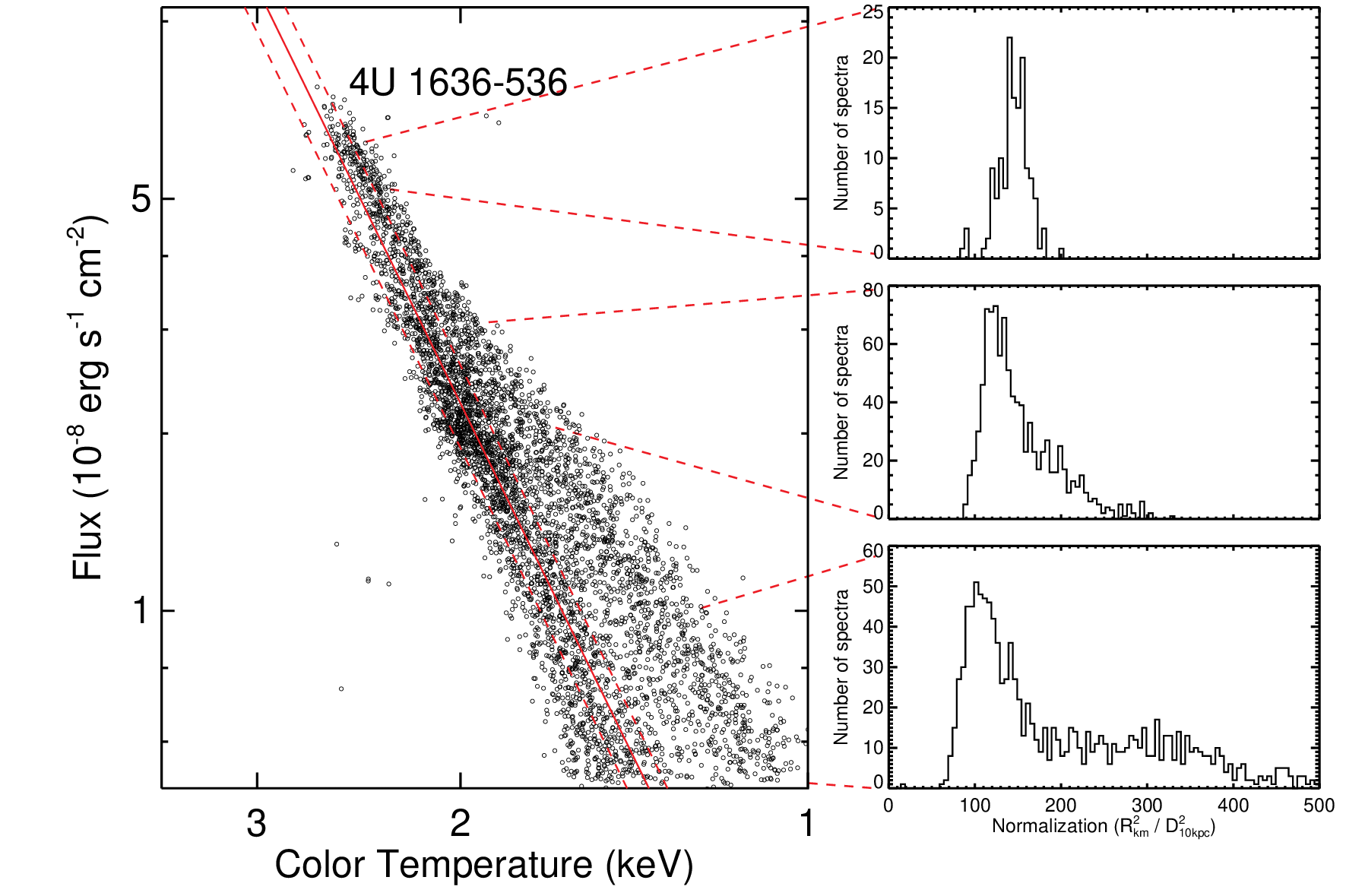}\\ 
   \caption{{\em  (Left)\/} The flux-temperature  diagram for  all the
     spectra   in   the   cooling    tails   of   bursts   from   {\em
       (top)\/}~KS~1731$-$260, {\em (Middle)\/}~4U~1728$-$34, and {\em
       (bottom)\/}~4U~1636$-$536  that  have statistically  acceptable
     values of  $\chi^2$/d.o.f.  The diagonal lines  correspond to the
     best-fit blackbody normalization and its uncertainty, as reported
     in  the   rightmost  column  of  Table~3.   {\em  (Right)\/}  The
     distribution  of measured normalization  values of  the blackbody
     spectra   in  three  of   the  flux   intervals  we   chose.  The
     normalization values for the vast majority of spectra fall within
     a narrowly  peaked distribution, with  only a number  of outliers
     towards   lower  (for   4U~1728$-$34)  or   higher   values  (for
     4U~1636$-$536)   of  the   normalization.   This  justifies   the
     assumption  that  the entire  neutron  star  surface is  emitting
     during  the cooling  tail of  a burst  with  marginal temperature
     variations in latitude or longitude.}
\label{fig:kt_flux_detail}
\end{figure*}

\subsection{Flux-Temperature Diagrams}

Figure~\ref{fig:kt_flux_detail} (left  panels) show the  dependence of
the  emerging flux on  color temperature  for all  the spectra  in the
cooling tails  of KS~1731$-$260, 4U~1728$-$34,  and 4U~1636$-$536 that
we consider to be statistically  acceptable (see \S3).  We chose these
three sources to  use as detailed examples because  of the high number
spectra obtained  for each and  the fact that  they span the  range of
behavior in cooling  tracks that we will discuss  below.  If the whole
neutron  star is emitting  as a  single-temperature blackbody  and the
color  correction factor  is  independent of  color temperature,  then
$F_{\rm  cool}$ should scale  as $T_{\rm  c}^4$.  Our  aim here  is to
investigate  the systematic  uncertainties on  the measurement  of the
apparent surface  area in  each source at  different flux  levels. We,
therefore, divided the data into a  number of flux bins and plotted in
the  same figure  (right  panels) the  distribution  of the  blackbody
normalization  values  for some  representative  bins.  The  blackbody
normalization for each spectrum is defined formally as $A\equiv F_{\rm
  cool}/\sigma_{\rm SB}T_{\rm c}^4$, although  in practice this is one
of the two measured parameters and  the flux is derived from the above
definition.   According  to  equation~(\ref{eq:rapp}),  the  blackbody
normalization is equal to  $A=f_c^{-4}R_{\rm app}^2/D^2$ and we report
it  in  units  of  (km/10~kpc)$^2$.  If  we do  not  correct  for  the
dependence of the  color correction factor on the  flux, we expect the
normalization $A$  to show a  dependence on color temperature  that is
the mirror symmetric of that shown in Figure~2.

The  flux-temperature  diagrams  of KS~1731$-$260,  4U~1728$-$34,  and
4U~1636$-$536   share  a   number   of  similarities   but  are   also
distinguished by a number of differences. In all three cases, the vast
majority of  data points  lie along a  very well  defined correlation.
This reproducibility of the cooling curves of tens of X-ray bursts per
source, combined  with the lack  of large amplitude  flux oscillations
during cooling  tails of bursts, provides the  strongest argument that
the thermal emission engulfs the entire neutron-star surface with very
small temperature variations at different latitudes and longitudes.

In 4U~1728$-$34, a deviation of the data points from the $F_{\rm
cool}\sim T_{\rm c}^4$ correlation is evident at high fluxes and color
temperatures ($T_{\rm c}\ge 2.5$~keV), which may be due to the
evolution of the color correction factor at high Eddington fluxes (see
the discussion in \S 4 and Fig.~\ref{fig:fcolor}). The same deviation
is not evident in KS~1731$-$260 or 4U~1636$-$536, for which the
highest temperatures encountered in the cooling tails were $\lesssim
2.5$~keV.

Finally,  in all  three sources,  a number  of outliers  exist  at the
lowest flux  levels, with normalizations  that deviate from  the above
correlation.   In   KS~1731$-$260  and  4U~1636$-$536,   the  outliers
correspond  to higher  normalization values,  whereas  in 4U~1728$-$34
they  correspond to  lower normalization  values with  respect  to the
majority of the data points.

Any combination  of the effects discussed  earlier in this  and in the
previous  section may  be responsible  for the  outliers.  Non-uniform
cooling of  the neutron star surface  will lead to  a reduced inferred
value for the  apparent surface area. Reflection of  the surface X-ray
emission  off  a  geometrically  thin  accretion disk  will  cause  an
increase in the inferred value for the apparent surface area. Finally,
Comptonization of  the surface  emission in a  corona may  have either
effect, depending on the Compton temperature of the radiation.

Our goal in this work is not to understand the origin of the outliers,
but rather to ensure that their presence does not introduce any biases
in the  measurements of the  apparent surface areas inferred  from the
vast majority of  spectra. Indeed, including the outliers  in a formal
fit  of  the  flux-temperature  correlation will  cause  a  systematic
increase  of  the  apparent  surface  area  with  decreasing  flux  in
KS~1731$-$260  and  in  4U~1636$-$536  and a  systematic  decrease  in
4U~1728$-$34 (c.f.\  Damen et al.\ 1990; Bhattacharyya  et al.\ 2010).
This issue  was largely  avoided in our  earlier work (\"Ozel  et al.\
2009; G\"uver et al.\ 2010a, 2010b) by considering only the relatively
early  time  intervals  in  the  cooling tail  of  each  burst,  which
correspond only  to the  brightest flux bins.   In order to  go beyond
this limitation here, we consider all spectral data for each burst and
employ  a Bayesian  Gaussian mixture  algorithm, which  is  a standard
procedure   for  outlier   detection  in   robust   statistics  (e.g.,
Titterington, Smith, \&  Makov 1985; McLachlan \& Peel  2000; Huber \&
Ronchetti 2009).

Our working  hypothesis that the  main peak of normalizations  in each
flux interval corresponds to the signal and the remainder are outliers
is shaped by two aspects  of the observations.  First, at high fluxes,
each histogram can be described  by a single Gaussian with no evidence
or room for a second distribution  of what we would call outliers.  At
lower  fluxes,  when  the   histograms  can  be  decomposed  into  two
Gaussians, the distribution with  the highest peak has properties that
smoothly connect  to those of  the single Gaussians at  higher fluxes.
Second,  the Gaussians  that we  call  our signal  always contain  the
majority of  the data points compared  to the distribution  of what we
call  the outliers. We  take these  as our  criteria for  defining our
signal.

The algorithm,  which we describe below  in detail, allows  us also to
measure the degree of intrinsic variation in the apparent surface area
for  each   source  that   is  consistent  with   the  width   of  the
flux-temperature  correlation.   Even  though  we compare  spectra  in
relatively narrow flux bins, the  different observing modes as well as
the different number of PCUs used in each observation result in a wide
range  of  count  rates  and,   hence,  in  a  wide  range  of  formal
errors. This necessitates the use of the Bayesian approach we describe
below  as   opposed  to,  e.g.,   performing  $\chi^2$  fits   of  the
distributions over blackbody normalization.

\subsection{Quantifying Systematic Uncertainties with a Bayesian 
Gaussian Mixture Algorithm for Outlier Detection}

Consider  first  the  situation  in  which  there  are  no  systematic
uncertainties  and  the inferred  blackbody  normalization from  every
spectrum  reflects the  apparent surface  area of  the  entire neutron
star. If this  were the case, the intrinsic  distribution of blackbody
normalizations  would   be  a  delta  function,   while  the  observed
distribution  would be  broader, with  a  width equal  to the  average
formal uncertainties of the measurements.

In a more realistic situation, there is an intrinsic range of the
inferred blackbody normalizations, caused by a combination of the
various effects discussed earlier. We assume that, in each flux bin,
the intrinsic distribution of normalization values is the Gaussian
\begin{equation}
  P_{\rm int}(A| A_{\rm int}, \sigma_{\rm int})
  =\frac{1}{\sqrt{2\pi \sigma_{\rm int}^2}}
    \exp\left[-\frac{(A-A_{\rm int}^2)}{2\sigma_{\rm int}^2}\right]\;,
    \label{eq:int_dist}
\end{equation}
where $A_{\rm int}$ is its mean and $\sigma_{\rm int}$ is its standard
deviation. The observed distribution of blackbody normalizations will
be again broader than the intrinsic distribution because of the formal
uncertainties in each measurement.  Moreover, at low flux levels, a
small number of outliers exists, which skews and introduces tails to
the observed distribution.  Our goal here is to quantify the most
probable value of the blackbody normalization, i.e., the parameter
$A_{\rm int}$ in the distribution~(\ref{eq:int_dist}), as well as the
intrinsic range of values in each flux bin, i.e., the parameter
$\sigma_{\rm int}$, while excluding the outliers.

\begin{figure*}[t]
\centering
   \includegraphics[scale=0.3]{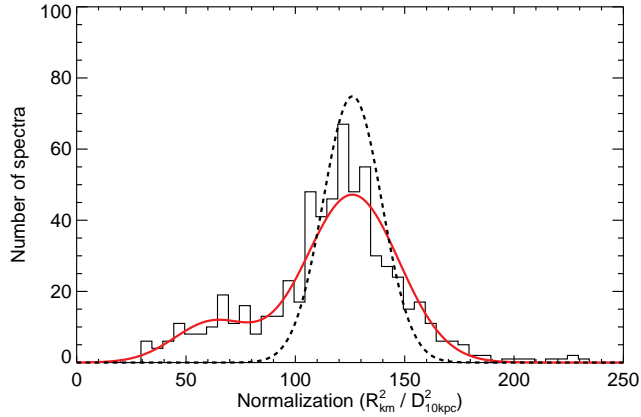} 
\caption{The  histogram shows the  distribution of  measured blackbody
  normalizations that  result from  fitting all the  available spectra
  during the cooling tails of bursts from 4U~1728$-$34, when the burst
  flux was in  the range $(1-2)\times 10^{-8}$~erg~cm$^{-2}$~s$^{-1}$.
  The  red solid  line  shows  the Gaussian  mixture  model that  best
  describes  the data.  The  black dashed  curve shows  the underlying
  Gaussian distribution of blackbody normalizations that gives rise to
  the Gaussian mixture model, when the observational uncertainties and
  the outliers  are taken into  account.  The width of  the underlying
  Gaussian  distribution reflects  the systematic  uncertainty  in the
  measurement of the blackbody  normalization for this flux interval.
\\}
\mbox{}
\label{fig:1728_detail}
\end{figure*}

When the inspection of an observed distribution requires an outlier
detection scheme (as, e.g., in the lower right panels of
Fig.~\ref{fig:kt_flux_detail}), we model the distribution of outliers
by a similar Gaussian, i.e.,
\begin{equation}
  P_{\rm out}(A| A_{\rm out}, \sigma_{\rm out})
  =\frac{1}{\sqrt{2\pi \sigma_{\rm out}^2}}
    \exp\left[-\frac{(A-A_{\rm out}^2)}{2\sigma_{\rm out}^2}\right]\;,
\end{equation}
with a mean $A_{\rm out}$ and a standard deviation $\sigma_{\rm out}$.
Our model distribution function is, therefore, in general the Gaussian
mixture
\begin{equation}
P_{\rm model}(A| A_{\rm int}, \sigma_{\rm int}, A_{\rm out},
\sigma_{\rm out},\eta)= P_{\rm int}(A| A_{\rm int}, \sigma_{\rm
  int})+\eta P_{\rm out}(A| A_{\rm out}, \sigma_{\rm out})\;,
\end{equation}
where $\eta$ measures the relative fraction of outliers in our sample. 

We are looking for the parameters in these distributions that maximize
the posterior probability $P(A_{\rm int}, \sigma_{\rm int}, A_{\rm
out}, \sigma_{\rm out}\vert{\rm data})$, which measures the likelihood
that a particular set of parameters is consistent with the data. Using
Bayes' theorem, this probability distribution is
\begin{equation}
P(A_{\rm int}, \sigma_{\rm int}, A_{\rm out}, \sigma_{\rm out},\eta\vert{\rm
  data})=C P({\rm data}\vert A_{\rm int}, \sigma_{\rm int}, A_{\rm
  out}, \sigma_{\rm out},\eta)
P(A_{\rm int})P(\sigma_{\rm int}) P(A_{\rm out}) P(\sigma_{\rm out})
P(\eta)\;,
\label{eq:post_prob}
\end{equation}
where $P({\rm  data}\vert A_{\rm int}, \sigma_{\rm  int}, A_{\rm out},
\sigma_{\rm  out},\eta)$   is  the  posterior   probability  that  the
particular set of normalization data have been measured for a given
set of parameters and the last five terms are the prior probabilities
for the model parameters; the constant $C$ ensures that the
probability density is normalized.  We consider a flat prior
probability distribution for the central value of each Gaussian in an
interval that extends beyond the observed range of normalization
values for each source. We also consider a flat prior probability
distribution for the standard deviation for each Gaussian from
$10^{-3}$~(km/10 kpc)$^2$ (which is practically zero) to a value equal
to the width of the range of normalizations. Finally, we take a flat
prior distribution of $\eta$ between zero and one.  Our results are
extremely weakly dependent on the ranges of parameter values we
considered.

Given  a set  of $N$  observed  normalization values  $A_i$ and  their
uncertainties  $\sigma_i$, we calculate  the posterior  probability of
the data given a set of model parameters as
\begin{equation}
P({\rm data}\vert A_{\rm int}, \sigma_{\rm int}, A_{\rm
  out} \sigma_{\rm out},\eta)=C _2 \prod_{i=1}^N\int dA 
\frac{1}{\sqrt{2\pi\sigma_i^2}}
\exp\left[-\frac{(A-A_{\rm i}^2)}{2\sigma_{\rm i}^2}\right]
\left[P_{\rm int}(A| A_{\rm int}, \sigma_{\rm int})
+\eta P_{\rm out}(A| A_{\rm out}, \sigma_{\rm out})\right]\;.
\label{eq:post_prob_data}
\end{equation}
Here, $A_i$  is the  value of each  measurement and $\sigma_i$  is its
corresponding uncertainty, which we assume to be normally distributed;
$C_2$ is  a normalization  constant. Using this  posterior probability
distribution,  we  identify the  most  probable  values  of the  model
parameters,  as  well   as  their  uncertainties,  following  standard
procedures.

Figure~\ref{fig:1728_detail} shows, as  an example, the application of
the  Gaussian  mixture  algorithm  to the  distribution  of  blackbody
normalizations obtained from fitting the spectra of 4U~1728$-$34, when
the     burst     flux    was     in     the    range     $(1-2)\times
10^{-8}$~erg~cm$^{-2}$~s$^{-1}$. The  histogram shows the distribution
of  blackbody  normalizations measured  in  this  flux interval.   The
parameters  of  the  Gaussian  mixture  that  maximize  the  posterior
probability           distribution           calculated          using
equation~(\ref{eq:post_prob}) correspond  to an intrinsic distribution
with     a    mean     and    standard     deviation     of    $A_{\rm
  int}=126.1$~(km/10~kpc)$^2$             and             $\sigma_{\rm
  int}=13.4$~(km/10~kpc)$^2$,  respectively,  and  a  distribution  of
outliers with a mean of $A_{\rm out}=62.3$~(km/10~kpc)$^2$, a standard
deviation  of $\sigma_{\rm  out}=6.1$~(km/10~kpc)$^2$, and  a relative
normalization of $\eta=0.2$. In  order to compare the Gaussian mixture
with  the  observed histogram,  we  first  convolve  it with  a  third
Gaussian  distribution  with  a  standard  deviation  of  $\sigma_{\rm
  formal}=16.5$~(km/10~kpc)$^2$  that is equal  to the  average formal
errors of all measurements in this flux interval.  The result is shown
as   a   red  solid   line   in  Figure~\ref{fig:1728_detail},   which
demonstrates that  our assumption of  a Gaussian distribution  for the
majority  of the  data and  for the  outliers is  consistent  with the
observations.

We can  appreciate the importance  of excluding the outliers  from the
sample using the Bayesian  Gaussian mixture algorithm in the following
way. If  we repeat the above  procedure for the same  flux interval in
4U~1728$-$34  but do  not  allow  for the  presence  of outliers,  the
parameters   of   the   intrinsic   distribution   will   be   $A_{\rm
  int}=118.6$~(km/10~kpc)$^2$             and             $\sigma_{\rm
  out}=26.4$~(km/10~kpc)$^2$,  which are significantly  different than
the results quoted  above. Moreover, had we assumed  that there are no
systematic  errors but that  the underlying  distribution was  a delta
function,   the  most   probable   value  would   have  been   $A_{\rm
  int}=117.1^{+0.2}_{-0.8}$~(km/10~kpc)$^2$.   The  formal  errors  in
such a  measurement would have been substantially  smaller compared to
the systematic uncertainties.

In the following,  we show in detail the  application of this Gaussian
mixture algorithm to all the flux intervals for the three sources.

\begin{figure}[t]
\centering
  \includegraphics[scale=0.25]{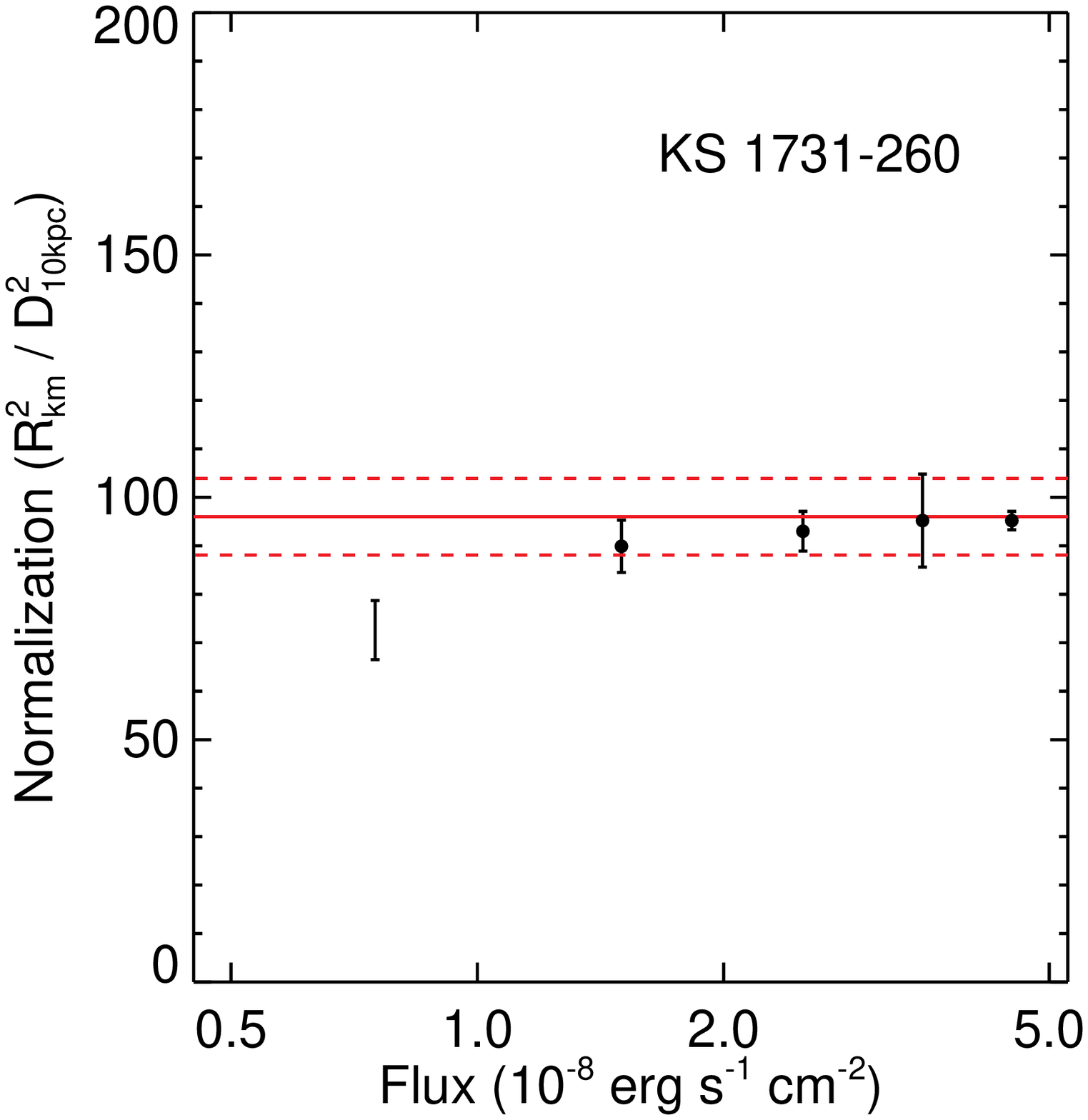}
  \includegraphics[scale=0.25]{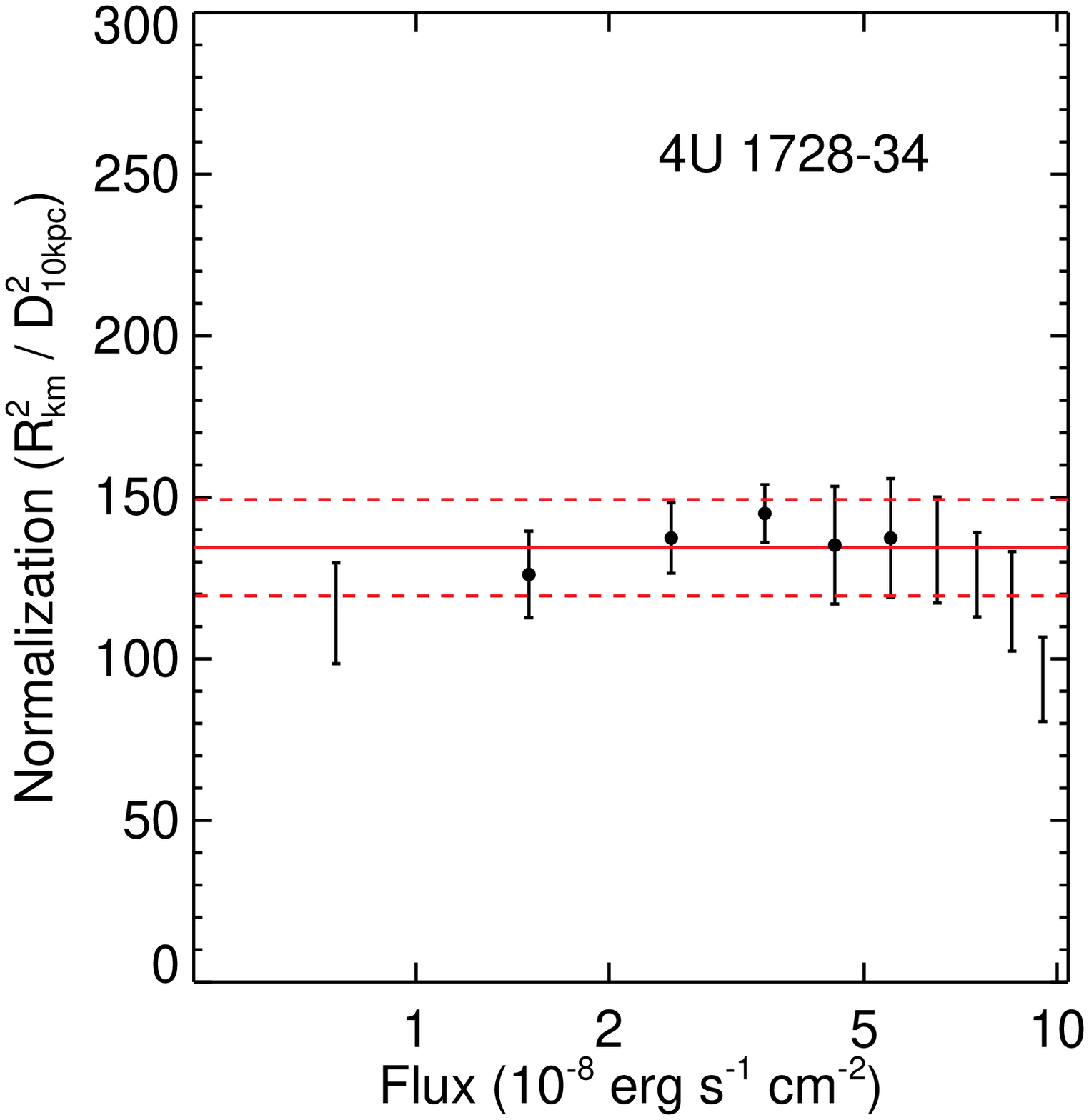}
  \includegraphics[scale=0.25]{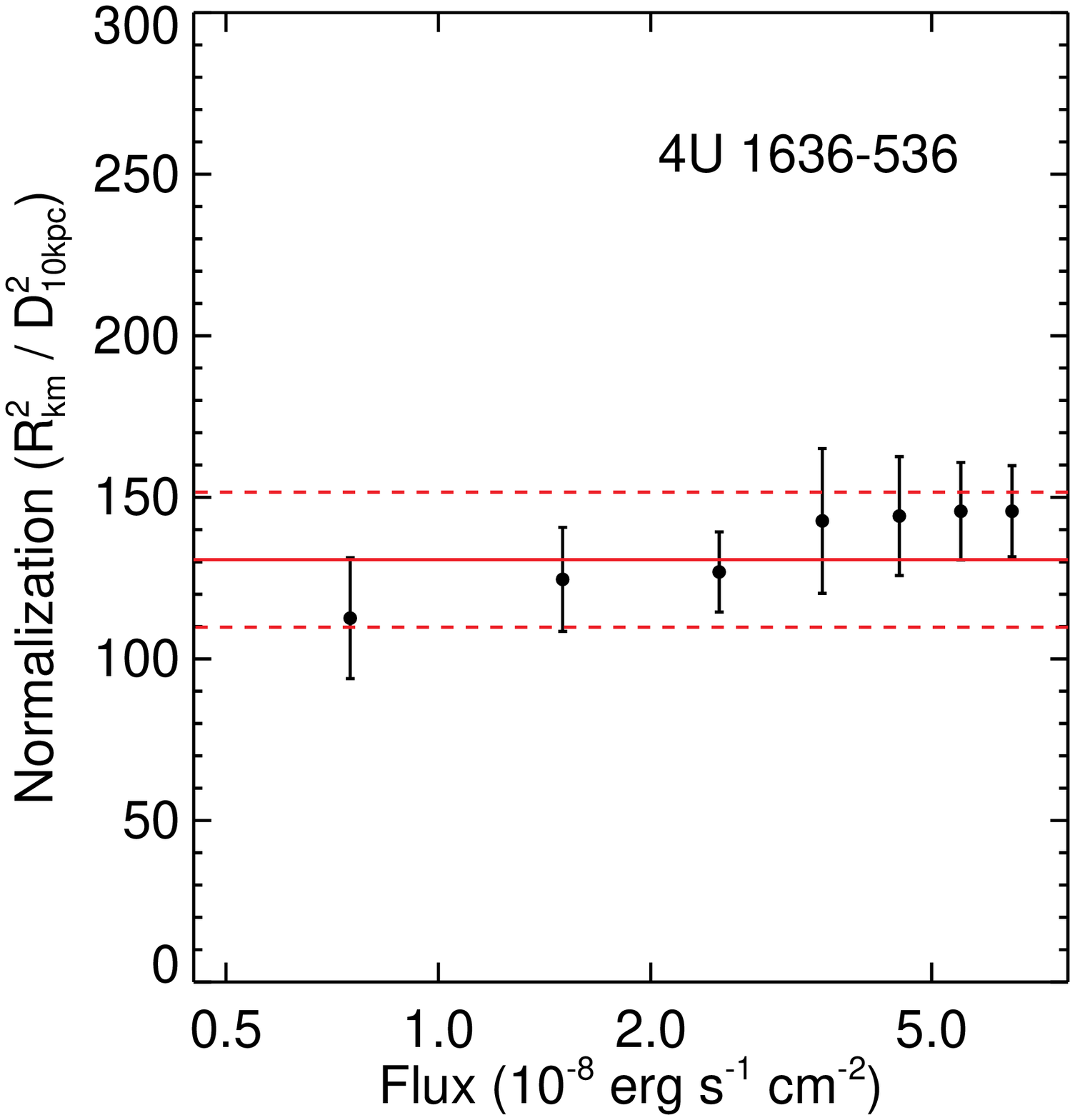}
  \caption{The   dependence  of  the   parameters  of   the  intrinsic
    distribution of blackbody normalizations  on X-ray flux during the
    cooling  tails  of thermonuclear  X-ray  bursts in  KS~1731$-$260,
    4U~1728$-$34,  and  4U~1636$-$536,  when  the outliers  have  been
    removed.   Each dot  represents the  most likely  centroid  of the
    intrinsic distribution,  while each error bar  represents its most
    likely width, as calculated  using the procedure outlined in \S4.2
    and depicted  in Figure~\ref{fig:1728_detail}. In  each panel, the
    solid and dashed horizontal  lines show the best-fit normalization
    and its  systematic uncertainty inferred using the  flux bins that
    do  not correspond  to near-Eddington  fluxes and  are  denoted by
    filled circles on the error bars.}  \mbox{}
   \label{fig:norm_vs_flux}
\end{figure}
\renewcommand{\thefootnote}{\alph{footnote}}
\begin{table*}
\centering
\caption{Blackbody Normalizations\tablenotemark{a}}
\begin{tabular}{lccccccccccc}
 \hline
 \hline
     & \multicolumn{10}{c}{Flux interval ($10^{-8}$~erg~s$^{-1}$~cm$^{-2}$)} & \\
Source & 0.5--1 & 1--2 & 2--3 & 3--4 & 4--5 & 5--6 & 6--7 & 7--8 & 8--9 & 
9--10 & Average\tablenotemark{b}\\
\hline
\hline
4U~1636$-$536 &  112.6 &    124.6 &    126.9 &    142.7 &    144.2 &    145.7 &    145.7 & 149.5  &  \nodata  & \nodata &    130.7\\
              & \er18.7& \er 16.1 & \er  12.4 & \er 22.4 & \er 18.4 & \er 15.1 & \er 14.1 & \er 4.6         &           &         & \er 20.9\\
 \hline
4U~1702$-$429 &  151.0 &    167.6 &    180.4 &    184.2 &    185.7 &    171.4 &    151.0 &    119.3 &  98.2  & \nodata &    176.6\\
              & \er13.9& \er 10.1 & \er  9.9 & \er  4.1 & \er 3.9 & \er  8.1 & \er 8.1 & \er 17.2 &   \er 12.4 &           & \er 11.6\\
\hline
4U~1705$-$44
             &     80.2 &     86.9 &     86.9 &     82.4 &  \nodata &  \nodata &  \nodata & \nodata  & \nodata  & \nodata &     83.9\\
             & \er  9.9 & \er  7.1 & \er 10.9 & \er  7.4  &         &          &          &          &          &         & \er  9.1\\
 \hline
4U~1724$-$307&    98.7 &    108.3 &    120.4 &    127.4 &    125.4 &     102.8 &  \nodata & \nodata  & \nodata  & \nodata &     113.8\\
             & \er 7.1 & \er  4.6 & \er 12.6 & \er  3.7  & \er  19.4 & \er 15.4 &          &          &          &         & \er  15.4\\
\hline
4U~1728$-$34 &   114.1 &    126.1 &    137.4 &    145.0 &    135.2 &    137.4 &    133.7 &    126.1 &     117.8 &    93.7 &    134.4\\
             & \er 15.6& \er 13.4 & \er 10.9 & \er  8.9 & \er 18.2 & \er 18.4 & \er 16.4 & \er 13.1 & \er 15.4 & \er 13.1 & \er 14.9\\
\hline
KS~1731$-$260
             &    72.6 &     89.9 &     93.0 &     95.2 &     95.2 &  \nodata &  \nodata & \nodata  & \nodata  & \nodata &    96.0\\
             & \er 6.1 & \er  5.4 & \er  4.1 & \er  9.6 & \er 1.9 &          &          &          &          &         & \er 7.9\\
\hline
4U~1735$-$44 & 71.6 & 71.6\tablenotemark{c} & 72.6\tablenotemark{c} & \nodata & \nodata & \nodata & \nodata & \nodata & \nodata & \nodata & 72.1\tablenotemark{c}\\
             & \er 4.6 & $^{+2.0}_{-1.5}$ & $^{+2.4}_{-1.9}$ &  &     &           &         &           &         &          & $^{+1.3}_{-1.0}$\\
\hline
4U~1746$-$37\tablenotemark{d}
             &    12.9 &     16.3 &     17.3 &     15.6 &     15.2 &     15.2 &  19.9 & \nodata  & \nodata  & \nodata &     15.7\\
             & \er 1.9 & \er  0.5 & \er  0.4 & \er 2.4  & \er  2.4 & \er  1.6 &  \er 6.4     &          &          &         & \er  2.4\\
\hline
SAX~J1748.9$-$2021
             &    92.7 &     87.7 &     91.7 &     87.7 &  \nodata &  \nodata &  \nodata & \nodata  & \nodata  & \nodata &     89.7\\
             & \er 11.9& \er  6.6 & \er  7.6 & \er 15.4  &         &          &          &          &          &         & \er  9.6\\
\hline
SAX~J1750.8$-$2900
             &   126.9 &    110.8 &    106.8 &     97.2 &     86.2 &  \nodata &  \nodata & \nodata  & \nodata  & \nodata &     93.2\tablenotemark{c}\\
             & \er 40.5& \er 19.9 & \er  10.1 & \er  8.1  & \er 8.6 &          &          &          &          &         & \er  9.4\\
 \hline
 \hline

\footnotetext[1]{~The  parameters  of  the intrinsic  distribution  of
  blackbody  normalizations in  different flux  intervals for  all the
  sources we considered in  this manuscript. All normalizations are in
  units of (km/10~kpc)$^{2}$.}

\footnotetext[2]{~Taking into account all flux intervals for which the
  color temperature  of the  spectrum is $\le  2.5$~keV; see  text for
  details.}

\footnotetext[3]{~The  range of normalizations  in this  flux interval
  are  consistent   with  no  systematic   uncertainties;  the  quoted
  uncertainties are purely statistical.}

\footnotetext[4]{~For 4U~1746$-$37, all flux intervals are in units of $10^{-9}$~erg~s$^{-1}$~cm$^{-2}$.}
\end{tabular}
\label{results}
\end{table*}

 \renewcommand{\thefootnote}{\arabic{footnote}}

\begin{table*}
\centering
\caption{The Apparent Radii of X-ray Bursters}
\begin{tabular}{lcc}
  \hline\hline
  Source &  R$_{km}$ / D$_{10kpc}$  \\
  \hline
  4U~1636$-$53          & 11.4 $\pm$ 1.0\\
  4U~1702$-$429         & 13.3 $\pm$ 0.4 \\
  4U~1705$-$44          &  9.2 $\pm$ 0.5 \\
  4U~1724$-$307         & 10.7 $\pm$ 0.7 \\
  4U~1728$-$34          & 11.6 $\pm$ 0.7\\
  KS~1731$-$260         &  9.8 $\pm$ 0.4 \\
  4U~1735$-$44          &  8.5$^{+0.08}_{-0.06}$ \\
  4U~1746$-$37         &  4.0 $\pm$ 0.3 \\
  SAX~J1748.9$-$2021    &  9.5 $\pm$ 0.5 \\
  SAX~J1750.8$-$2900    &  9.6 $\pm$ 0.5 \\
\hline
\label{radii_table}
\end{tabular}
\end{table*}

\subsection{Dependence of Blackbody Normalizations on X-ray Flux}

Figure~\ref{fig:norm_vs_flux}   shows  the   result  of   the  outlier
detection  procedure  as applied  to  the  various  flux intervals  of
KS~1731$-$260,    4U~1728$-$35,    and    4U~1636$-$536   (see    also
Table~\ref{results}). Each dot represents  the most likely centroid of
the  intrinsic distribution  of blackbody  normalizations,  while each
error bar represents its most likely width.

The normalization of the blackbody in the case of 4U~1728$-$34 shows a
strong   dependence  on   flux  when   the  source   is   emitting  at
near-Eddington  rates.   This  is   not  seen  in   KS~1731$-$260  and
4U~1636$-$536,   for  which  the   color  temperature   never  reached
temperatures  above 2.5~keV.  In  fact, in  our  observed sample,  the
strong  dependence of the  normalization on  the flux  is seen  in all
three  sources with color  temperatures in  excess of  2.5~keV, namely
4U~1728$-$34, 4U~1702$-$429, and  4U~1724$-$307 (see Fig.~6 below) but
is  absent from  the other  sources.  The evolution  of the  blackbody
normalization   at  near-Eddington   fluxes  depends   on   the  local
gravitational acceleration ($g\sim M/R^2$) on the neutron-star surface
and  on  its composition.   Therefore,  modeling  the  decline of  the
blackbody normalization at large  fluxes would allow us, in principle,
to  measure the  combination $M/R^2$  of the  mass and  radius  of the
neutron star, as has been attempted by Majczyna \& Madej (2005) and by
Suleimanov et al.\ (2011).

\begin{figure*}
\centering
\includegraphics[scale=0.25]{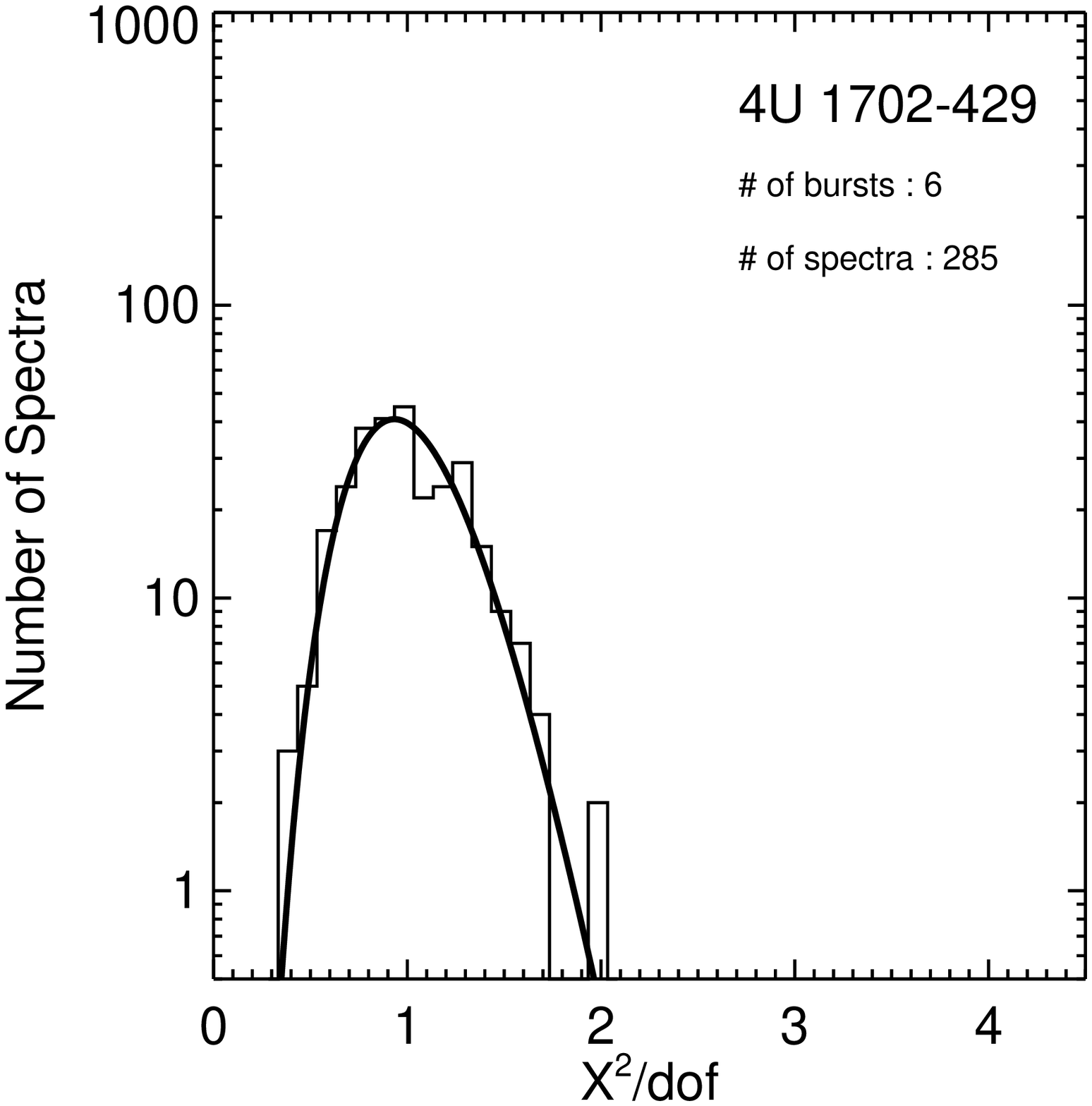}
\includegraphics[scale=0.25]{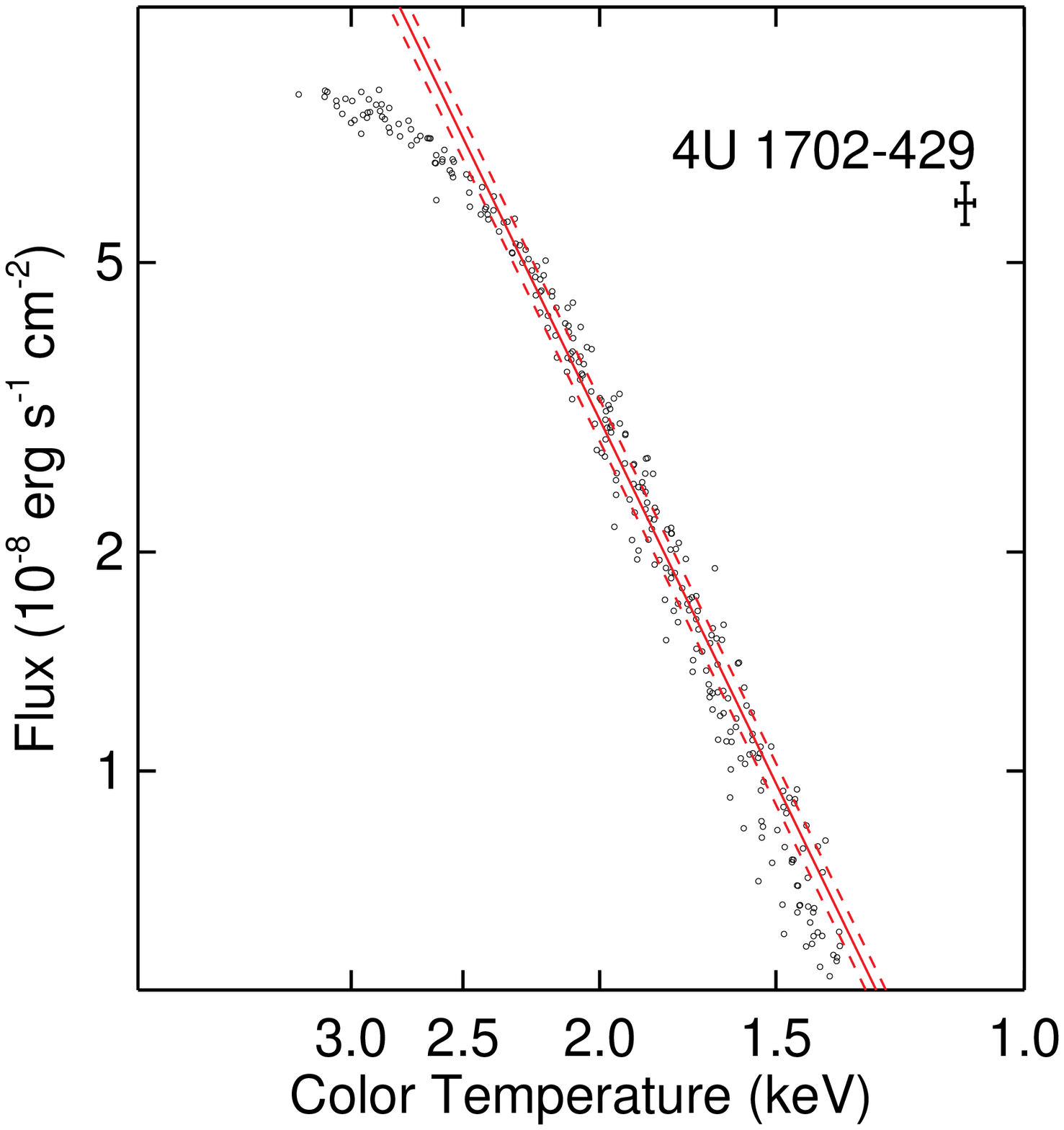}
\includegraphics[scale=0.25]{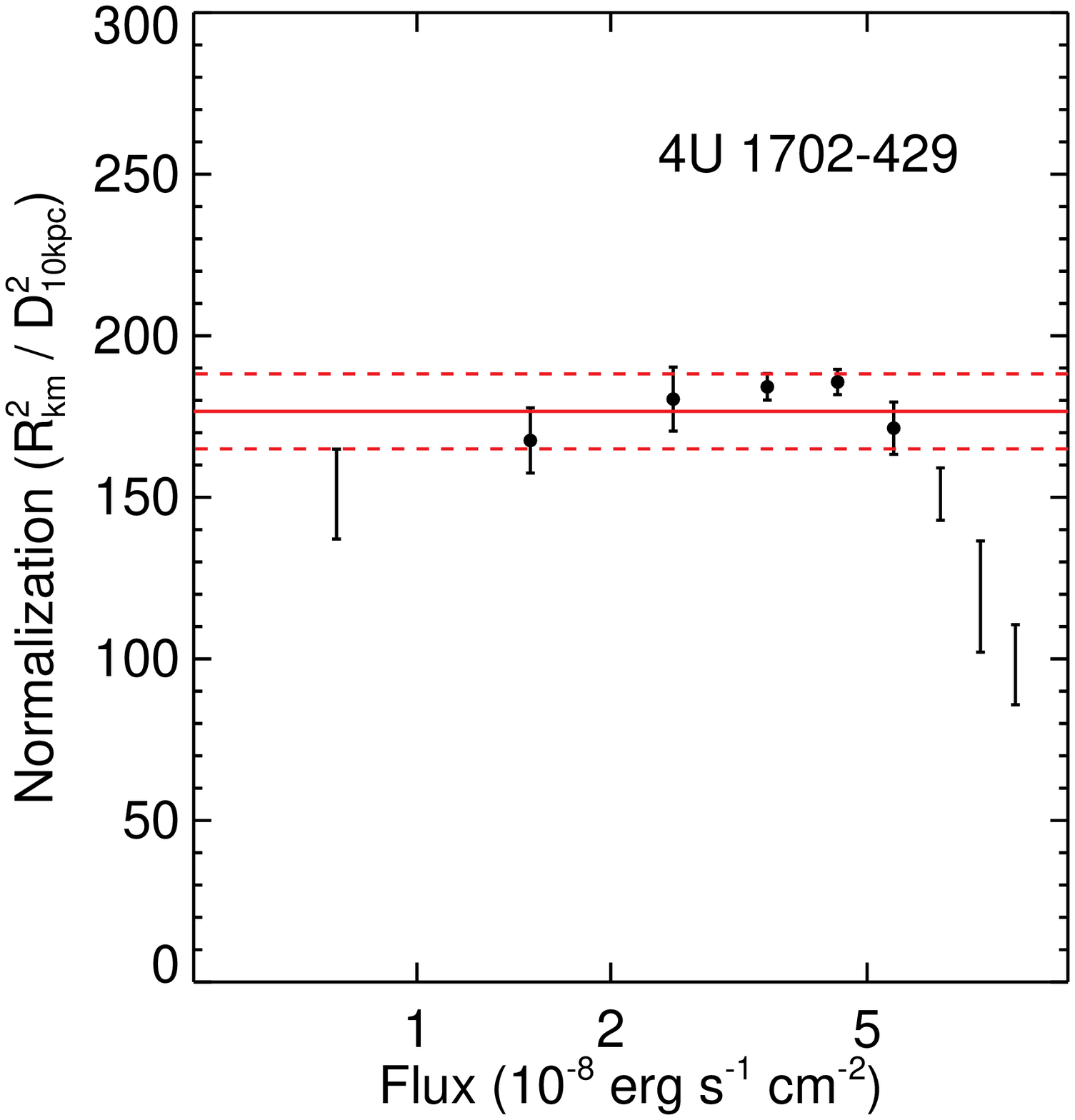}\\

\includegraphics[scale=0.25]{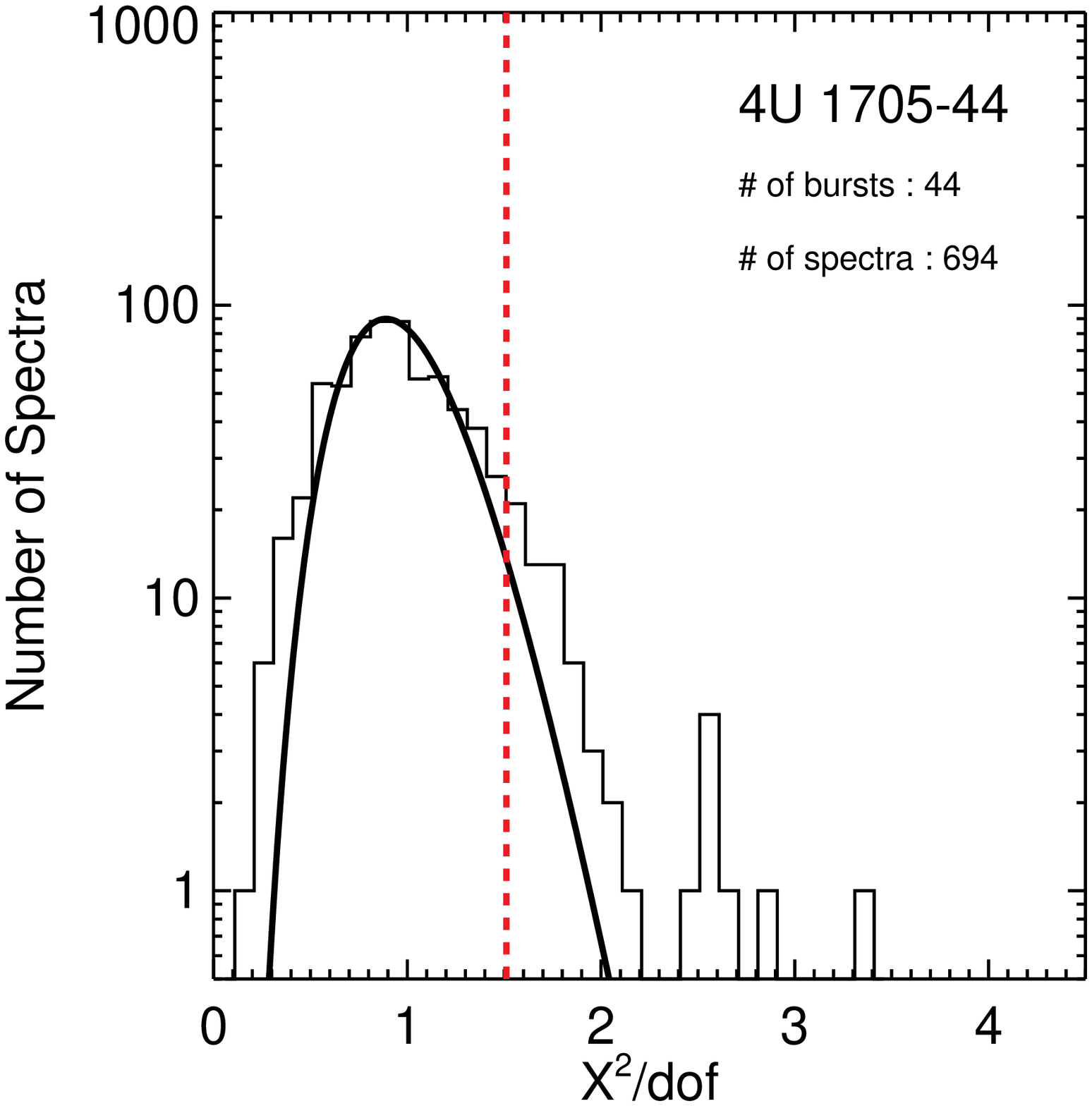}
\includegraphics[scale=0.25]{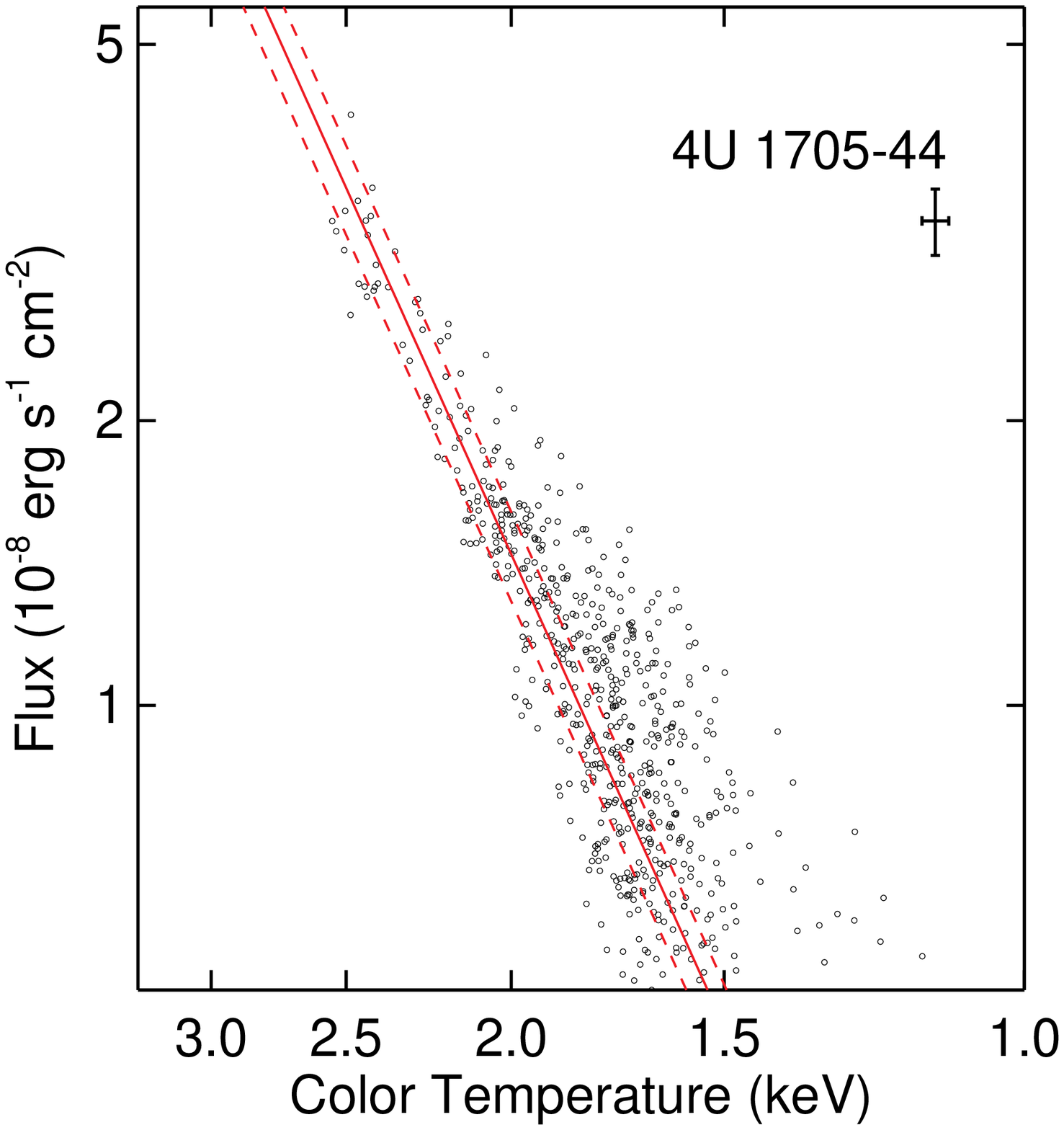}
\includegraphics[scale=0.25]{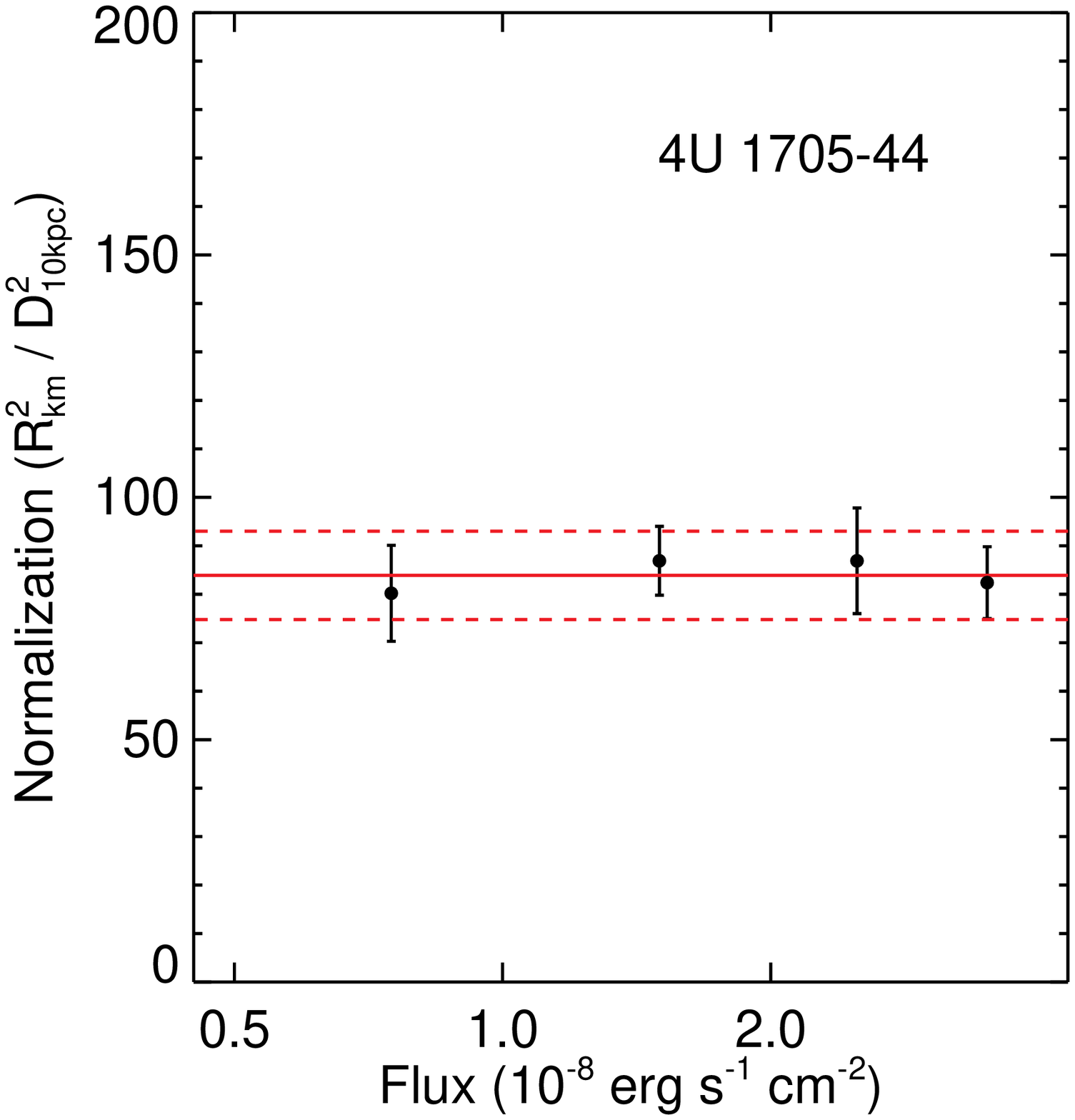}\\

\includegraphics[scale=0.25]{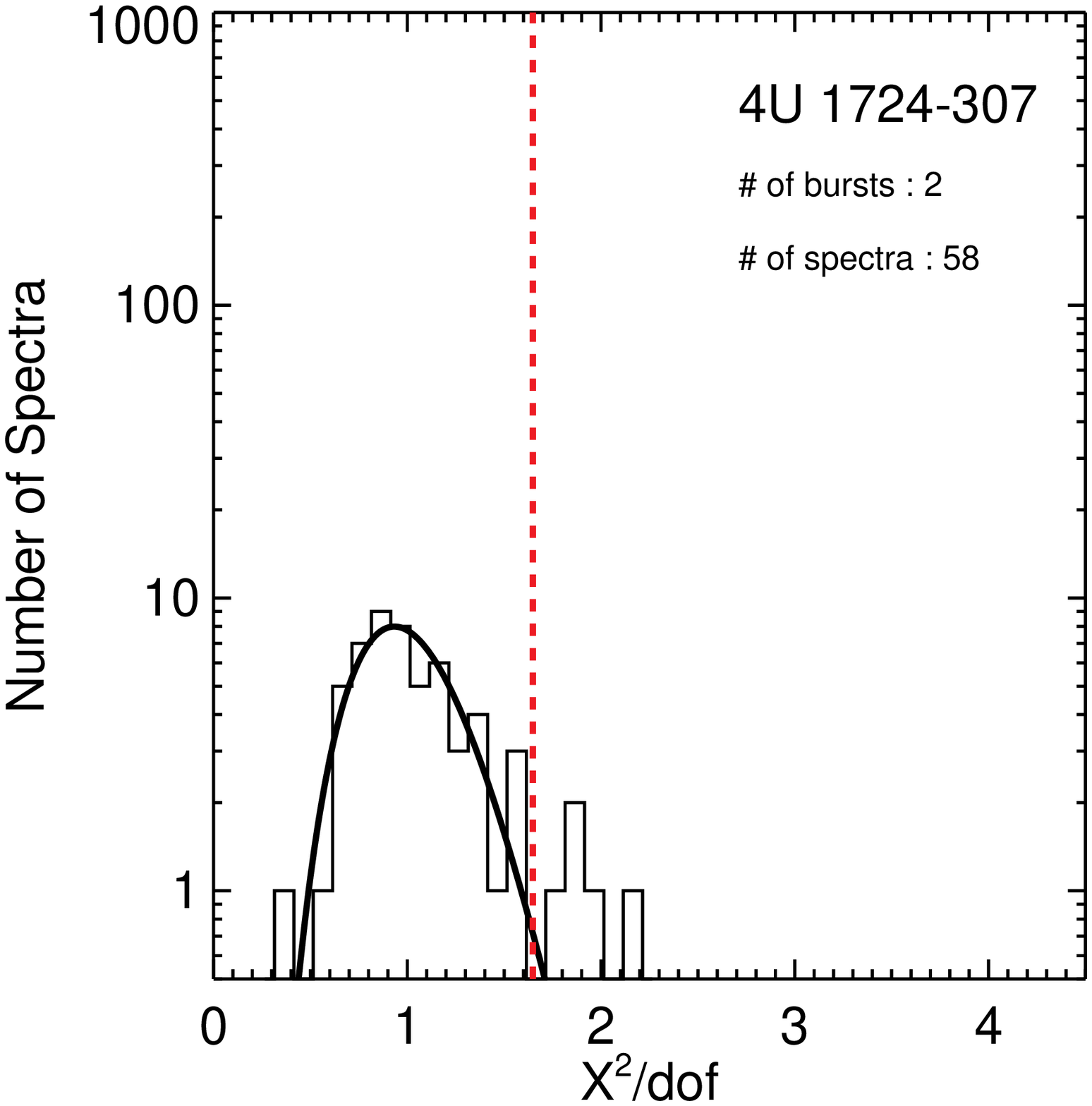}
\includegraphics[scale=0.25]{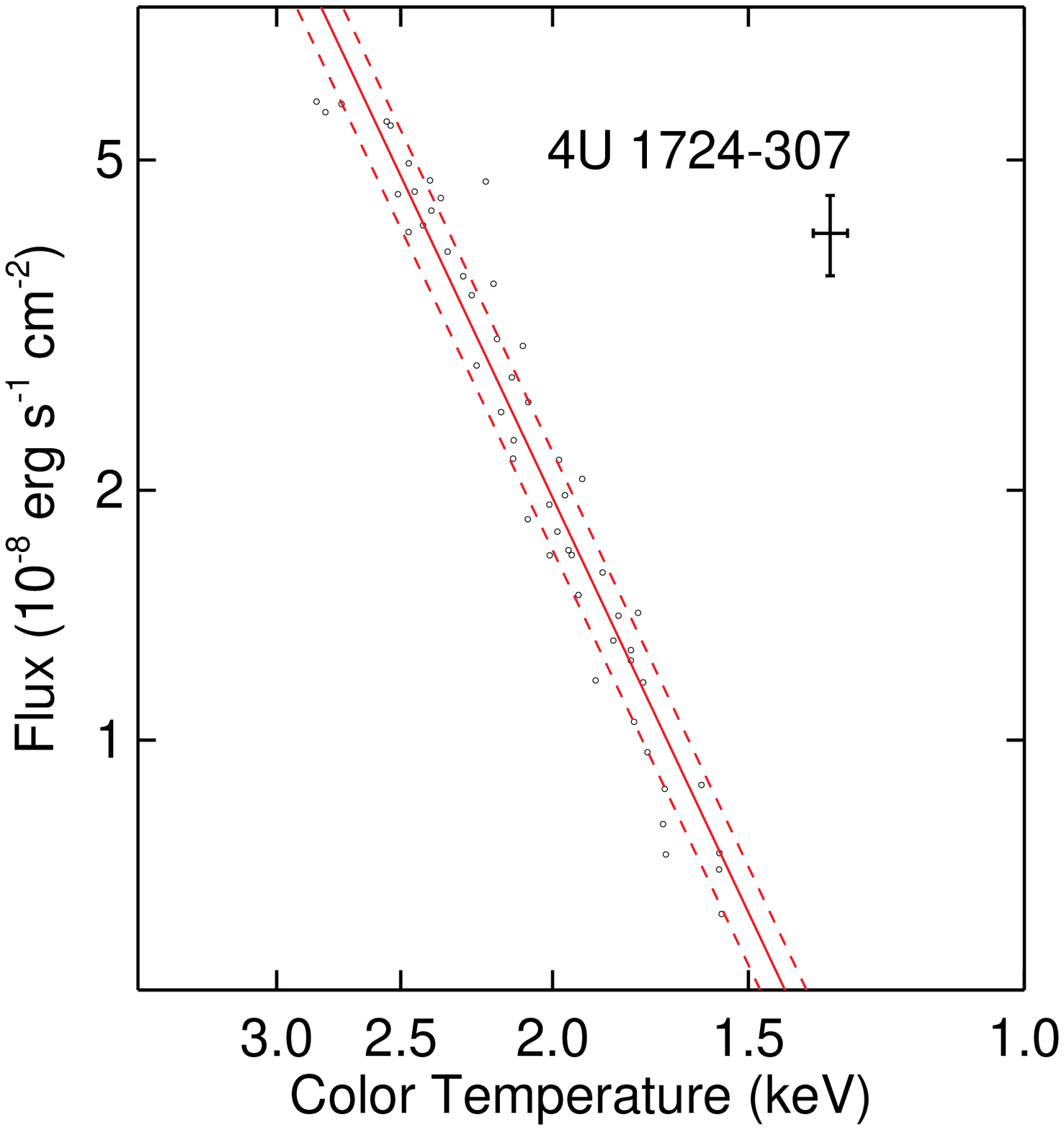}
\includegraphics[scale=0.25]{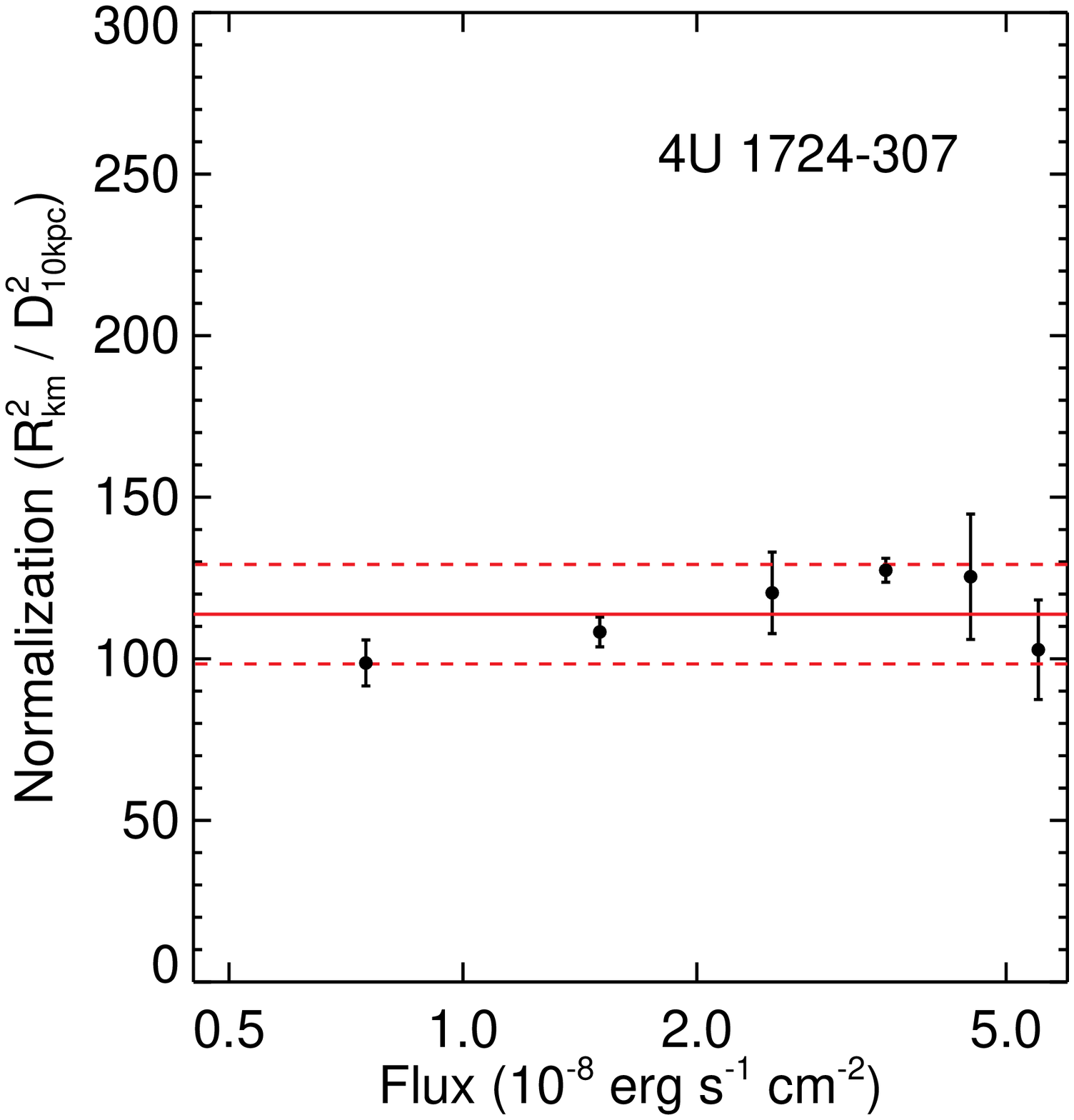}\\

\caption{{\em  (Left)\/}  The  distribution  of the  $X^2$/dof  values
  obtained from fitting the  spectra during the tails of thermonuclear
  X-ray bursts observed  from the sources 4U~1702$-$429, 4U~1705$-$44,
  and   4U~1724$-$307  together   with   the  theoretically   expected
  distribution; the  vertical dashed line  marks the maximum  value of
  $X^2$/dof  beyond  which  we  consider  the blackbody  model  to  be
  inconsistent  with the data.  {\em (Middle)\/}  The flux-temperature
  diagrams of the  cooling tails of bursts from  the same sources; the
  solid and dashed line correspond  to the most probable values of the
  blackbody normalizations throughout  the bursts and their systematic
  uncertainties.  {\em (Right)\/} The  dependence of the parameters of
  the intrinsic  blackbody normalization on X-ray flux;  the solid and
  dashed  lines  correspond  to   the  most  probable  values  of  the
  normalizations and their systematic  uncertainties for the flux bins
  that  are marked  by  a  filled circle.  Error  bars without  filled
  circles appear  at near-Eddington fluxes where  the color correction
  factor   increases,   causing    the   apparent   decline   in   the
  normalization.}
\label{fig:radii1}
\end{figure*}

\begin{figure*}
\centering
\includegraphics[scale=0.25]{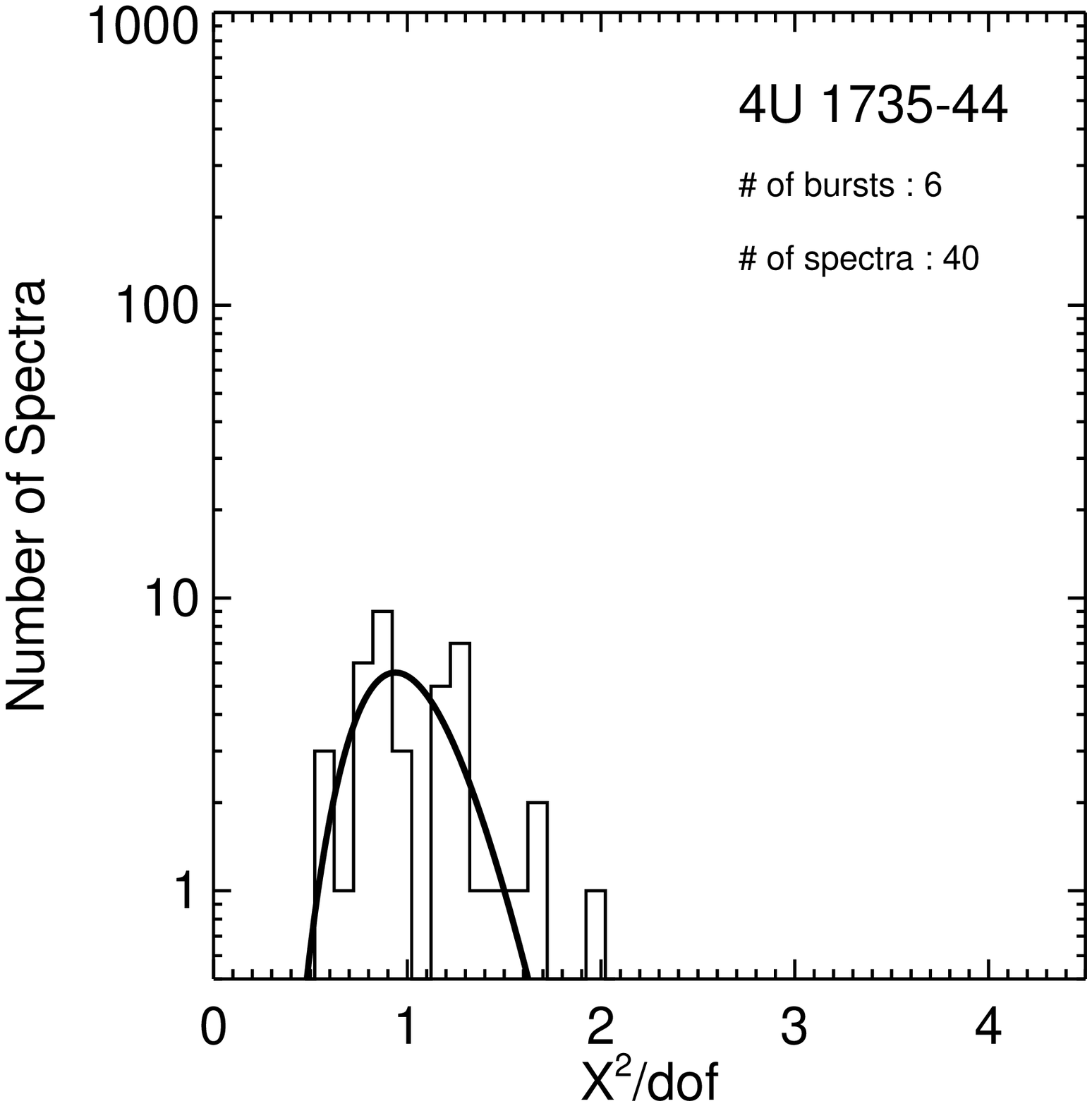}
\includegraphics[scale=0.25]{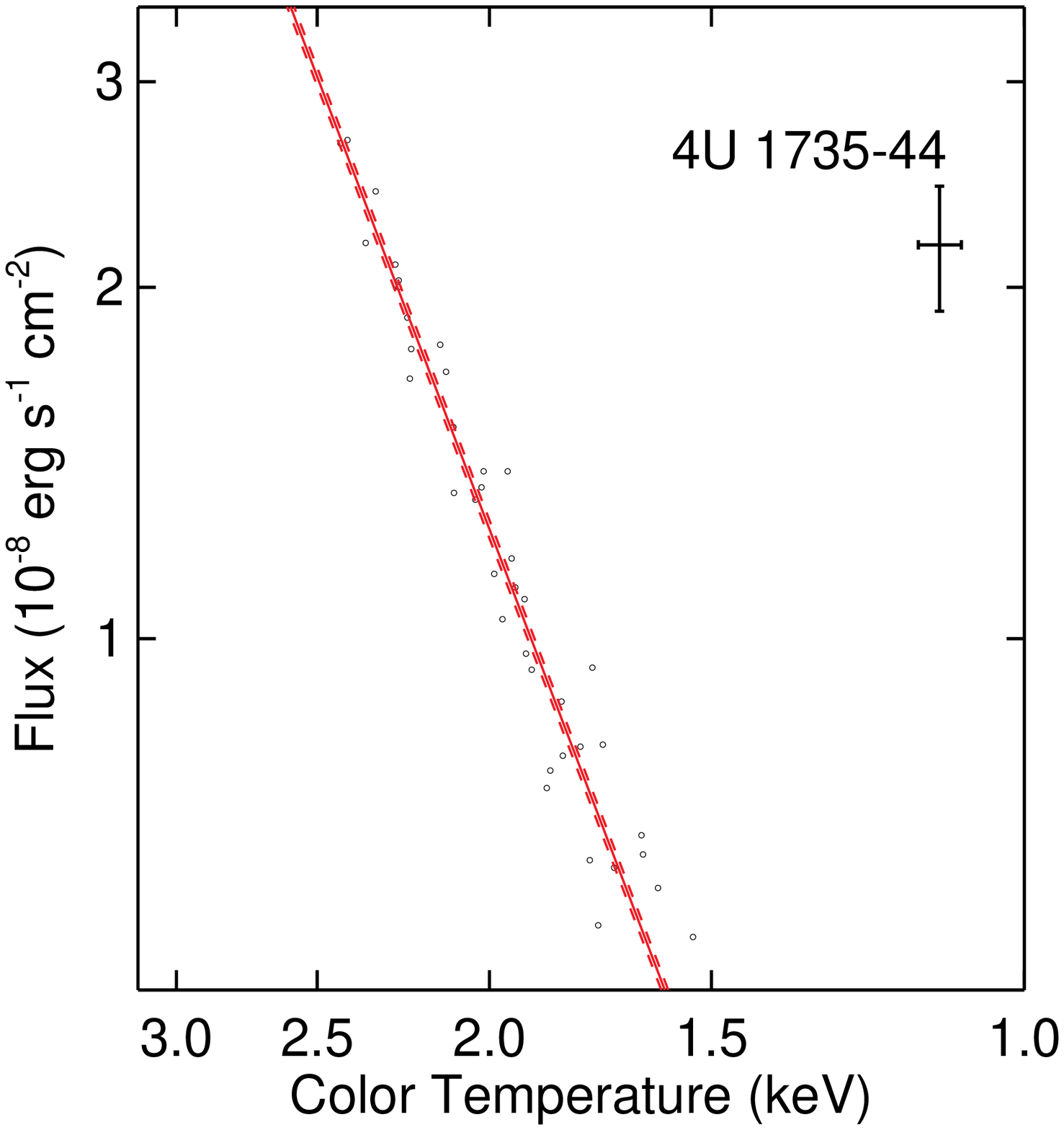}
\includegraphics[scale=0.25]{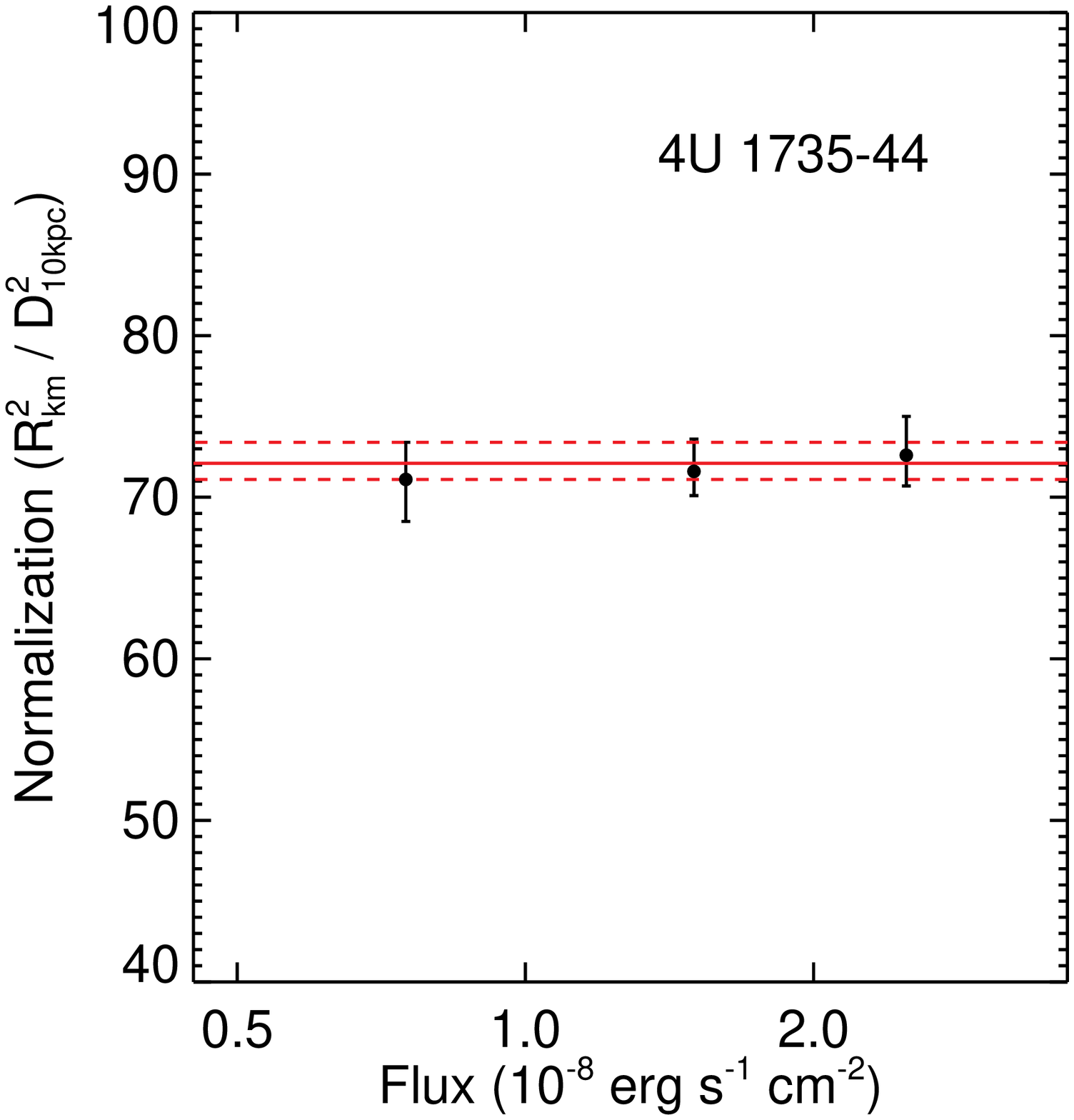}\\

\includegraphics[scale=0.25]{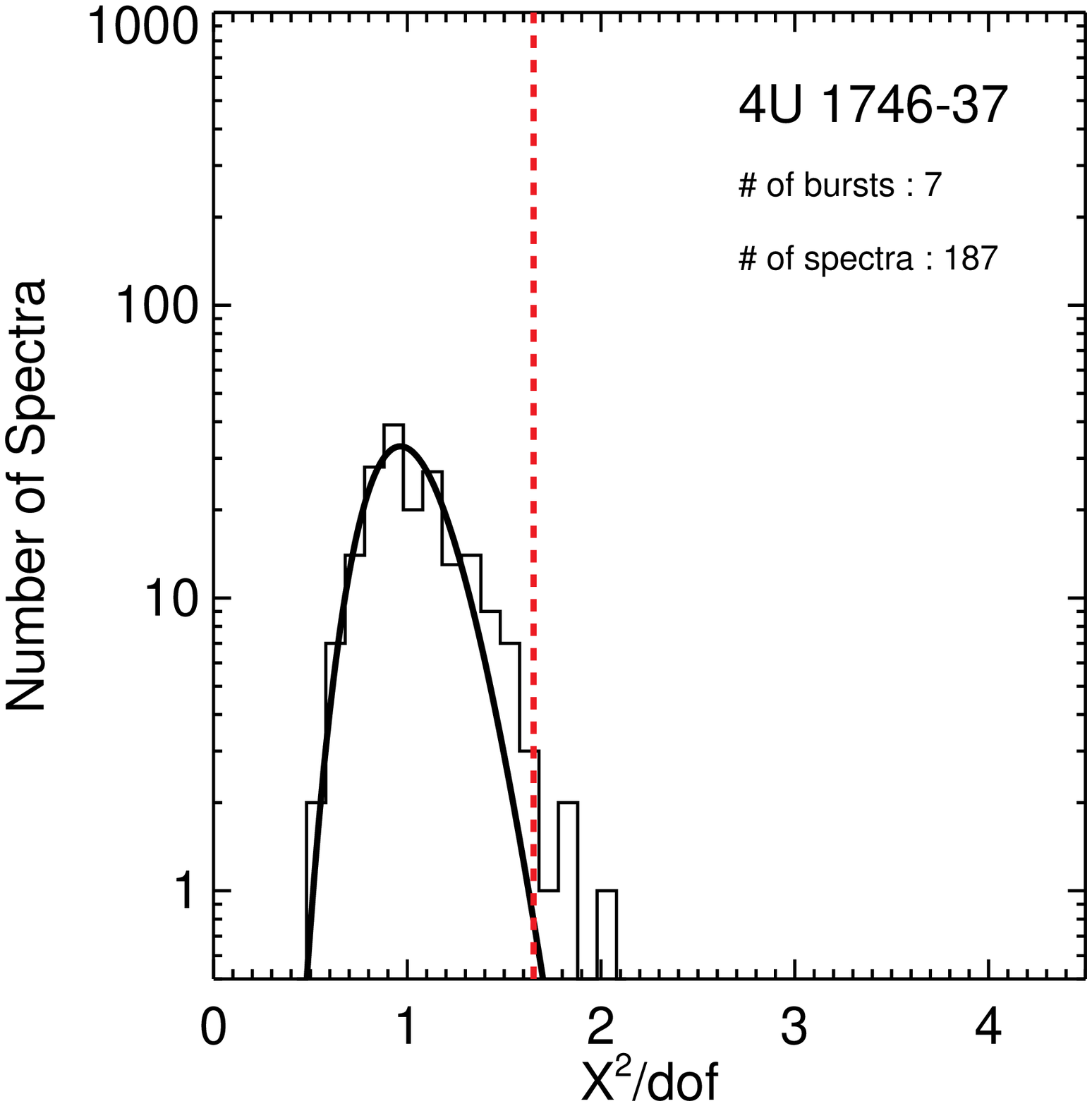}
\includegraphics[scale=0.25]{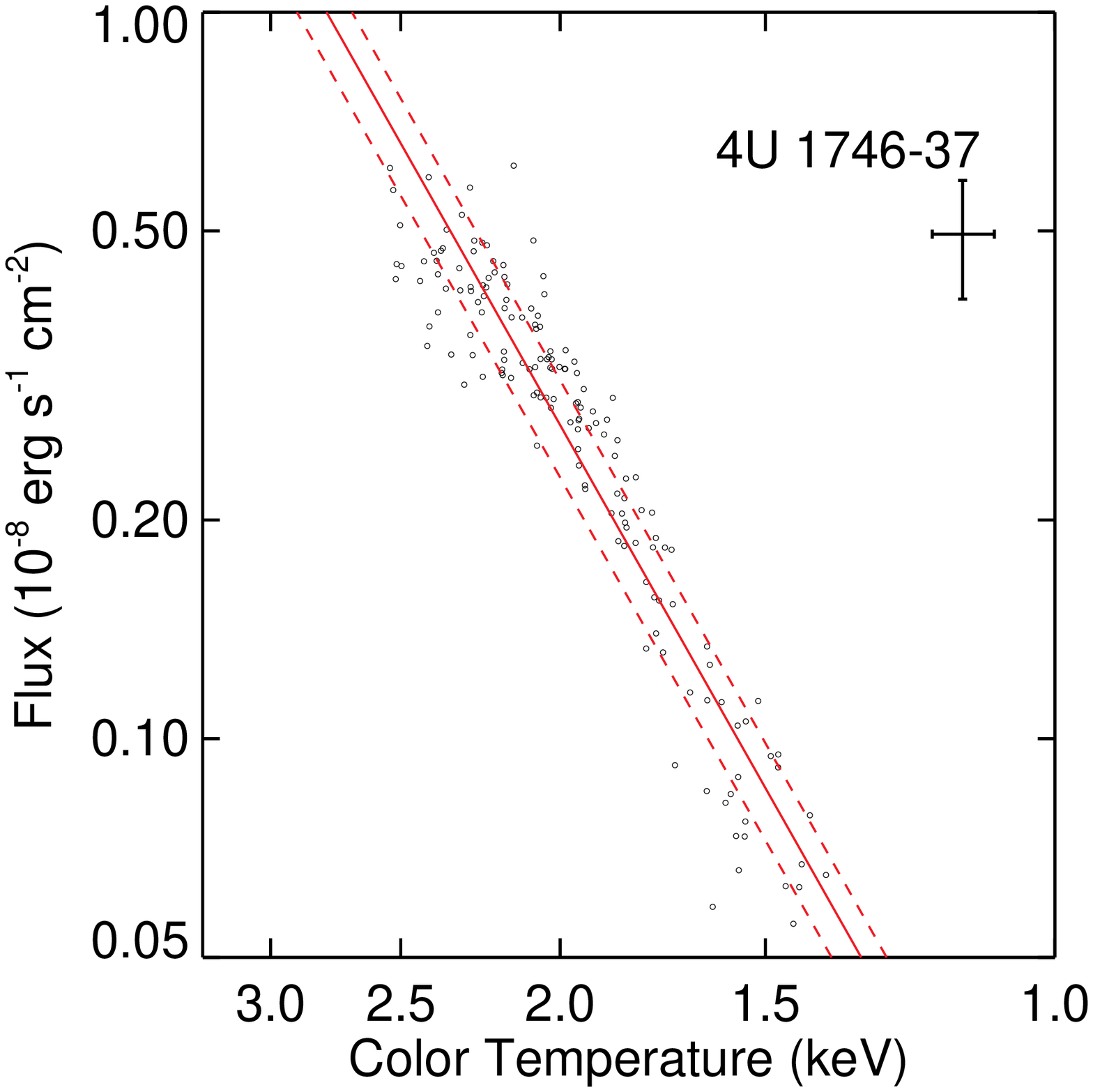}
\includegraphics[scale=0.25]{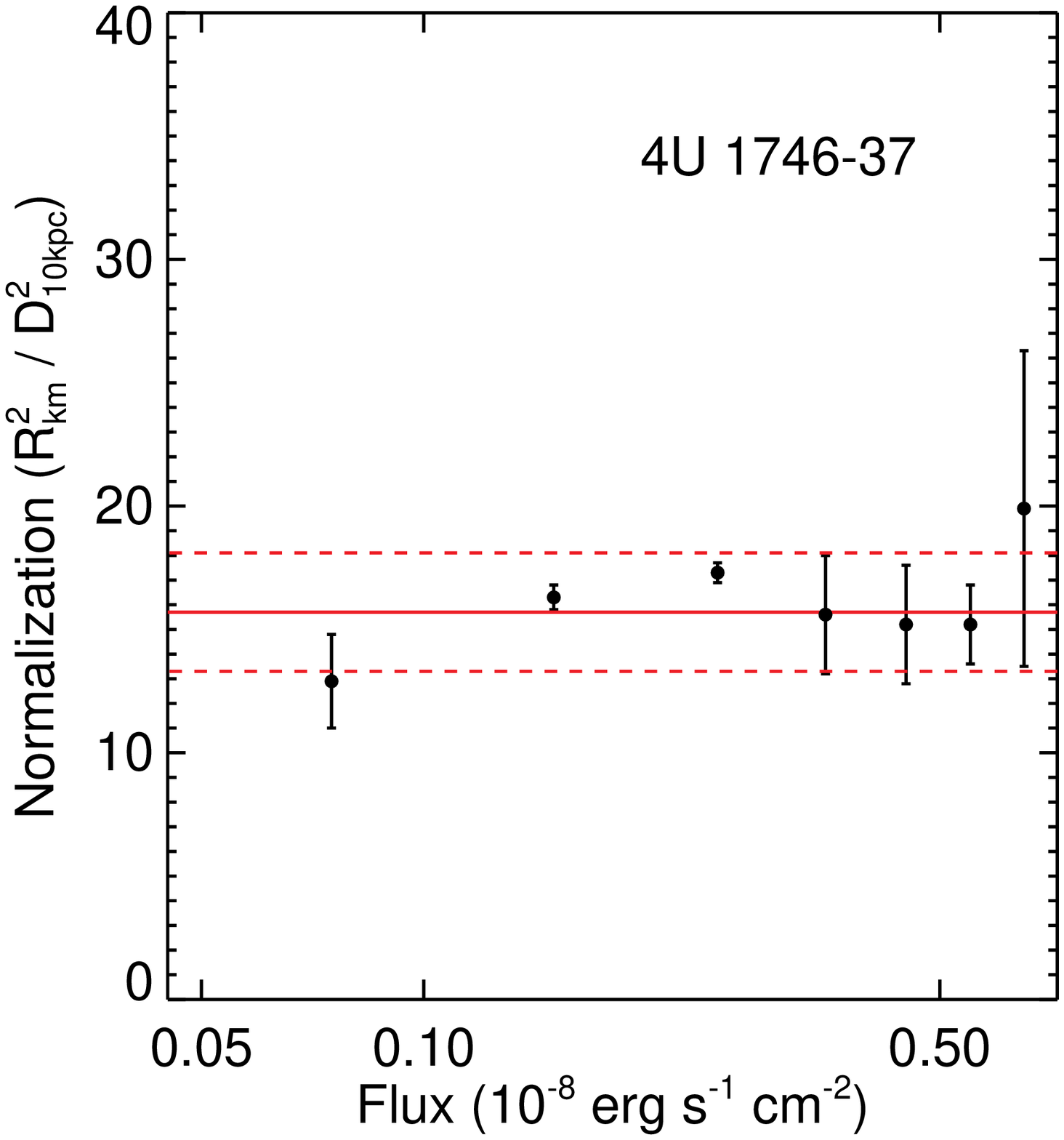}\\

\includegraphics[scale=0.25]{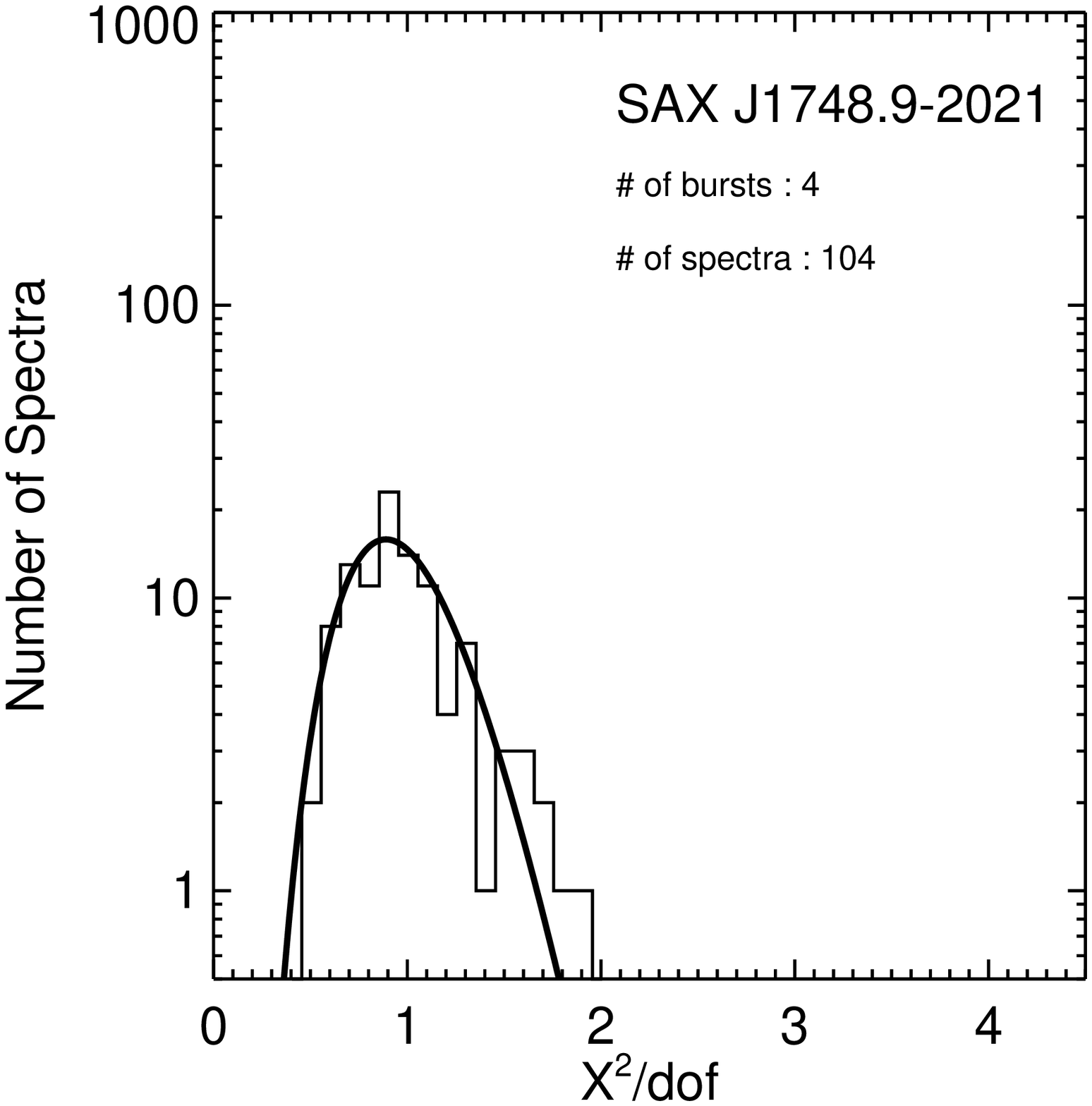}
\includegraphics[scale=0.25]{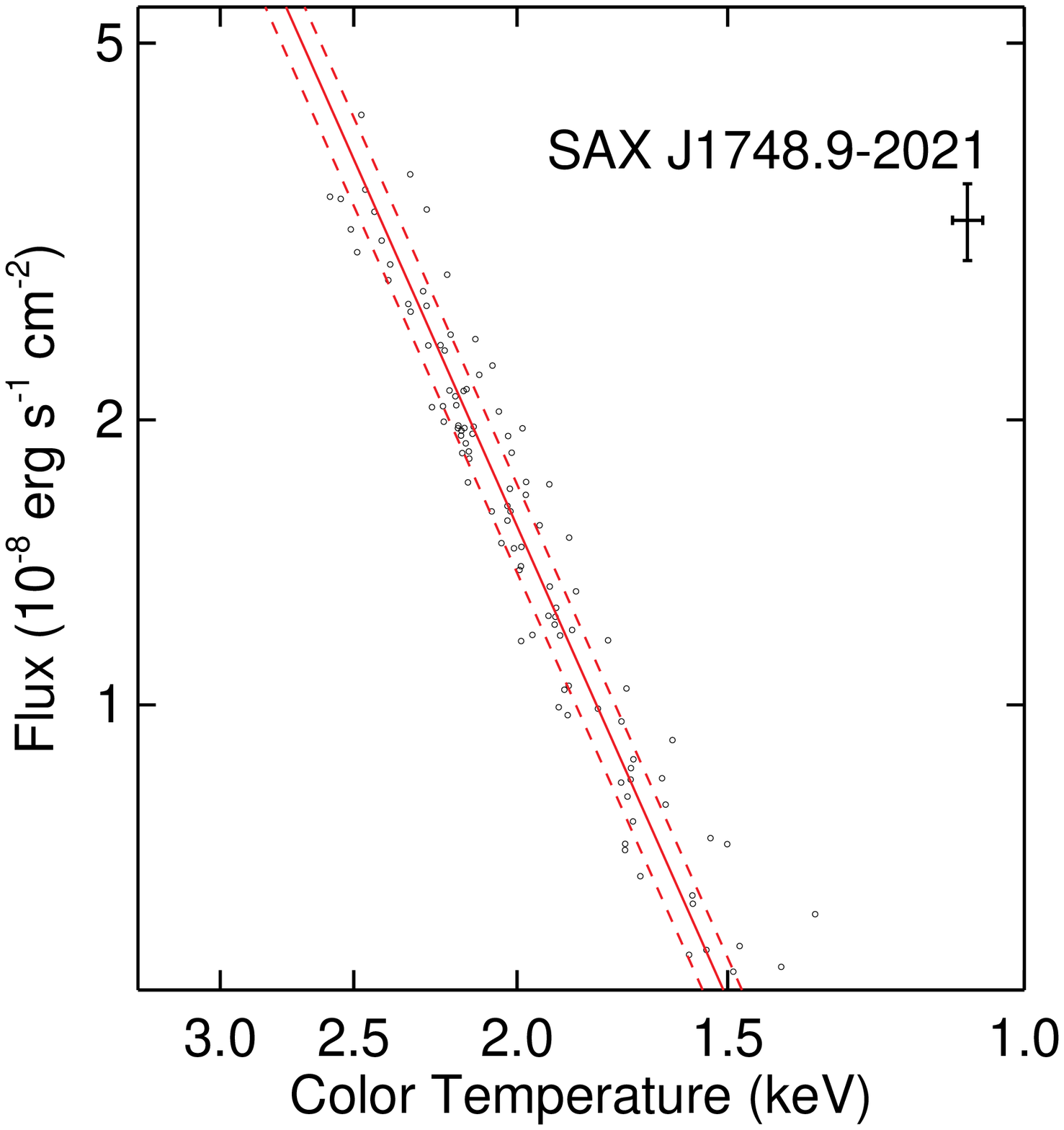}
\includegraphics[scale=0.25]{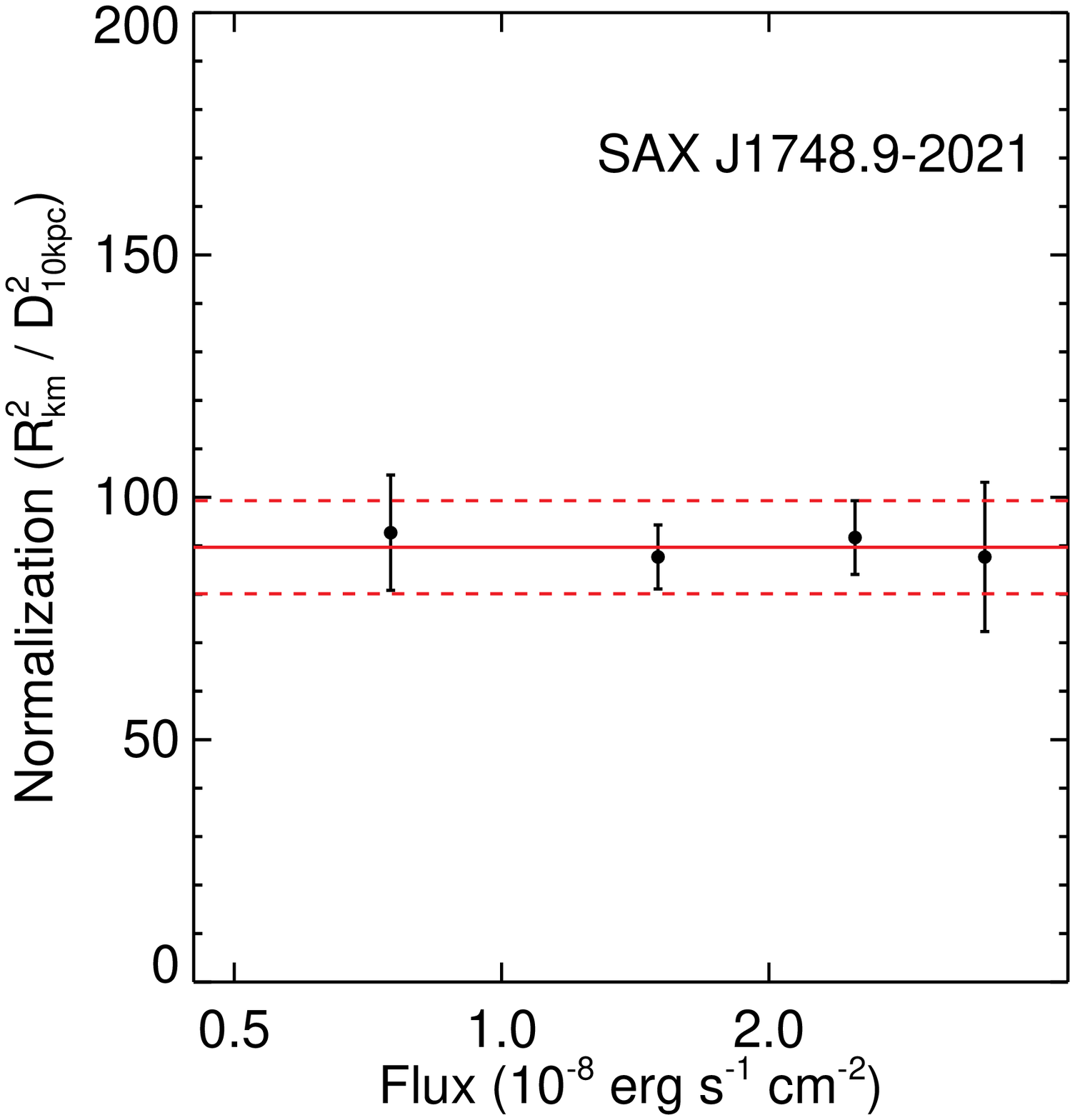}\\
   \caption{Same as Figure~\ref{fig:radii1} but for the
     sources 4U~1735$-$44, 4U~1746$-$37 and SAX~J1748.9$-$2021.}
\label{fig:radii2}
\end{figure*}

\begin{figure*}
\centering
   \includegraphics[scale=0.25]{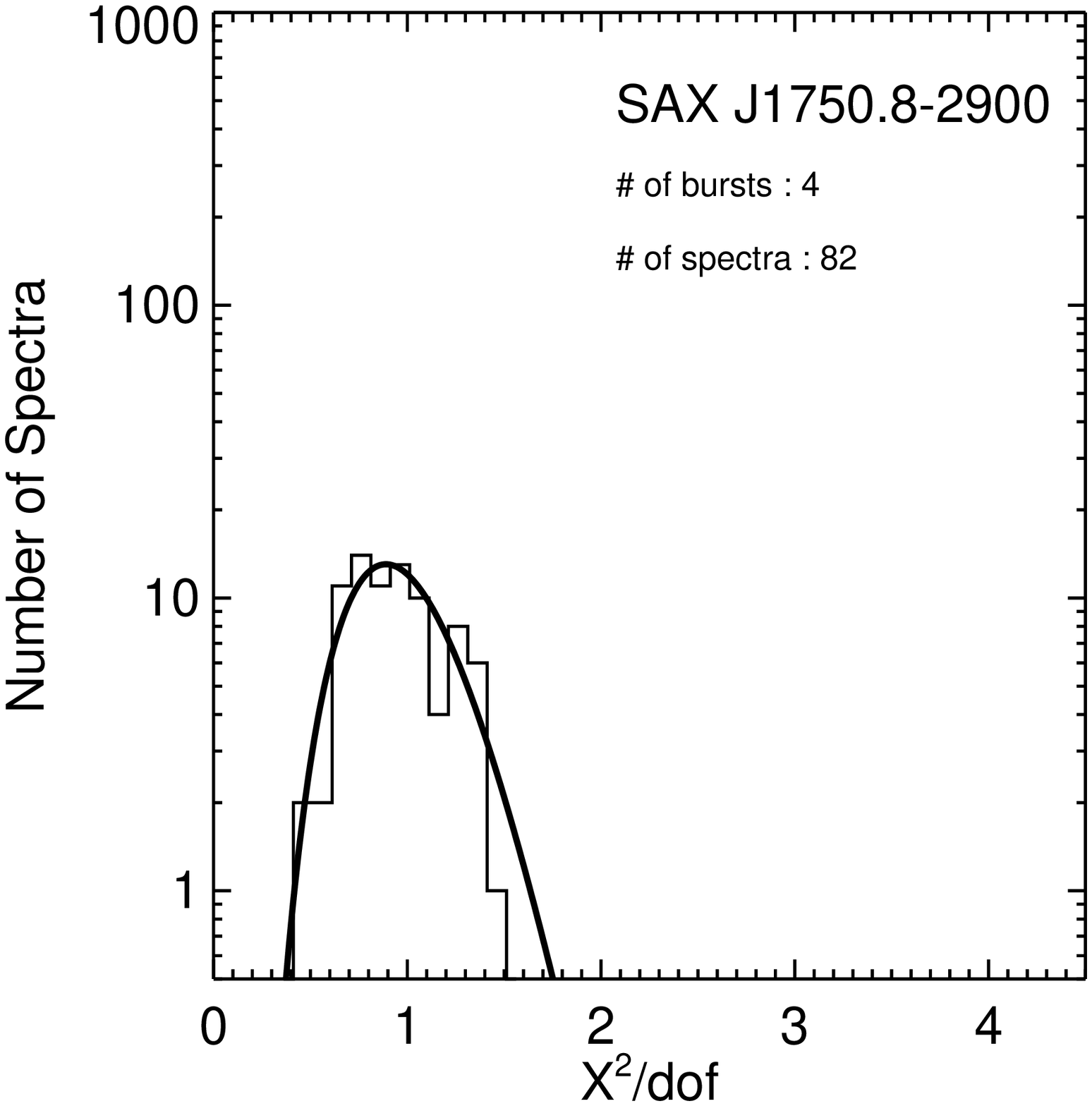}
   \includegraphics[scale=0.25]{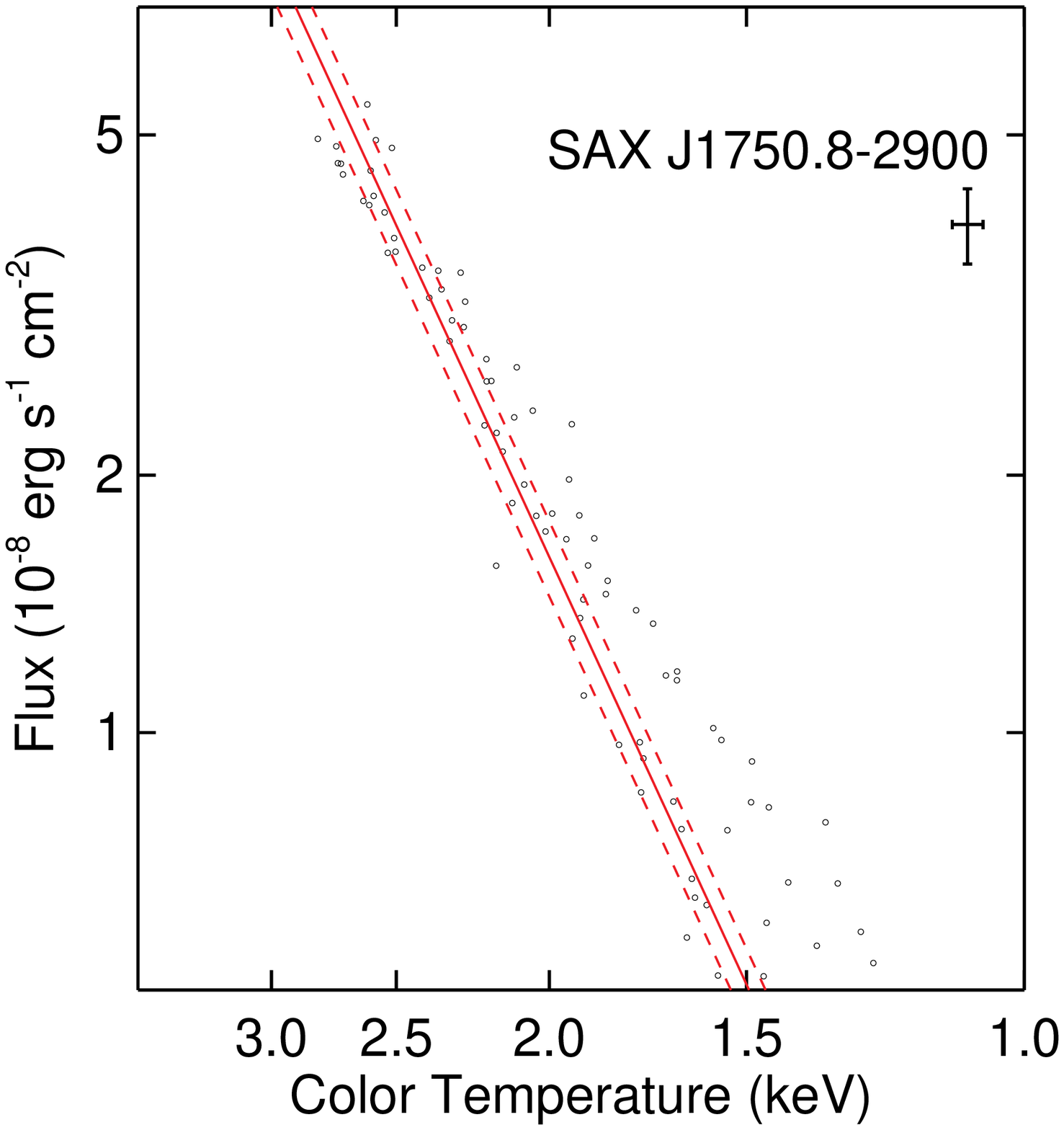}
   \includegraphics[scale=0.25]{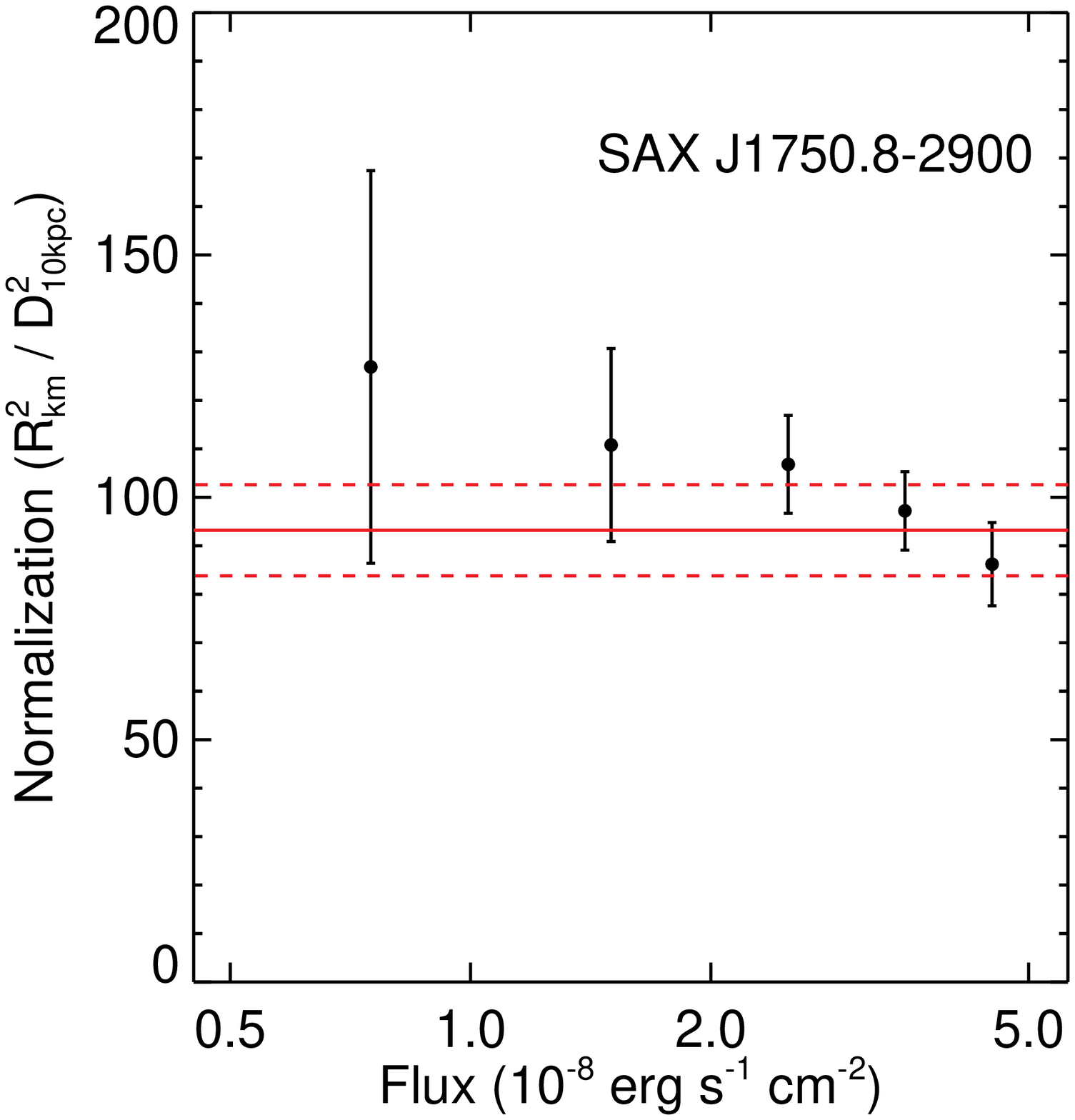}

   \caption{Same  as   Figure~\ref{fig:radii1}  but  for   the  source
     SAX~J1750.8$-$2900. Note that, for this source, the low number of
     spectra  precluded the  clear identification  of outliers  in the
     individual  flux  intervals  and  resulted  in  large  systematic
     uncertainties. However,  when all flux  intervals were considered
     together,  a population  of outliers  (coming primarily  from one
     burst) became  clear and the intrinsic distribution  was found to
     be consistent with having no systematic uncertainties.}
\label{fig:radii3}
\end{figure*}

At low burst  fluxes, the observed normalizations show  a small albeit
statistically significant  trend towards lower values. In  the case of
4U~1728$-$34,  where  the  potential  decrease  is  the  largest,  the
normalization   changed   from   $133.7\pm  16.4$~(km/10~kpc)$^2$   to
$114.0\pm  15.6$~(km/10~kpc)$^2$ as  the flux  declined  from $6\times
10^{-8}$~erg~s$^{-1}$~cm$^{-2}$              to             $0.5\times
10^{-8}$~erg~s$^{-1}$~cm$^{-2}$.   This  corresponds  to  a  $\lesssim
15$\% reduction in the normalization. In other sources, the decline is
even smaller. This weak dependence seen in the data argues against the
models with solar composition, as shown in Figure~\ref{fig:fcolor}.

The decline in  the normalization can, in principle,  be accounted for
with a  $\lesssim 4$\%  increase in the  color correction  factor (see
eq.~[\ref{eq:rapp}])  towards  low temperatures.  As  can  be seen  in
Figure~\ref{fig:fcolor}, current atmosphere models do not predict such
an  evolution. However, as  we discussed  in the  beginning of  \S4, a
number of effects related to the physics of bursts on the neutron star
surface  (uneven  cooling and  burst  oscillations),  as  well as  the
reduced sensitivity of the PCA at low energies, are capable of causing
the observed trend in  the normalization. Moreover, the $\lesssim 4\%$
effect  in  the data  we  would  aim to  model  is  comparable to  the
uncertainty  in inferring  theoretically the  color  correction factor
from the atmosphere models and  is also comparable to the deviation of
the  observed and  theoretical  spectra from  blackbodies. The  latter
concern can be remedied  by fitting directly theoretical model spectra
to the data, but reducing the theoretical uncertainties present in the
models  to  less  than   $\sim$  few  percent  is  significantly  more
challenging.

An alternative  approach is  to allow  for a range  of values  for the
color  correction   factor  that   span  the  spread   of  theoretical
uncertainties,  metallicities, and  fluxes of  the cooling  tails.  In
earlier work (G\"uver et al.\ 2010a, 2010b) we allowed for a 7\% range
in the color correction factor (from 1.3 to 1.4), which is adequate to
account for the evolution in the blackbody normalizations shown by the
data.  Following this  approach,  we  need to  identify  the range  of
blackbody normalizations  we obtained from the data  at different flux
intervals as a systematic uncertainty  in the measurement. In order to
achieve this,  we applied the  Bayesian Gaussian mixture  algorithm to
the  combined data set  for each  source for  all flux  intervals that
correspond to a color temperature $\le 2.5$~keV. This way, we computed
the  most probable  value  for the  blackbody  normalization for  each
source in a wide flux range as well as the systematic uncertainties in
that  measurement, which  account  for the  potential  decline of  the
normalization with decreasing flux.

In  Figure~\ref{fig:norm_vs_flux}, we identify  the flux  intervals we
used for each source with a  filled circle in the middle of each error
bar. We also  depict with a solid line the most  probable value of the
normalization in this wide flux  range and with dashed lines the range
of systematic  uncertainties. Our  results are summarized  in Table~3.
The  blackbody  normalizations  for KS~1731$-$260,  4U~1728$-$34,  and
4U~1636$-$536             are            96.0$\pm$7.9~(km/10~kpc)$^2$,
134.4$\pm$14.9~(km/10~kpc)$^2$,   and  130.7$\pm$20.9~(km/10~kpc)$^2$,
respectively,   with   the   uncertainties   dominated   entirely   by
systematics.

\section{The Apparent Radii of X-ray Bursters}

Figures~\ref{fig:radii1}--\ref{fig:radii3}                          and
Tables~\ref{chi2table}--\ref{results} show  the results obtained after
applying  the procedure  outlined in  \S4 to  seven more  sources from
Table~\ref{sourcestable}.   The  last  two sources,  4U~0513$-$40  and
Aql~X-1  show large  variations  in the  cooling  tails of  individual
bursts and are  discussed in detail in the  Appendix. Moreover, in the
Appendix we also  discuss a number of bursts  from 4U~1702$-$429 and a
long  burst  from 4U~1724$-$307,  which  we  did  not include  in  the
analysis.

As in the case of the three sources discussed in the previous section,
the vast majority of the  spectra observed during the cooling tails of
X-ray bursts are very well described by blackbody functions, with only
marginal allowance for  systematic variations. In the flux-temperature
diagrams,  the  cooling  tails  largely  follow  a  well  defined  and
reproducible track.   Finally, in each flux interval  of most sources,
the range of blackbody normalizations  obtained from a large number of
bursts is consistent with a small degree of systematic uncertainties.

The  measurements  of  the  average blackbody  normalizations  in  all
sources are dominated by  systematic uncertainties (the main exception
is 4U~1735$-$44). However, these uncertainties are small, ranging from
$\simeq 6$\% in the case of 4U~1702$-$429 to $\simeq 16$\% in the case
of 4U~1636$-$536. The  apparent radius of each neutron  star scales as
the  square root  of the  blackbody  normalization. As  a result,  the
errors  in  the  spectroscopic  determination  of  neutron-star  radii
introduced by systematic effects in  the cooling tails of X-ray bursts
are in  the range  $\simeq 3-8$\% (see  Table~\ref{radii_table}). Such
small  errors by  themselves  do not  preclude distinguishing  between
different equations of state of neutron-star matter.

\acknowledgements{We  thank Duncan  Galloway for  numerous discussions
  and his significant contribution to  the data analysis.  We thank an
  anonymous  referee  for  insightful  comments and  bringing  to  our
  attention  the effect of  the deadtime  corrections in  the spectral
  analysis. DP was  supported by the NSF CAREER  award NSF 0746549 and
  Chandra Theory grant TMO-11003X.   FO acknowledges support from NASA
  ADAP  grant NNX10AE89G  and  Chandra Theory  grant TMO-11003X.  This
  research  has  made  use  of  data obtained  from  the  High  Energy
  Astrophysics Science Archive  Research Center (HEASARC), provided by
  NASA's  Goddard Space  Flight Center,  and of  the  SIMBAD database,
  operated at CDS, Strasbourg, France.}

\appendix

\begin{figure*}
\centering
   \includegraphics[scale=0.35]{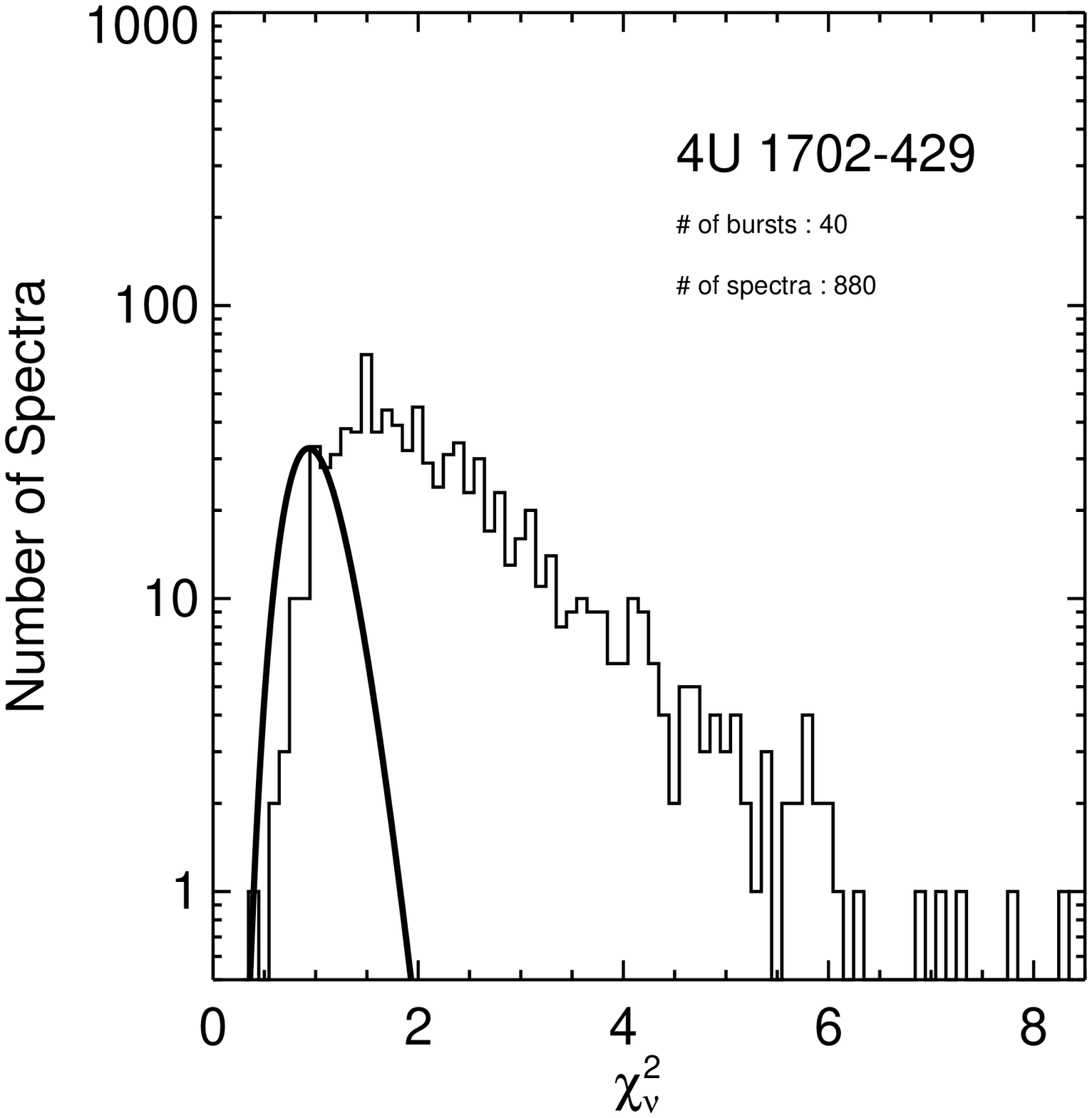}
   \includegraphics[scale=0.35]{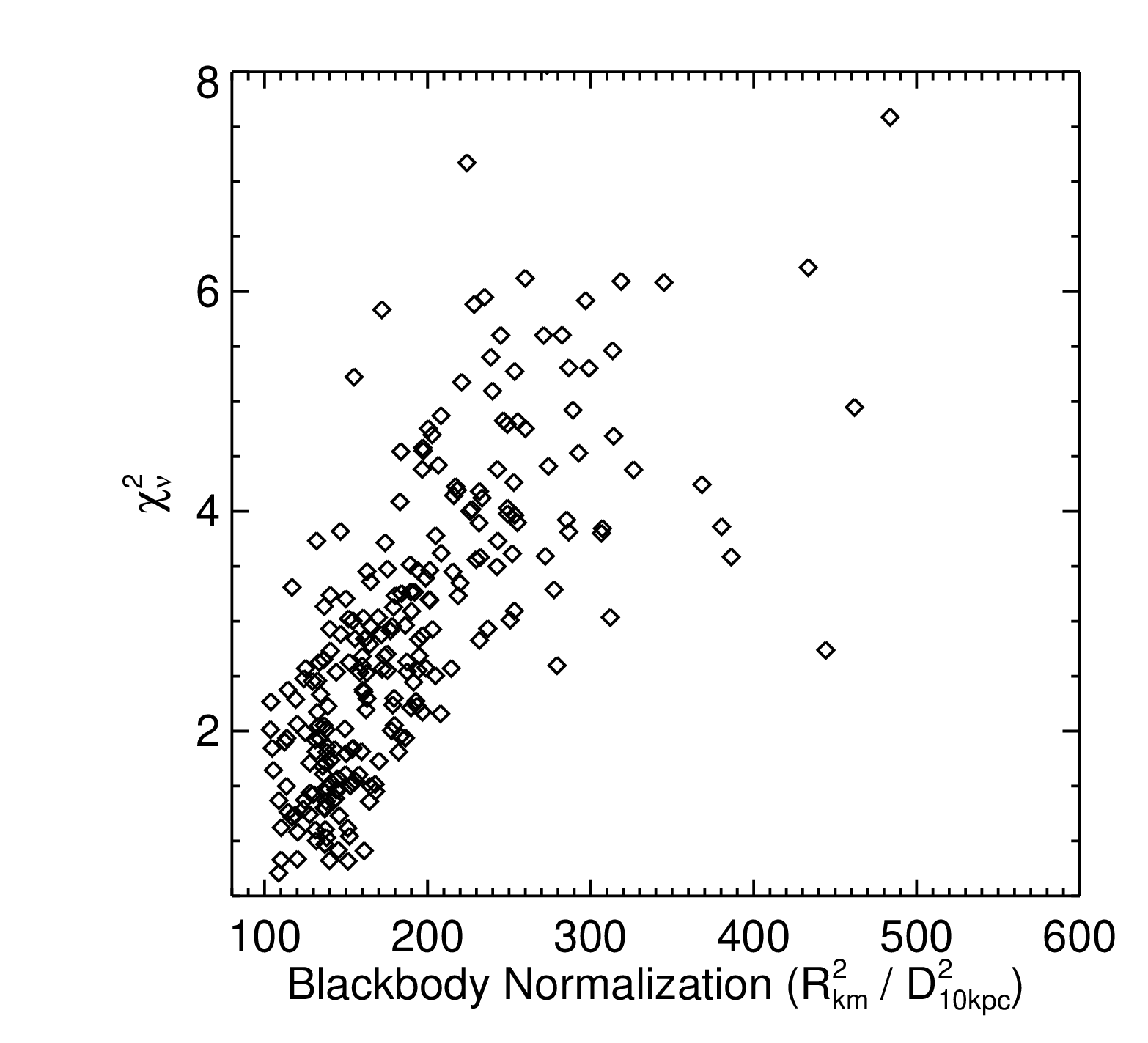}
   \caption{{\em (Left)\/}  The distribution of  $\chi^2$/dof obtained
     from fitting  the spectra of  40 weak X-ray bursts  observed from
     4U~1702$-$429; the solid line shows the expected distribution for
     the  same  number of  degrees  of  freedom.  {\em (Right)\/}  The
     relation  between the  inferred blackbody  normalization  and the
     $\chi^2$/dof  observed during the  cooling tails  of the  40 weak
     bursts observed from 4U~1702$-$429.}
\label{fig:1702_appendix}
\end{figure*}

\begin{figure*}
\centering
   \includegraphics[scale=0.35]{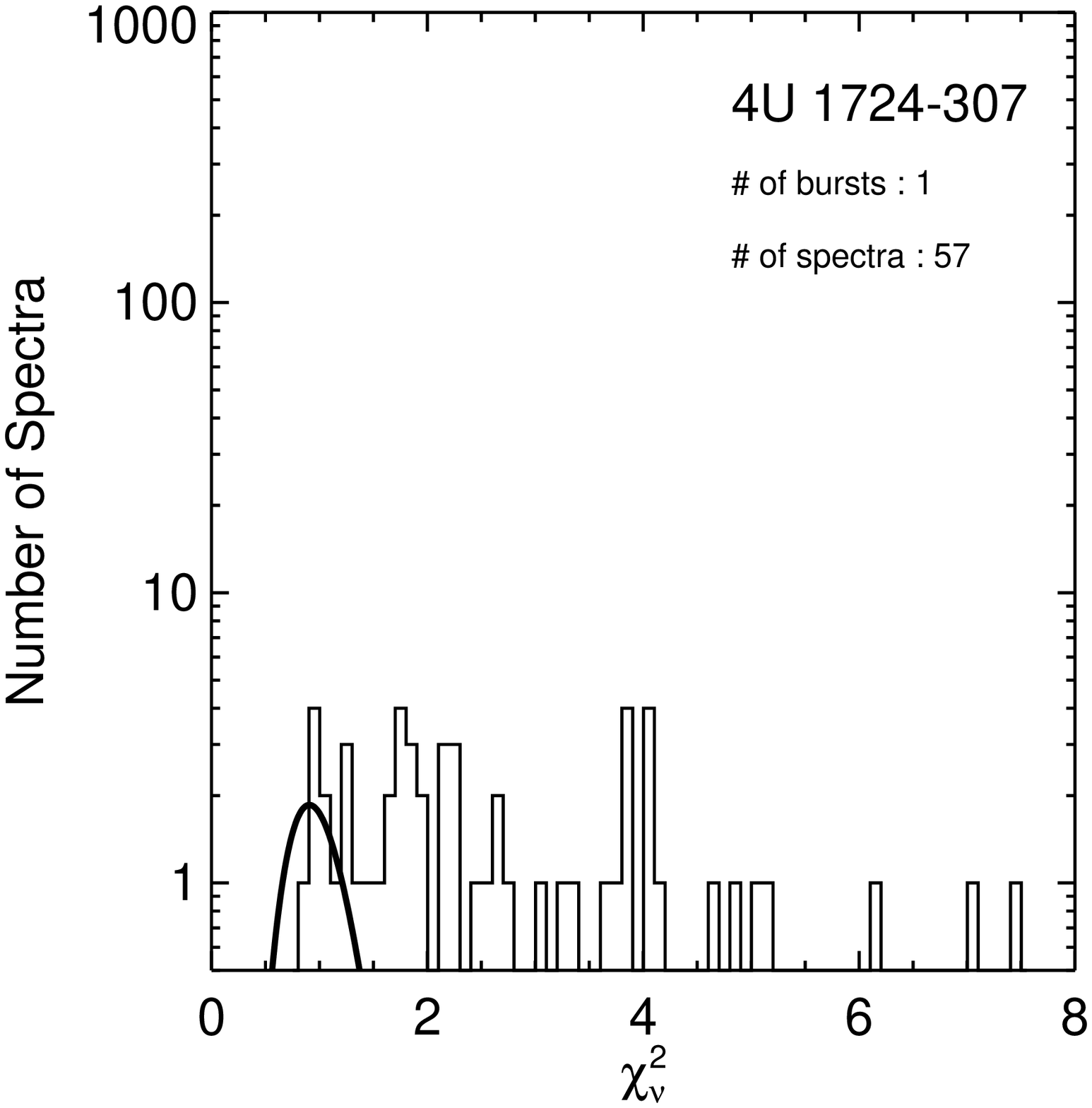}
   \includegraphics[scale=0.35]{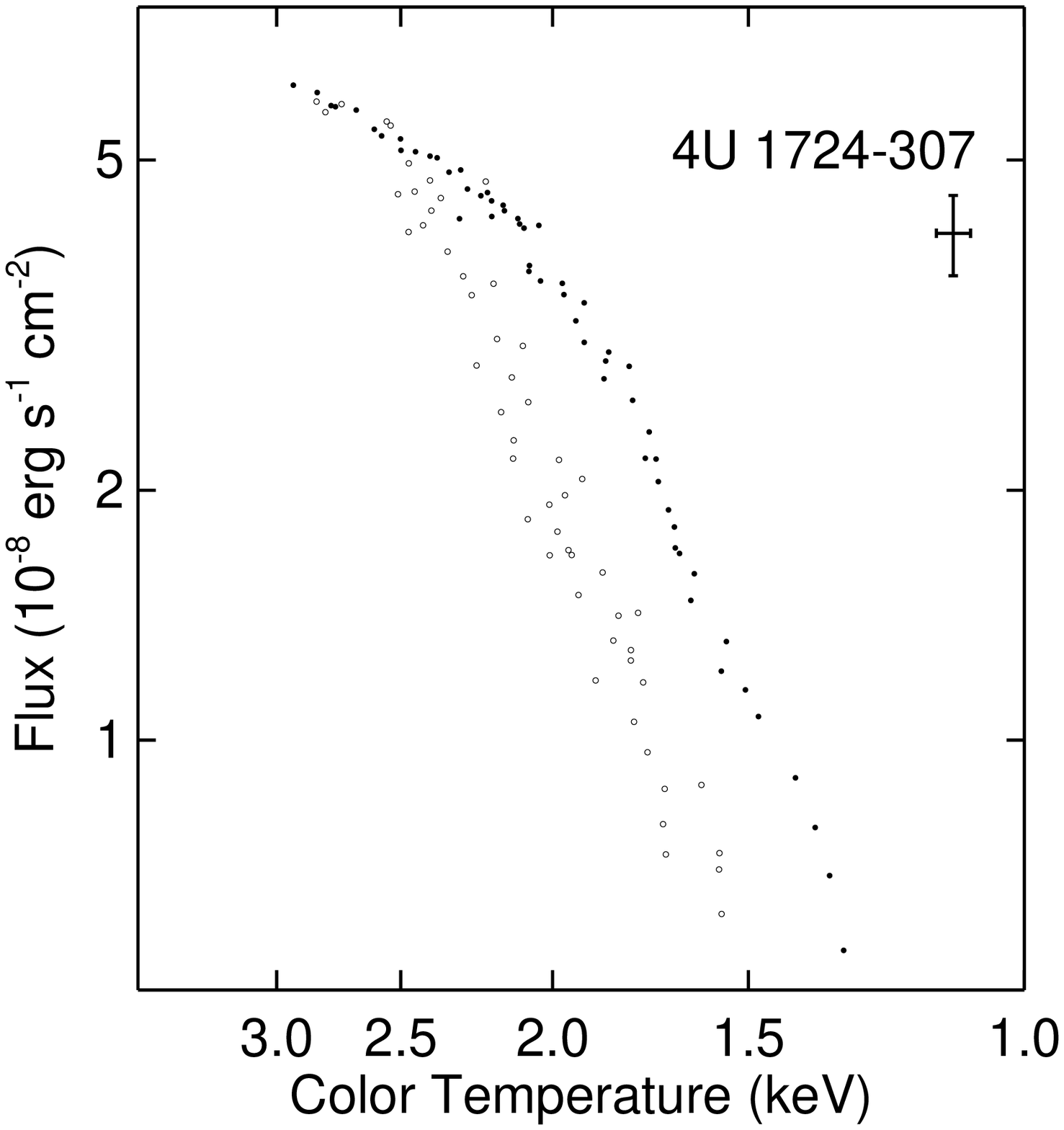}\\
   \caption{{\em (Left)\/}  The distribution of  $\chi^2$/dof obtained
     from fitting  the spectra in the  cooling tail of  one long X-ray
     burst  observed  from 4U~1724$-$307;  the  solid  line shows  the
     expected distribution for the  same number of degrees of freedom.
     {\em  (Right)} The  flux-temperature  diagram of  the long  burst
     observed from 4U~1724$-$307 (filled  circles) compared to that of
     the short bursts discussed in the main text (open circles).}
\label{fig:1724_appendix}
\end{figure*}

\section{4U~1702$-$429}

A total of 46 bursts have been observed from the source 4U~1702$-$429.
Six of  these bursts  reach high fluxes  and have been  categorized as
Photospheric Radius Expansion bursts  by Galloway et al.\ (2008a). The
remaining 40 bursts typically reach  lower fluxes. In the main body of
this paper,  we focused on  the 6 bright  bursts from this  source and
exclude  the  remaining  for  two  reasons that  we  explore  in  this
appendix.

The left  panel of Figure~\ref{fig:1702_appendix}  shows the histogram
of $\chi^2$/dof values for the 880 spectra observed during the cooling
tails of the 40 bursts  and compares them to the expected distribution
given the number  of degrees of freedom. It  is evident that blackbody
functions provide  statistically unacceptable fits to  the majority of
the spectra.

Had we not considered these  spectra unacceptable, we would find that,
approximately 4 seconds after the bursts start, the inferred blackbody
normalizations  increase rapidly  to  large values.   The increase  in
normalization  is,  in  fact,  correlated  with  an  increase  in  the
$\chi^{2}$/dof   values,   as   shown    in   the   right   panel   of
Figure~\ref{fig:1702_appendix},  rendering  them  even  more  suspect.
These two arguments  strongly suggest that the spectra  of the 40 weak
X-ray  bursts from  4U~1702$-$429  are not  dominated  by the  thermal
emission from the neutron star.

\section{4U~1724$-$307}

The first X-ray burst observed  from 4U~1724$-$307 on 1996 November 08
during observation  10090-01-01-02 is  very different compared  to the
other two X-ray bursts that we  used in the main text. The first burst
is unusually  long ($\simeq  120$~s), while the  other two  bursts are
much  shorter ($\simeq  15-20$~s range,  as  is typical  for the  vast
majority of  the Type-I bursts observed). Moreover,  while the spectra
of the  short bursts are  accurately modeled with  blackbody functions
(see top panels of  Fig.~\ref{fig:radii1}), fitting the spectra during
the  cooling tail  of the  long  burst results  in unacceptably  large
values  of  $\chi^2$/dof  and   the  distritubiton  of  the  resulting
$\chi^2$/dof values  do not follow the  expected $\chi^2$ distribution
(see  left panel of  Fig.~\ref{fig:1724_appendix}).  Adding  a certain
amount of systematic  uncertainty to the data can  in principle result
in a  decrease in the  individual $\chi^2$/dof values;  however,it can
not change the fact  that the resulting $\chi^2$/dof distribution does
not follow the expected distribution.

If we went ahead and further analyzed these statistically unacceptable
spectra, we would have obtained values for the blackbody normalization
that are  larger compared  to those of  the shorter bursts  (see right
panel of Fig.~\ref{fig:1724_appendix}).

The large values and the seemingly random distribution of $\chi^2$/dof
suggest that the  X-ray spectra in the cooling tail  of the long burst
from 4U~1724$-$307 are not dominated  by the thermal emission from the
neutron star. In fact, in't Zand \& Weinberg (2010) found evidence for
atomic edges and a reflection  component in the spectra of this burst,
which they attributed to the presence  of heavy metals in the ashes of
previous  bursts  that  were  exposed  by  the  long  radius-expansion
episode.  For these reasons, we excluded the long X-ray burst observed
from 4U~1724$-$307  from our analysis.   Note that Suleimanov  et al.\
(2011) chose  the spectra from this  burst in order to  infer the mass
and radius of  the neutron star in 4U~1724$-$307.  Had they chosen the
other  two bursts  from  the same  source,  the spectra  of which  are
actually  well  described  by  blackbody functions,  they  would  have
obtained a radius  of the neutron star that  is $\simeq 30$\% smaller,
making their result  consistent with the radii inferred  for the other
sources using this method (\"Ozel et al.\ 2009; G\"uver et al.\ 2010a,
2010b) as well as with  radii inferred from quiescent neutron stars in
globular clusters  (e.g., Webb \&  Barret 2007; Guillot,  Rutledge, \&
Brown 2010).

\section{4U~0513$-$40}

\begin{figure*}
\centering
   \includegraphics[scale=0.35]{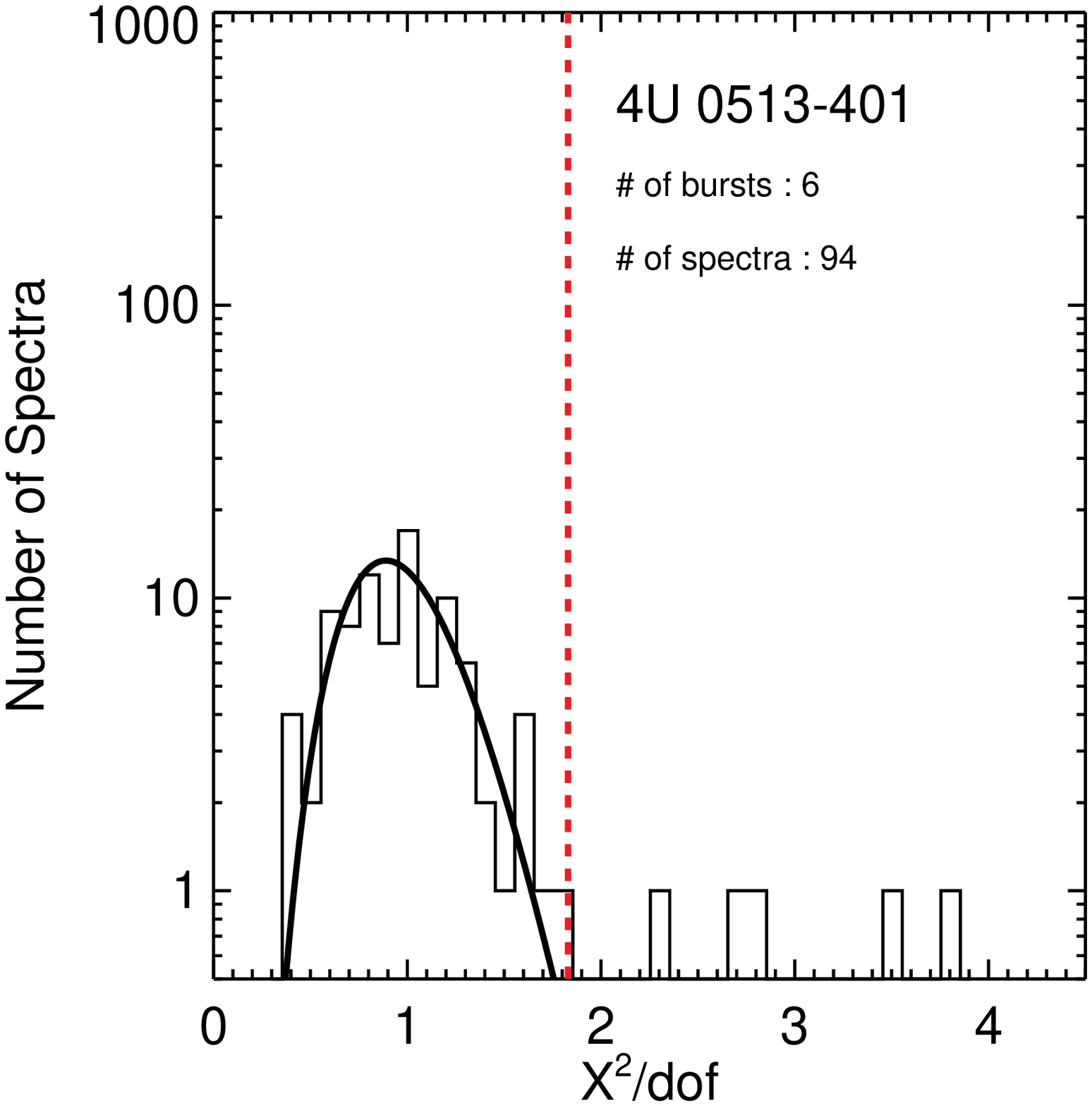}
   \includegraphics[scale=0.35]{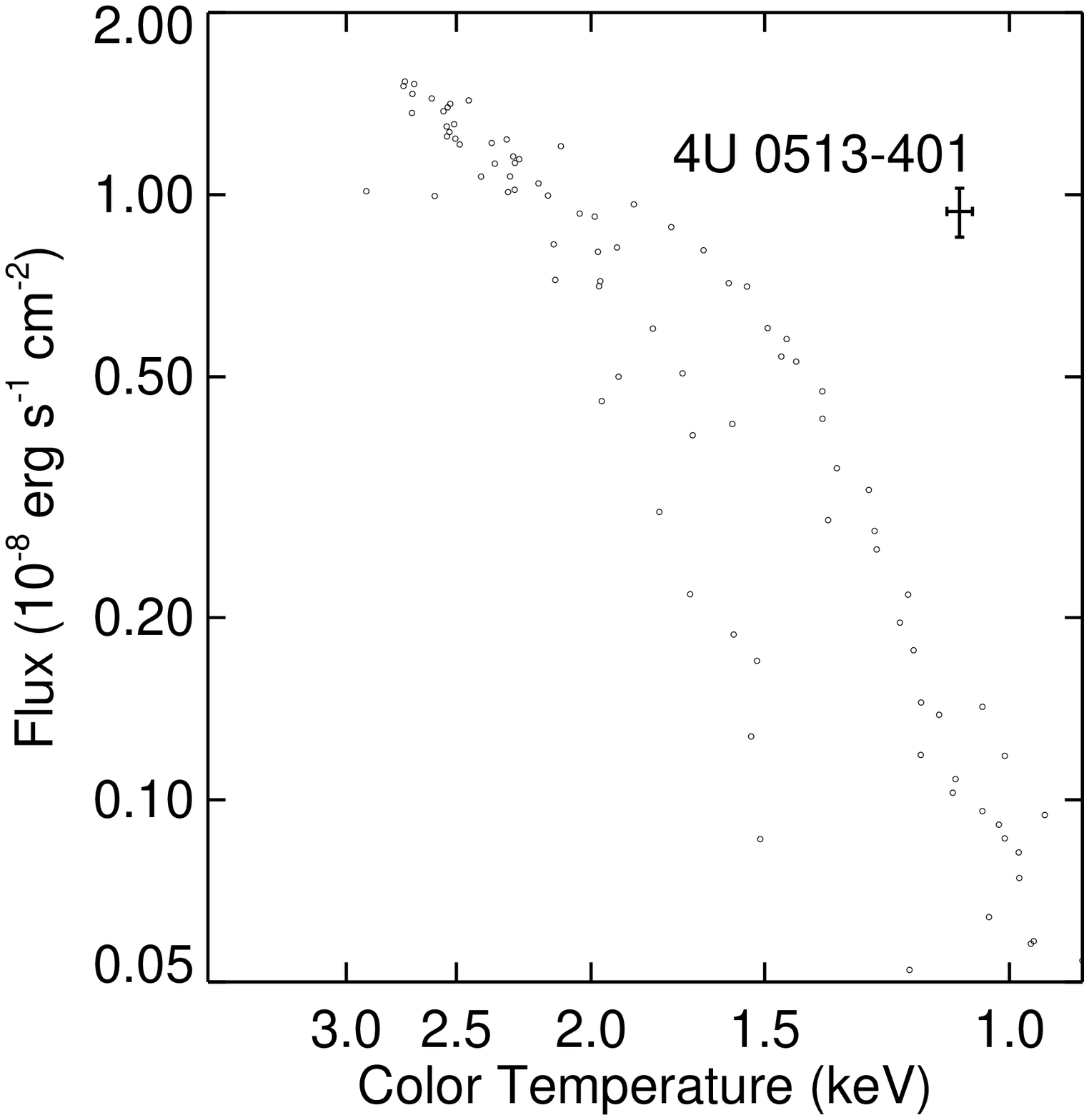}\\
   \caption{Same as Figure~\ref{fig:radii1} but for the
     source 4U~0513$-$401.}
\label{fig:0514}
\end{figure*}

Six bursts have been observed from the source 4U~0513$-$40. While most
of the X-ray  spectra during the cooling tails of  these bursts can be
described    well   by    blackbody   functions    (left    panel   of
Fig.~\ref{fig:0514}),  the  flux-temperature  diagrams show  irregular
behaviour for the cooling  tails (right panel of Fig.~\ref{fig:0514}).
In  particular,  two  of  the  bursts  show a  cooling  tail  that  is
reminiscent  of the  long  burst from  4U~1724$-$307 discussed  above:
early in the cooling phase  the temperature of the blackbody decreases
while the flux remains constant  and, when the flux starts decreasing,
the cooling  occurs with blackbody normalizations that  are on average
larger compared to the other bursts.

It is plausible that the spectra of these bursts from 4U~0513$-$40 are
also affected by the presence  of atomic lines, but the relatively low
flux  of 4U~0513$-$40  (which  is  a factor  of  $\simeq 3-5$  smaller
compared  to  that of  4U~1724$-$307  at  similar temperatures)  still
allows us to fit the observed spectra with blackbody functions.  We do
not consider  this source suitable  for a radius measurement  based on
the burst data.

\begin{figure*}
\centering
\includegraphics[scale=0.32]{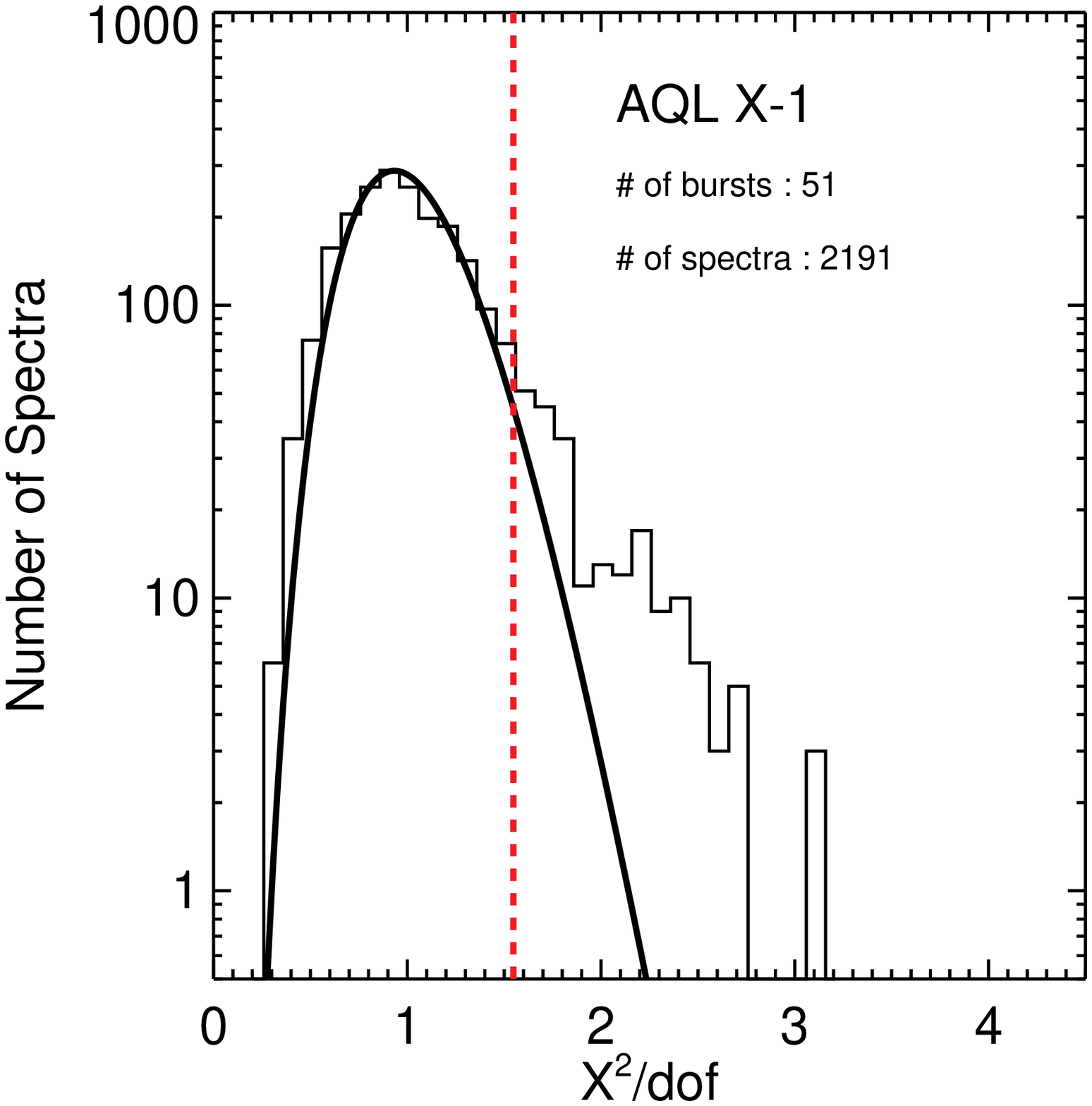}
\includegraphics[scale=0.32]{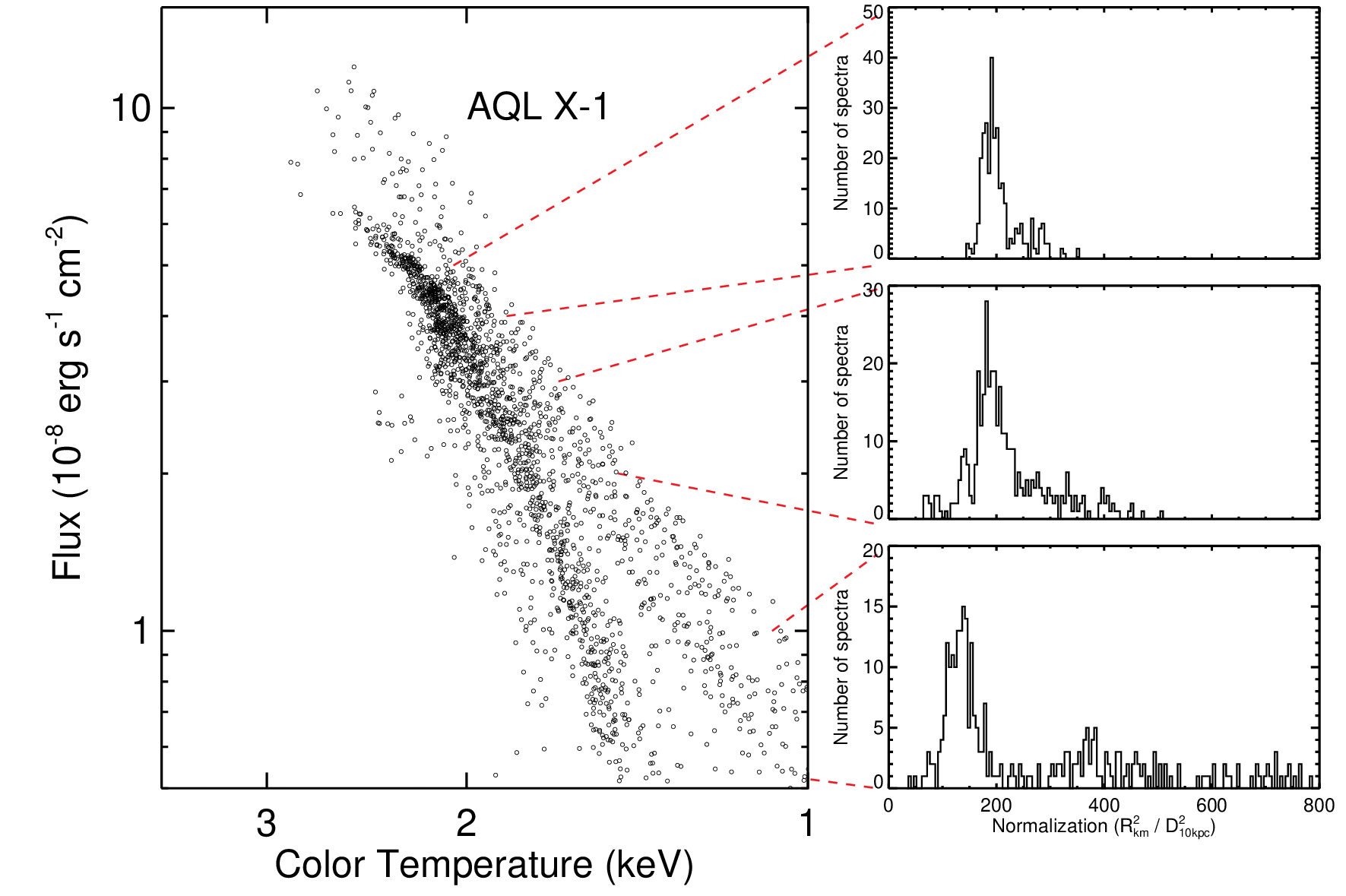}\\
   \caption{Same as Figure~\ref{fig:radii1} but for the
     source Aql~X-1.}
\label{fig:aqlx1}
\end{figure*}

\section{Aql~X-1}

The transient source Aql~X-1 has  shown 51 X-ray bursts, from which we
extracted  2191  spectra.  Most  of  these  spectra  are well  fit  by
blackbody functions, but their  cooling tracks in the flux-temperature
diagram  depend  strongly  on   the  properties  of  the  bursts  (see
Fig.~\ref{fig:aqlx1}).  In  particular, a  large number of  bursts are
relatively short and reach  modest fluxes (typically $\lesssim 5\times
10^{-8}$~erg~s$^{-1}$~cm$^{-2}$)   with    cooling   tracks   in   the
flux-temperature diagram  that are  reproducible.  On the  other hand,
many  bursts, which  are relatively  longer, reach  fluxes that  are a
factor  of  $\simeq  2$  higher,  are often  characterized  by  radius
expansion  episodes, and  follow  a  range of  cooling  tracks in  the
flux-temperature  diagram  with   blackbody  normalizations  that  are
typically larger than those of the other bursts.

One plausible  exlanation is  related to the  fraction of  the neutron
star surface  that is  engulfed by the  thermonuclear flash.   In weak
flashes, the burning  front may not propage across  the entire surface
and, therefore, only a fraction  of the stellar surface contributes to
the normalization  of the  blackbody. On the  other hand,  in stronger
flashes, a larger fraction of the neutron star surface is covered.

The source  Aql~X-1 is one of  the two main examples  (the other being
the source 4U~1636$-$536) discussed by Bhattacharrya et al.\ (2010) as
evidence  for  systematic  variations   in  the  burning  area  during
thermonuclear  X-ray  bursts.   It  is important  to  emphasize  here,
however,  that the case  of Aql~X-1  is very  unusual compared  to the
remaining sources  that we  analyzed and is,  perhaps, related  to the
presence of  a weak but  dynamically important magnetic field  in this
source, as  inferred from  the observation of  intermittent persistent
pulsations (Casella et al.\ 2008).

\end{document}